\documentclass[12pt, book, titlepage]{llncs}
\usepackage{llncsdoc}

\usepackage{color}
\usepackage[usenames,dvipsnames]{xcolor}

\usepackage{amsmath}
\usepackage{amssymb}
\usepackage{amsfonts}

\usepackage[cp1251]{inputenc}
\usepackage[T2A]{fontenc}

\usepackage[margin=4cm]{geometry}

\usepackage{listings}

\usepackage{caption}
\DeclareCaptionFont{black}{\color{black}}
\DeclareCaptionFormat{listing}{\rule{\textwidth}{0.5pt} \linebreak \parbox{\textwidth}{\hspace{15pt}#1#2#3}}
\usepackage{algorithm}
\usepackage{algorithmicx}
\usepackage[noend]{algpseudocode}
\newcommand*\Let[2]{\State #1 $\gets$ #2}

\usepackage{graphicx}
\graphicspath{{}{./figures/}}

\usepackage{caption}
\usepackage{subcaption}

\usepackage{url}
\usepackage{hyperref}
\hypersetup{colorlinks=true}

\usepackage{indentfirst}
\usepackage{nameref}

\usepackage{soul}

\usepackage{enumitem}

\usepackage{array}
\newcolumntype{x}[1]{>{\centering\arraybackslash\hspace{0pt}}m{#1}}

\begin{document}

\newgeometry{left=3cm,bottom=2cm,right=3cm,top=3cm}

\begin{titlepage}
\begin{center}

\includegraphics[height=0.13\textheight]{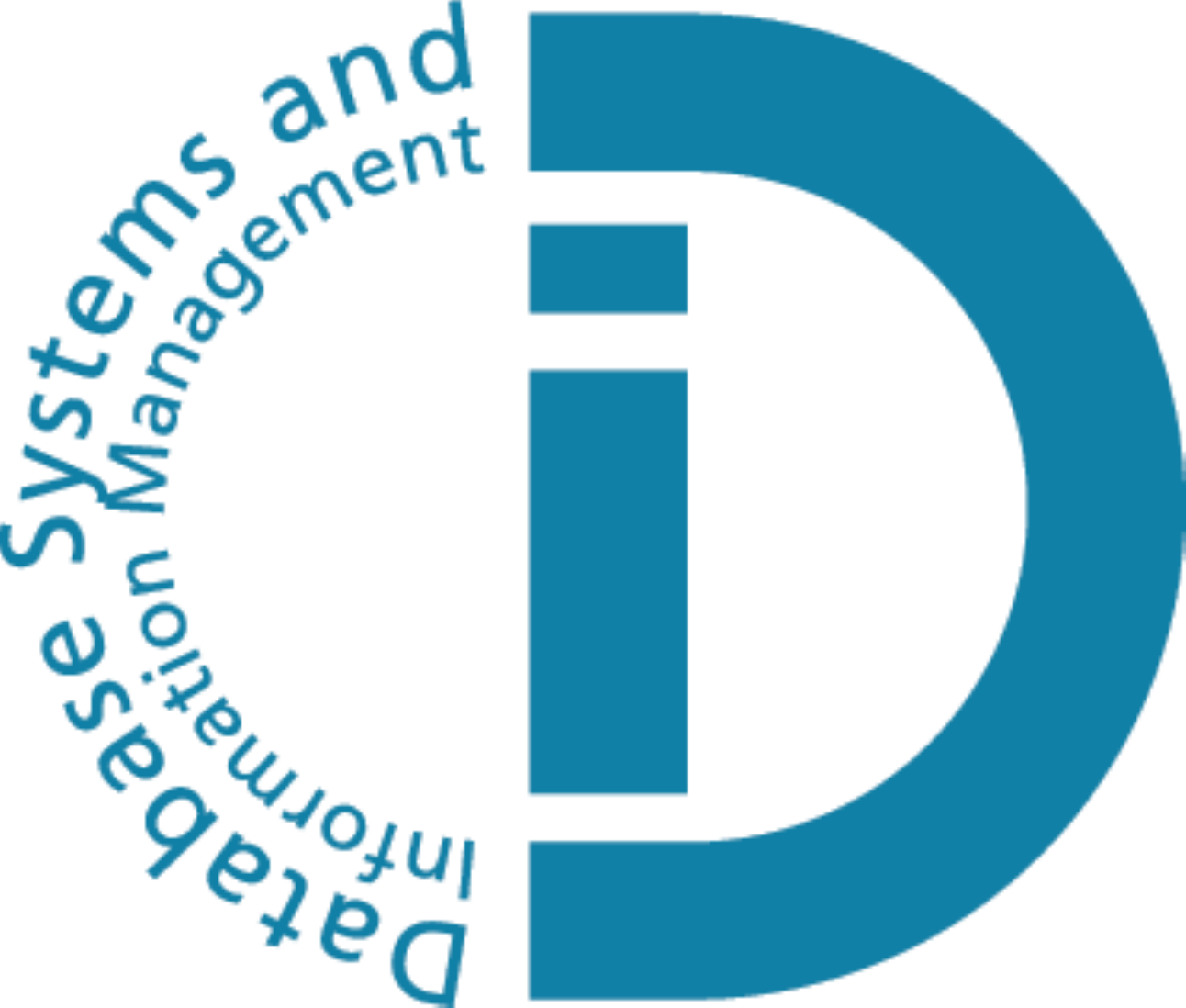}
\hfill
\includegraphics[height=0.15\textheight]{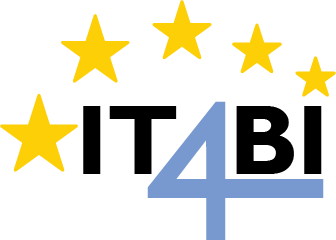}
\hfill
\includegraphics[height=0.13\textheight]{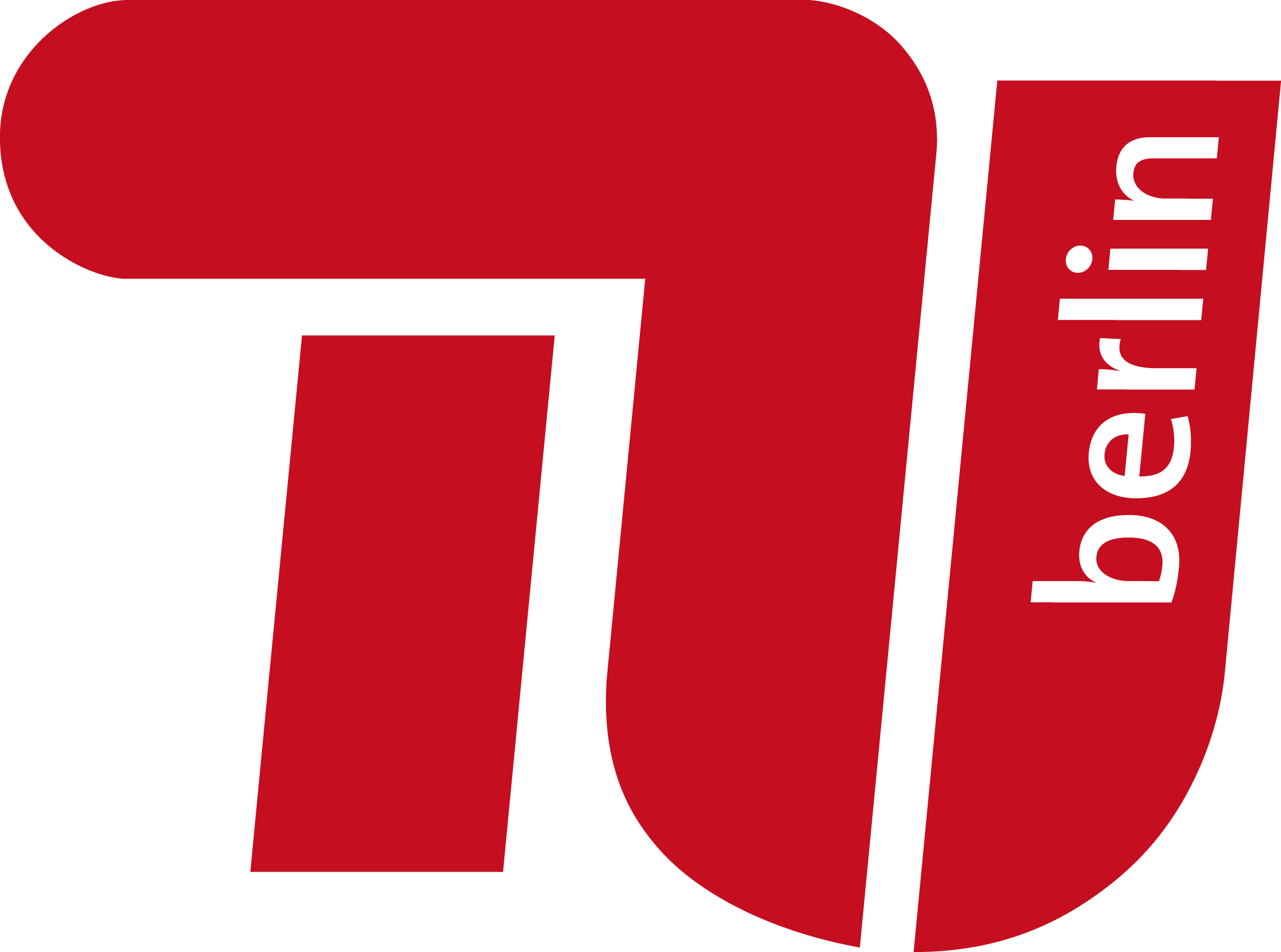}

\vspace*{1.5cm}

\LARGE
\textsc{Identifier Namespaces in Mathematical Notation}

\vspace{1cm}

\Large \textsc{Master Thesis}

\vspace{0.2cm}

by

\vspace{0.4cm}

\textbf{Alexey \textsc{Grigorev}}

\vspace{1.0cm}

\vfill

\large
Submitted to the Faculty IV, Electrical Engineering and Computer Science
Database Systems and Information Management Group
in partial fulfillment of the requirements for the degree of

\textbf{Master of Science in Computer Science}

as part of the \textsc{Erasmus Mundus} IT4BI programme

at the

\textsc{Technische Universit\"{a}t Berlin} \\

July 31, 2015

\vfill

\begin{flushright}
\normalsize
\emph{Thesis Advisors:}\\
Moritz \textsc{Schubotz} \\
Juan \textsc{Soto} \\

\ \\

\emph{Thesis Supervisor:}\\
Prof. Dr. Volker \textsc{Markl}\\
\end{flushright}

\end{center}
\end{titlepage}

\restoregeometry

\newpage

\ \\

\textbf{Eidesstattliche Erkl\"arung}

Ich erkl\"are an Eides statt, dass ich die vorliegende Arbeit selbstst\"andig
verfasst, andere als die angegebenen Quellen/Hilfsmittel nicht benutzt,
und die den benutzten Quellen w\"ortlich und inhaltlich entnommenen Stellen
als solche kenntlich gemacht habe.

\ \\

\textbf{Statutory Declaration}

I declare that I have authored this thesis independently, that I have not used
other than the declared sources/resources, and that I have explicitly marked
all material which has been quoted either literally or by content from the used sources.

\ \\ \ \\

\begin{flushright}

Berlin, July 31, 2015

\ \\

Alexey \textsc{Grigorev}

\end{flushright}

\newpage

\section*{Abstract}

In Computer Science, namespaces help to structure source code and
organize it into hierarchies. Initially, the concept of namespaces did not
exist for programming languages, and programmers had to manage the source code 
 themselves
to ensure there were  no name conflicts. However, nowadays, namespaces are
adopted by the majority of modern programming languages.

The concept of namespaces is beneficial for mathematics as well:
In mathematics, short one-symbol identifiers are very common,
and the meaning of these identifiers is hard to understand immediately.
By introducing namespaces to mathematics, we will be able to organize mathematical
identifiers. Also, mathematicians will also benefit from a hierarchical organization of
knowledge in the same way programmers do. In addition, the structure
will make it easier to understand the meaning of each identifier in a document.

In this thesis, we look at the problem of assigning each identifier of a document
to a namespace. At the moment, there does not exist a special dataset where
all identifiers are grouped 
 to namespaces, and therefore we need to create
such a dataset ourselves.

Namespaces are hard to prepare manually: building them requires a lot of time and effort.
However, it can be done automatically, and we propose
a method for automatic namespace discovery from a collection of documents.

To do that, we need to find groups of documents that use identifiers in the same way.
This can be done with cluster analysis methods.
We argue that documents can be represented 
by the identifiers they contain, and this
approach is similar to representing textual information in the Vector Space Model.
Because of this, we can apply traditional document clustering techniques for namespace
discovery.

To evaluate the results, we use the category information, and look for pure
``namespace-defining'' clusters: clusters where all documents are from the same category.
In the experiments, we look for algorithms that discover as many namespace-defining
clusters as possible.

Because the problem is new, there is no gold standard dataset, and it is
hard to evaluate the performance of our method. To overcome it, we first
use Java source code as a dataset for our experiments, since it contains the namespace
information. We verify that our method can partially recover namespaces
from source code using  only information about identifiers. 

The algorithms are evaluated on the English Wikipedia, and the proposed method
can extract namespaces on a variety of topics. After extraction, the namespaces
are organized into a hierarchical structure by using existing classification schemes
such as MSC, PACS and ACM.
We also apply it to the Russian Wikipedia, and the results are consistent across
the languages.

To our knowledge, the problem of introducing namespaces to mathematics has not
been studied before, and prior to our work there has been no dataset where identifiers
are grouped into namespaces. Thus, our result is not only a good start,
but also a good indicator that automatic namespace discovery is possible.

\section*{Zusammenfassung}
Namespaces in Informatik helfen um den Quellcode zu organisieren.
Namespaces sind auch f\"ur Mathematik vorteilhaft. Durch die Einfьhrung von Namespaces in der Mathematik, kann mann mathematische Identifikatoren in einer hierarchy organisieren.

In dieser Masterarbeit betrachten wir das Problem der Zuordnung jeder Identifikator eines Dokuments zu einem Namespaces. Im Moment gibt es keine spezielle Datenmenge in der alle Mathematische Identifikatoren zu Namespaces zugeordnet sind. Deshalb m\"ussen wir eine Datenmenge selbst erstellen. Wir schlagen eine automatisierte methode vor f\"ur die Entdeckung von der Zuordnung von Namespaces zu Identifikatoren aus einer Sammlung von Dokumenten.

Um dies zu erreichen, m\"ussen wir Gruppen von Dokumenten finden, die Identifikatoren in der gleichen Weise verwenden.
Das kann mit Clustering-Algorithmen durchgefьhrt werden. Wir schlagen vor dass Dokumente repr\"a\-sentiert werden k\"onnen durch die Identifikatoren die sie enthalten. Dieser Ansatz ist \"ahnlich zu der Abbildung von Textinformationen in dem Vektorraummodell.
Aus diesem Grund k\"onnen wir traditionelle Dokument Clustering-Algorithmen f\"ur Namespace Entdeckung gelten.

Um die Ergebnisse zu bewerten, verwenden wir die Kategorieinformationen, und suchen wir nach reinen ``namespace-definierenden'' Cluster: Cluster, in dem alle Dokumente zu der gleichen Kategorie geh\"oren.
In die Experimenten suchen wir nach Algorithmen, die so viele reine Cluster wie mцglich entdecken.

Die Algorithmen sind auf der Englischen Wikipedia ausgewertet.
Unsere Methode kann Namespaces auf einer Vielzahl von Themen extrahieren.
Nach der Extraktion, organisieren wir die Namespaces in einer Hierarchie.
Wir setzen unsere Methode auch auf der Russischen Wikipedia und finden \"anlische Ergebnisse.
Unser Ergebnis zeigt, dass die automatische Namespaces Entdeckung m\"oglich ist.

\section*{Acknowledgements}

This thesis addresses a topic that has not been studied previously, and
it was challenging, but extremely interesting and I learned a lot while working
on it. I would like to thank everybody who made it possible.

First, I would like to express my gratitude to my thesis advisor, Moritz Schubotz, who not
only introduced me to the topic of namespace discovery, but also guided me through
the thesis with  useful comments and enlightening discussions.

Secondly, I thank the IT4BI committee who selected me among other candidates
and allowed me to pursue this master's degree.
I thank all my teachers who gave me enough background to
successfully complete the thesis. I especially would like to thank
Dr. Ver\'onika Peralta  and Dr. Patrick Marcel, the teachers of
Information Retrieval course at Universit\'e Francois Rabelais,
Prof. Arnaud Giacometti, the teacher of Data Mining class at Universit\'e Francois Rabelais,
and finally, Prof. Klaus-Robert M\"uller, the teacher of Machine Learning
class at Technische Universit\"at Berlin.

I am also grateful to Yusuf Ameri and Juan Soto for their suggestions on improving 
the language of this work. 

Last, but not least, I would like to thank my wife for supporting me for the duration
of the master program.

\newpage
\setcounter{tocdepth}{2}
\tableofcontents
\newpage

\section{Introduction}

\subsection{Motivation}

In computer science, a \emph{namespace} refers to a collection of terms
that are grouped, because they share functionality or purpose,
typically for providing modularity and resolving name conflicts \cite{duval2002metadata}.
For example, XML uses namespaces to prefix element names to ensure uniqueness and remove ambiguity between them \cite{xmlnamespaces}, and the Java programming language uses packages to organize classes into namespaces for modularity \cite{gosling2014java}.

In this thesis, we extend the notion of namespaces to mathematical formulae. 
In mathematics, there is a convention of choosing identifier names in
\emph{mathematical notation} \cite{wikinotation}. Because of the notation, 
when people write ``$E=mc^2$'', the meaning of this expression is recognized among scientists. 
However, the same identifier may be used in different areas, but denote 
different things: For example, ``$E$\,'' may refer to ``energy'', ``expected value'' or 
``elimination matrix'', depending on the domain of the article.
We can compare this problem with the problem of name collision in computer science 
and introduce namespaces of identifiers in mathematical notation to overcome it. 

In this work we aim to discover namespaces of identifiers in mathematical notation. 
However, the notation only exists in the documents where it is used, and it does 
not exist in isolation. 
It means that the identifer namespaces should be discovered from the documents 
with mathematical formulae. 
Therefore, the goal of this work is to \textbf{automatically discover a set of identifier
namespaces given a collection of documents}.


We expect the namespaces to be meaningful, in the sense that they can be related to real-world areas of knowledge, such as physics, linear algebra or statistics.

Once such namespaces are found, they can give good categorization of scientific 
documents based on formulas and notation used in them. We believe that this may 
facilitate better user experience: when learning a new area it will help the users 
familiarize with the notation faster.
Additionally, it may also help to locate the usages of a particular identifier and 
refer to other documents where the identifier is used. 

Namespaces also give a way to avoid ambiguity. If we refer to an identifier from a 
particular namespace, then it is clear what the semantic meaning of this 
identifier. For example, if we say that ``$E$\,'' belongs to a namespaces 
about physics, it gives additional context and makes it clear that ``$E$\,''
means ``energy'' rather than ``expected value''.

Finally, using namespaces is beneficial for relating identifiers to 
definitions. Thus, as an application of namespaces, we can use them for better 
definition extraction. It will help to overcome some of the current problems in 
this area, for example, the problem of dangling identifiers~-- identifiers that are 
used in  formulas but never defined in the document \cite{pagael2014mlp}. 
Such identifiers may be defined in other documents that share the same namespace, 
and thus we can take the definition from the namespace and assign it to the dangling identifier.

\subsection{Thesis Outline}

The thesis is organizes as follows: 

\begin{description}
\item[Chapter \ref{sec:background} -- \nameref{sec:background}] \hfill \\
  In this chapter, we do a survey of the related work. We discuss how namespaces are
  used in Computer Science. Secondly, we review how definitions for identifiers 
  used in mathematical formulae can be extracted from the natural language text
  around the formulae. Finally, we review the Vector Space Model~-- a way of 
  transforming texts to vectors, and then discuss how these vectors can be 
  clustered. 

\item[Chapter \ref{sec:namespace-discovery-chap} -- \nameref{sec:namespace-discovery-chap}] \hfill \\
  In chapter \ref{sec:namespace-discovery-chap} we introduce the problem of namespaces 
  in mathematical notation, discuss its similarities with namespaces in Computer Science,
  and propose an approach to namespace discovery by using document clustering techniques. 
  We also extend the Vector Space Model to represent identifiers and suggest several 
  ways to incorporate definition information to the vector space. 

\item[Chapter \ref{sec:implementation} -- \nameref{sec:implementation}] \hfill \\
  Chapter \ref{sec:implementation} describes how the proposed approach is implemented. 
  It includes the description of the data sets for our experiments, and the details of 
  implementation of definition extraction and document cluster analysis algorithms. 
  Additionally, we propose a way of converting document clusters to namespaces.

\item[Chapter \ref{sec:evaluation} -- \nameref{sec:evaluation}] \hfill \\
  In chapter \ref{sec:evaluation}, we describe the parameter selection procedure,
  and we present the results of the best performing method. Once the clusters are 
  discovered, they are mapped to a hierarchy, and we summarize our findings 
  by analyzing the most frequent namespaces and most frequent identifier-definition
  relations in these namespaces.

\item[Chapter \ref{sec:conclusions} -- \nameref{sec:conclusions}] \hfill \\
  Chapter \ref{sec:conclusions} summarizes the findings.

\item[Chapter \ref{sec:future-work} -- \nameref{sec:future-work}] \hfill \\
  Finally, in chapter \ref{sec:future-work} we discuss the possible areas of improvements.
  We conclude this chapter by identifying the questions that are not resolved and 
  present challenges for future research on identifier namespace discovery.

\end{description}

\section{Background and Related Work} \label{sec:background}

In this chapter we explain the related work: we give a short overview 
about existing approaches and relevant methods. 

First, we describe the concept of a namespace from the Computer Science point
of view in section~\ref{sec:namespaces-cs}. Then we discuss how identifier 
definitions can be extracted, and for that we first introduce Part-of-Speech 
Tagging and its application to mathematical texts in section~\ref{sec:postagging},
and then review the extraction methods in section~\ref{sec:definition-extraction-methods}.

Next, section~\ref{sec:vsm}  describes the Vector Space Model, a traditional
way of representing a collection of documents as vectors, and then 
section~\ref{sec:similarity-distance} reviews common similarity and distance 
functions that are useful for document clustering. 

Finally, document clustering techniques are described in section~\ref{sec:doc-clustering},
and the Latent Semantic Analysis method for revealing semantic information from the 
document corpus is in section~\ref{sec:lsa}.

\subsection{Namespaces in Computer Science} \label{sec:namespaces-cs}

In computer science, a \emph{namespace} refers to a collection of terms
that are grouped because they share functionality or purpose,
typically for providing modularity
and resolving name conflicts \cite{duval2002metadata}.

\textbf{Example 1.} Namespaces are used in XML (eXtensible Markup Language), which
is a framework for defining markup languages \cite{moller2006introduction}.
However, different XML languages may use the same names for elements and attributes.
For example, consider two XML languages: XHTML for specifying the layout of web
pages, and some XML language for describing furniture. Both these languages have
the \verb|<table>| elements there, in XHTML table is used to present some data in
a tabular form, while the second one uses it to describe a particular piece of
furniture in the database.

The \verb|<table>| elements have very different semantics in these languages
and there should be a way to distinguish between these two elements.
In XML this problem is solved with XML namespaces~\cite{xmlnamespaces}:
the namespaces are used to ensure the uniqueness of attributes and resolve ambiguity.
It is done by binding a short namespace alias with some uniquely defined URI
(Unified Resource Identifier), and then appending the alias to
all attribute names that come from this namespace. In the example above,
we can bind an alias \verb|h| with XHTML's URI \url{http://www.w3.org/TR/xhtml1}
and then use \verb|<h:table>| to refer to XHTML's table. Likewise,
in the furniture database language the element names can be prepended
with a prefix \verb|d|, where \verb|d| is bound to some URI, e.g.
\url{http://www.furniture.de/2015/db}.


\textbf{Example 2.} Namespaces are also used in programming languages for organizing
variables, procedures and other identifiers into groups and
for resolving name collisions. In programming languages without
namespaces the programmers have to take special care to avoid
naming conflicts. For example, in the PHP programming language
prior to version 5.3 \cite{mcarthur2008php6} there is no notion of namespace, and
the namespaces have to be emulated to ensure that the names
are unique, and 
\verb|Zend_Search_Lucene_Analysis_Analyzer|\footnote{\url{http://framework.zend.com/apidoc/1.7/Zend_Search_Lucene/Analysis/Zend_Search_Lucene_Analysis_Analyzer.html}}
and other long names is the result.

Other programming languages have the notion of namespaces built in
from the very first versions. For example, the Java programming
language~\cite{gosling2014java} uses packages to organize identifiers into
namespaces, and packages solve the problem of ambiguity. For example,
in the standard Java API there are two classes with the name \texttt{Date}:
one in the package \texttt{java.util} and another in the package \texttt{java.sql}.
To be able to distinguish between them, the classes are referred by their
\emph{fully qualified name}: an unambiguous name that uniquely specifies the class
by combining the package name with the class name. Thus, to refer to a particular
\texttt{Date} class in Java  \texttt{java.util.Date} or  \texttt{java.sql.Date}
should be used.

It is not always convenient to use the fully qualified name in the code to
refer to some class from another package. Therefore in Java it is possible to
\emph{import} the class by using the import statement which associates
a short name alias with its fully qualified name.
For example, to refer to \texttt{java.sql.Date} it is possible to import
it by using \texttt{import java.sql.Date} and then refer to it by the alias
\texttt{Date} in the class \cite{gosling2014java}.

Although there are no strict rules about how to organize the classes into
packages, it is a good software design practice to put
related objects into the same namespace and by doing this achieve
better modularity. There are design principles that tell software engineers
how to best organize the source code: classes in a well designed system
should be grouped in such a way that namespaces
exhibit low \emph{coupling} and high \emph{cohesion}~\cite{larman2005applying}.
Coupling describes the degree of dependence between namespaces, and
low coupling means that the interaction between classes of different
namespaces should be as low as possible. Cohesion, on the other hand,
refers to the dependence within the classes of the same namespace,
and the high cohesion principle says that the related classes
should all be put together in the same namespace.

\subsection{Math-aware POS tagging} \label{sec:postagging}
Part-of-Speech Tagging (POS Tagging) is a typical Natural Language Processing
task which assigns a POS Tag to each word in a given text \cite{jurafsky2000speech}.
While the POS Tagging task is mainly a tool for text processing, it can
also be applicable to scientific documents with mathematical expressions,
and can be adjusted to dealing with formulae \cite{schoneberg2014pos}
\cite{pagael2014mlp}.

A \emph{POS tag} is an abbreviation that corresponds to some
part of speech. Penn Treebank POS Scheme \cite{santorini1990part} is
a commonly used POS tagging scheme which defines a set of part-of-speech tags
for annotating English words.
For example, \texttt{JJ} is an adjective (``big''), \texttt{RB} as in adverb,
\texttt{DT} is a determiner (``a'', ``the''), \texttt{NN} is a
noun (``corpus'') and \texttt{SYM} is used for symbols (``$>$'', ``$=$'').

However the Penn Treebank scheme does not have special tags for mathematics,
but it is flexible enough and can be extended to include additional tags.
For example, we can include a math-related tag \texttt{MATH}.
Usually it is done by first applying traditional POS taggers (like Stanford
CoreNLP \cite{manning2014stanford}), and then
refining the results by re-tagging math-related tokens of text as \texttt{MATH}
\cite{schoneberg2014pos}.

For example, consider the following sentence:
``The relation between energy and mass is
described by  the mass-energy equivalence formula $E = mc^2$,
where $E$ is energy, $m$ is mass and $c$ is the speed of light''.
In this case we will assign the \verb|MATH| tag to ``$E = mc^2$'', ``$E$'',
``$m$'' and ``$c$''

However we can note that for finding identifier-definition relations
the \texttt{MATH} tag alone is not sufficient: we need to distinguish
between complex mathematical expressions and stand-alone identifiers -
mathematical expressions that contain only one symbol: the identifier.
For the example above we would like to be able to distinguish the
expression ``$E = mc^2$'' from identifier tokens ``$E$'',
``$m$'' and ``$c$''. Thus we extend the Penn Treebank scheme even more
and introduce an additional tag \texttt{ID} to denote stand-alone identifiers.

Thus, in the example above ``$E = mc^2$'' will be assigned the \texttt{MATH} tag
and ``$E$'', ``$m$'' and ``$c$'' will be annotated with \texttt{ID}.

In the next section we discuss how this can be used to find identifier-definition
relations.

\subsection{Mathematical Definition Extraction} \label{sec:definition-extraction-methods}

In Natural Language Processing, Word Sense Disambiguation is a problem of
identifying in which sense a polysemous word is used \cite{jurafsky2000speech}.
Analogously, the Identifier Disambiguation problem is a problem of
determining the meaning of an identifier in a mathematical formula. This
problem is typically solved by extracting definitions from the natural
language description that surrounds the formula.

For example, given the sentence ``The relation between energy and mass is
described by  the mass-energy equivalence formula $E = mc^2$,
where $E$ is energy, $m$ is mass and $c$ is the speed of
light''\footnote{\url{https://en.wikipedia.org/wiki/Mass\%E2\%80\%93energy\_equivalence}}
the goal is to extract the following identifier-definition relations:
($E$, ``energy''), ($m$, ``mass'') and ($c$, ``the speed of light'').

Formally, a phrase that defines a mathematical expression consists of three parts \cite{kristianto2012extracting}:

\begin{itemize}
\itemsep1pt\parskip0pt\parsep0pt
  \item \emph{definiendum} is the term to be defined: it is a mathematical expression
      or an identifier;
  \item \emph{definiens} is the definition itself: it is the word or phrase that defines the definiendum in a definition;
  \item \emph{definitor} is a relator verb that links definiendum and definiens.
\end{itemize}

In this work, we are interested in the first two parts: \emph{definiendum} and
\emph{definiens}. Thus we define a \emph{relation} as a pair
(definiendum, definiens). For example, ($E$, ``energy'') is a relation where
``$E$'' is a definiendum, and ``energy'' is a definiens. We refer to definiendum as
identifier, and to definiens as definition, so relations are identifier-definition
pairs.

There are several ways of extracting the identifier-definition relations.
Here we will review the following:

\begin{itemize}
\itemsep1pt\parskip0pt\parsep0pt
  \item Nearest Noun
  \item Pattern Matching
  \item Machine-Learning based methods
  \item Probabilistic methods
\end{itemize}

\subsubsection{Nearest Noun Method} \ \\

The Nearest Noun  \cite{grigore2009towards} \cite{yokoi2011contextual}
is the simplest definition extraction method.
It assumes that the definition is a combination of ad
It finds definitions by looking for combinations of adjectives and nouns
(sometimes preceded by determiners) in the text before the identifier.

I.e. if we see a token annotated with \texttt{ID}, and then a sequence
consisting only of adjectives (\texttt{JJ}), nouns (\texttt{NN}, \texttt{NNS})
and determiners (\texttt{DET}), then we say that this sequence is
the definition for the identifer.

For example, given the sentence ``In other words, the bijection $\sigma$ normalizes
$G$ in ...'' we will extract a relation ($\sigma$, "bijection").

\subsubsection{Pattern Matching Methods} \label{sec:pattern-matching} \ \\

The Pattern Matching method \cite{quoc2010mining} is an extension of the
Nearest Noun method: In Nearest Noun, we are looking for one specific patten
where identifier is followed by the definition, but we can define several such
patterns and use them to extract definitions.

For example, we can define the following patterns:

\begin{itemize}
\itemsep1pt\parskip0pt\parsep0pt
  \item \texttt{IDE} \texttt{DEF}
  \item \texttt{DEF} \texttt{IDE}
  \item let$|$set \texttt{IDE} denote$|$denotes$|$be \texttt{DEF}
  \item \texttt{DEF} is$|$are denoted$|$\texttt{def}ined$|$given as$|$by \texttt{IDE}
  \item \texttt{IDE} denotes$|$denote$|$stand$|$stands as$|$by \texttt{DEF}
  \item \texttt{IDE} is$|$are \texttt{DEF}
  \item \texttt{DEF} is$|$are \texttt{IDE}
  \item and many others
\end{itemize}

In this method \texttt{IDE} and \texttt{DEF} are placeholders that are
assigned a value when the pattern is matched against some subsequence
of tokens.  \texttt{IDE} and DEF need to
satisfy certain criteria in order to be successfully matched: like in the
Nearest Noun method we assume that \texttt{IDE} is some token annotated with
\texttt{ID} and \texttt{DEF} is a phrase containing adjective (\texttt{JJ}),
nouns (\texttt{NN}) and  determiners (\texttt{DET}). Note that the first patten corresponds
to the Nearest Noun pattern.

The patterns above are combined from two lists: one is extracted from a
guide to writing mathematical papers in English (\cite{trzeciak1995writing}),
and another is extracted from ``Graphs and Combinatorics'' papers from Springer
\cite{kristianto2012extracting}.

The pattern matching method is often used as the baseline method
for identifier-definition extraction methods \cite{kristianto2012extracting}
\cite{kristianto2014extracting} \cite{pagael2014mlp}.

\subsubsection{Machine Learning Based Methods}

\ \\

The definition extraction problem can be formulated as a binary classification
problem: given a pair (identifier, candidate-definition), does this pair
correspond to a real identifier-definition relation?

To do this we find all candidate pairs: identifiers are tokens
annotated with \texttt{ID}, and candidate defections are nouns and
noun phrases from the same sentence as the definition.

Once the candidate pairs are found, we extract the following features
\cite{yokoi2011contextual} \cite{kristianto2014extracting}:

\begin{itemize}
\itemsep1pt\parskip0pt\parsep0pt
  \item boolean features for each of the patterns from
    section~\ref{sec:pattern-matching} indicating if the pattern is matched,
  \item indicator if there's a colon or comma between candidate and identifier,
  \item indicator if there's another math expression between candidate and identifier,
  \item indicator if candidate is inside parentheses and identifier is outside,
  \item distance (in words) between the identifier and the candidate,
  \item the position of candidate relative to identifier,
  \item text and POS tag of one/two/three preceding and following tokens around the candidate,
  \item text of the first verb between candidate and identifier,
  \item many others.
\end{itemize}

Once the features are extracted, a binary classifier can be trained to predict
if an unseen candidate pair is a relation or not.
For this task the popular choices of classifiers are Support Vector Machine
classifier with linear kernel \cite{kristianto2014extracting} \cite{yokoi2011contextual}
and Conditional Random Fields \cite{kristianto2014extracting},
but, in principle, any other binary classifier can be applied
as well.

\subsubsection{Probabilistic Approaches} \label{sec:mlp} \ \\

In the Mathematical Language Processing approach \cite{pagael2014mlp}
a definition for an identifier is extracted by ranking
candidate definitions by the probability of definining
the identifier, and only the most probable candidates are retained.

The main idea of this approach is that the definitions occur very closely
to identifiers in sentences, and the closeness can be used to
model the probability distribution over candidate definitions.

The candidates are ranked by the following formula:
$$R(n, \Delta, t, d) = \cfrac{\alpha \, R_{\sigma_d}(\Delta) + \beta \, R_{\sigma_s}(n) + \gamma \, \text{tf}(t)}{\alpha + \beta + \gamma}$$
where $\Delta$ is the number of tokens between the identifier and the definition
candidate, and $R_{\sigma_d}(\Delta)$ is a Gaussian that models this distance, parametrized
with $\sigma_d$; $n$ is the number of sentences between the definition candidate
and the sentence where the identifier occurs for the first time, and is a Gaussian
parameterized with $\sigma_s$; finally $\text{tf}(t)$ is a frequency of term $t$
in a sentence. All these quantities are combined together and $\alpha, \beta, \gamma$ are weighting parameters.

The following weighting parameters $\alpha, \beta, \gamma$ are proposed in
\cite{pagael2014mlp}: $\alpha = \beta = 1$ and $\gamma = 0.1$.

\subsection{Vector Space Model} \label{sec:vsm}

Vector Space Model is a statistical model for representing documents
in some vector space. It is an Information Retrieval
model \cite{manning2008introduction}, but it is also used for various
Text Mining tasks such as Document Classification \cite{sebastiani2002machine}
and Document Clustering \cite{oikonomakou2005review} \cite{aggarwal2012survey}.

In Vector Space Model we make two assumptions about the data:
(1) \emph{Bag of Words assumption}: the order of words is not important,
only word counts;
(2) \emph{Independence assumption}: we treat all words as independent.
Both assumptions are quite strong, but nonetheless this method often
gives good results.

\begin{figure}[h]
\centering\includegraphics[width=0.5\textwidth]{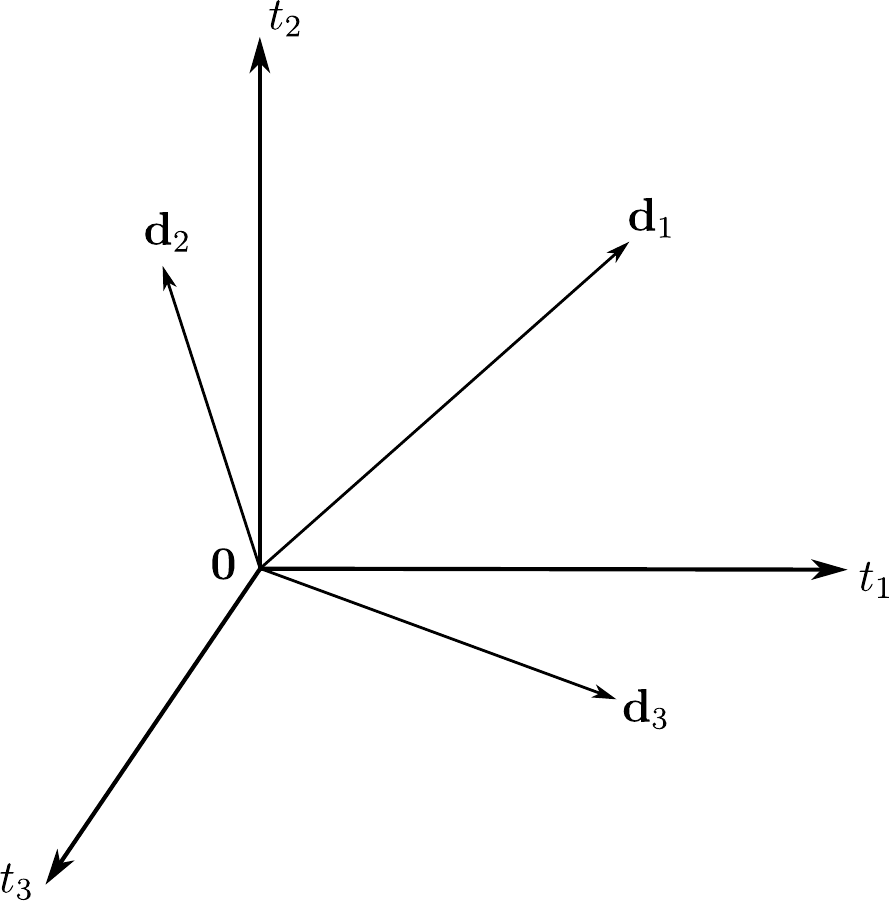}
\caption{Documents $\mathbf d_1, \mathbf d_2, \mathbf d_3$
in a document space with dimensions $t_1, t_2, t_3$.}
\label{fig:document-vsm}
\end{figure}

Let $\mathcal V = \{t_1, t_2, \ ... \ , t_m \}$ be a set of $n$ terms.
Then documents can be represented as $m$-vectors
$\mathbf d_i = (w_1, w_2, \ ... \ , w_m)$, where $w_j$ is the weight
of term $t_j$ in the document $\mathbf d_i$,
and the document collection can be represented by a \emph{term-document matrix}
$D$, where columns of $D$ are document vectors
$\mathbf d_1, \mathbf d_2, \ ... \ , \mathbf d_n$
and rows of $D$ are indexed by terms $t_1, t_2, \ ... \ , t_m$
(see fig.~\ref{fig:document-vsm}).

There are the following term weighting schemes \cite{manning2008introduction}:

\begin{itemize}
\itemsep1pt\parskip0pt\parsep0pt
  \item binary: 1 if a term is present, 0 otherwise;
  \item term frequency (TF): number of occurrences of the term in a document;
  \item document frequency (DF): number of documents containing the terml
  \item TF-IDF: combination of TF and inverse DF.
\end{itemize}

\textbf{Term Frequency (TF)} weights terms by local frequency in the document.
That is, the term is weighed by how many times it occurs in the document.
Sometimes a term is used too often in a document, and we want to
reduce its influence, and this is typically done by applying some
sublinear transformation to TF, for instance, a square root or a logarithms.

\textbf{Document Frequency (DF)} weights terms by their global frequency
in the collection, which is the number of documents that contain the token.
But more often we are interested in domain specific words than in neutral words,
and these domain specific words tent to occur less frequently and they usually
have more discriminative power: that is, they are better in telling one document apart from another. So we use \textbf{Inverse Document Frequency (IDF)} to give more
weight to rare words rather than to frequent words.

A good weighting system gives the best performance when it assigns
more weights to terms with high TF, but low DF \cite{salton1988term}.
This can be achieved by combining both TF and IDF
schemes. Usually a sublinear TF is used to avoid the dominating effect of
words that occur too frequently. As the result, terms appearing
too rarely or too frequently are ranked low.
The TF and IDF are combined together in \textbf{TF-IDF} weighting scheme:
$$\text{tf-idf}(t, \mathbf d) = (1 + \log \text{tf}(t, \mathbf d)) \cdot \log \cfrac{n}{\text{df}(t)} \, ,$$
where $\text{tf}(t, \mathbf d)$ is term frequency of term $t$ in document
$\mathbf d$ and $\text{df}(t)$ is the document frequency of term $t$ in
the document collection.

\subsection{Similarity Measures and Distances} \label{sec:similarity-distance}

Once the documents are represented in some vector space, we need to
define how to compare these documents to each other. There are two
ways of doing this: using a similarity function that computes how similar
two objects are (the higher values, the more similar the objects),
or using a distance function, sometimes called ``dissimilarity function'',
which is the opposite of similarity (the higher the values, the less similar
the objects).

We consider Euclidean distance, inner product, cosine similarity and
Jaccard coefficient.

\subsubsection{Euclidean Distance} \ \\

The Euclidean distance function (also called length or $L_2$ norm) 
is the most commonly used distance function in vector spaces. 
Euclidean distance corresponds to the geometric distance between two data 
points in the vector space. Let $\mathbf x, \mathbf y \in \mathbb R^n$, then 
the Euclidean distance between $\mathbf x$ and $\mathbf y$ is defined 
as $\| \mathbf x - \mathbf y \| = \sqrt{\sum_{i = 1}^n (x_i - y_i)^2}$.

This distance is useful for low-dimensional data, but it does not always work
well in high dimensions, especially with sparse vector such as
document vectors \cite{ertoz2003finding}, and this effect is often called 
``Curse of Dimensionality'' \cite{beyer1999nearest}.

\subsubsection{Inner product} \ \\

The inner product between two vectors can be used as a similarity function:
the more similar two vectors are, the larger is their inner product.
Geometrically the inner product between two vectors $\mathbf x$ and $\mathbf y$
is defined as
$\mathbf x^T \mathbf y = \|\mathbf x \| \, \| \mathbf y \| \, \cos \theta$
where $\theta$ is the angle between vectors $\mathbf x$ and $\mathbf y$.
In Linear Algebra, however, the inner product
is defined as a sum of element-wise products of two vectors:
given two vectors $\mathbf x$ and $\mathbf y$, the inner product is
$\mathbf x^T \mathbf y = \sum_{i = 1}^n x_i \, y_i$ where $x_i$ and $y_i$
are $i$th elements of $\mathbf x$ and $\mathbf y$, respectively.
The geometric and algebraic definitions are equivalent \cite{huges2013calculus}.

\subsubsection{Cosine Similarity} \label{sec:cosine} \ \\


Inner product is sensitive to the length of vectors, and thus
it may make sense to consider only the angle between them:
the angle does not depend on the magnitude, but it is still
a very good indicator of vectors being similar or not.

The angle between two vectors can be calculated from the geometric
definition of inner product:
$\mathbf x^T \mathbf y = \|\mathbf x \| \, \| \mathbf y \| \, \cos \theta$.
By rearranging the terms we get
$\cos \theta = \mathbf x^T \mathbf y \, / \, (\|\mathbf x \| \, \| \mathbf y \|)$.

We do not need the angle itself and can use the cosine directly
\cite{manning2008introduction}.
Thus can define \emph{cosine similarity} between two documents $\mathbf d_1$ and
$\mathbf d_2$ as
$$\text{cosine}(\mathbf d_1, \mathbf d_2) = \cfrac{\mathbf d_1^T \mathbf d_2}{\|\mathbf d_1 \| \, \| \mathbf d_2 \|} \ .$$
If the documents have unit lengths, then cosine similarity is the same as
dot product: $\text{cosine}(\mathbf d_1, \mathbf d_2) = \mathbf d_1^T \mathbf d_2$.

The cosine similarity can be converted to a distance function.
The maximal possible cosine is 1 for two identical documents.
Therefore we can define \emph{cosine distance} between two vectors
$\mathbf d_1$ and $\mathbf d_2$ as
$d_c(\mathbf d_1, \mathbf d_2) = 1 - \text{cosine}(\mathbf d_1, \mathbf d_2)$.
The cosine distance is not a proper metric \cite{korenius2007principal},
but it is nonetheless useful.

The cosine distance and the Euclidean distance are connected \cite{korenius2007principal}.
For two unit-normalized vectors $\mathbf d_1$ and $\mathbf d_2$ the Euclidean distance
between them is $\| \mathbf d_1 - \mathbf d_2 \|^2 = 2 - 2 \, \mathbf d_1^T \mathbf d_2 =2 \, d_c(\mathbf d_1, \mathbf d_2)$. Thus we can use Euclidean distance on
unit-normalized vectors and interpret it as cosine distance.

\subsubsection{Jaccard Coefficient} \ \\

Finally, the Jaccard Coefficient is a function that compares how similar
two sets are. Given two sets $A$ and $B$, it is computed as
$J(A, B) = \frac{|A \cap B|}{|A \cup B|}$.
It is also applicable to document vectors with binary weights, and it can
be defined as $J(\mathbf d_1, \mathbf d_2) =
\frac{\mathbf d_1^T \mathbf d_2}{\| \mathbf d_1^T \|^2 + \| \mathbf d_2^T \|^2 - \mathbf d_1^T \mathbf d_2}$ \cite{manning2008introduction}.

\subsection{Document Clustering Techniques} \label{sec:doc-clustering}

Cluster analysis is a set of techniques for organizing collection
of items into coherent groups. In Text Mining clustering is often
used for finding topics in a collection of document \cite{aggarwal2012survey}.
In Information Retrieval clustering is used to assist the users and group
retrieved results into clusters \cite{cutting1992scatter}.

There are several types of clustering algorithms:
hierarchical (agglomerative and divisive), partitioning,
density-based, and others.

\subsubsection{Agglomerative clustering} \label{sec:clustering-heierarchical} \ \\

The general idea of agglomerative clustering algorithms is to start with
each document being its own cluster and iteratively merge clusters based
on best pair-wise cluster similarity.

Thus, a typical agglomerative clustering algorithms consists of the following steps:

\begin{enumerate}
\itemsep1pt\parskip0pt\parsep0pt
  \item Let each document be a cluster on its own;
  \item Compute similarity between all pairs of clusters an store the
      results in a similarity matrix;
  \item Merge two most similar clusters;
  \item Update the similarity matrix;
  \item Repeat until everything belongs to the same cluster.
\end{enumerate}

These algorithms differ only in the way they calculate similarity between
clusters. It can be \textbf{Single Linkage}, when the clusters are merged based
on the closest pair; \textbf{Complete Linkage}, when the clusters are merged
based on the worst-case similarity -- the similarity between the most
distant objects on the clusters; \textbf{Group-Average Linkage}, based
on the average pair-wise similarity between all objects in the clusters;
and \textbf{Ward's Method} when the clusters to merge are chosen to
minimize the within-cluster error between each object and its centroid
is minimized \cite{oikonomakou2005review}.

Among these algorithms only Single Linkage is computationally feasible
for large data sets, but it doesn't give good results compared to other
agglomerative clustering algorithms. Additionally, these algorithms
are not always good for document clustering because they tend to
make mistakes at early iterations that are impossible to correct
afterwards \cite{steinbach2000comparison}.

\subsubsection{$K$-Means} \label{sec:kmeans} \ \\

Unlike agglomerative clustering algorithms, K-Means is an iterative
algorithm, which means that it can correct the mistakes made
at earlier iterations. Lloyd's algorithm is the most popular way
of implementing K-Means \cite{xu2005survey}: given a desired number of clusters $K$,
it iteratively improves the Euclidean distance between each data
point and the centroid, closest to it.

Let $\mathcal D = \{  \mathbf d_1, \mathbf d_2, \ ... \ , \mathbf d_n \}$
be the document collection, where documents $\mathbf d_i$ are represented
is a document vector space $\mathbb R^m$ and $K$ is the desired
number of clusters. Then we define $k$ cluster centroids $\boldsymbol \mu_j$ that are
also in the same document vector space $\mathbb R^m$.
Additionally for each document $\mathbf d_i$ we maintain the assignment
variable $c_i \in \{ 1, 2, \ ... \ , k \}$, which specifies to what
cluster centroid $\boldsymbol \mu_1, \boldsymbol \mu_2, \ ... \ , \boldsymbol \mu_k$
the document $\mathbf d_i$ belongs.

The algorithms consists of three steps: (1) seed selection step,
where each $\boldsymbol \mu_j$ is randomly assigned some value,
(2) cluster assignment step, where we iterate over all document vectors
$\mathbf d_i$ and find its closest centroid, and (3)  move centroids step,
where the centroids are re-calculated. Steps (2) and (3) are repeated
until the algorithm converges. The pseudocode for $K$-Means is presented
in the listing~\ref{algo:k-means}.

\begin{algorithm}
\caption{Lloyd's algorithm for $K$-Means}
\label{algo:k-means}

\begin{algorithmic}[0]
  \Statex
  \Function{K-Means}{no. clusters $k$, documents $\mathcal D$}
    \For{$j \leftarrow 1 \ .. \ k$} \Comment{random seed selection}
      \Let{$\boldsymbol \mu_j$}{random $\mathbf d \in \mathcal D$}
    \EndFor

    \While{not converged}
      \For{each $\mathbf d_i \in \mathcal D$} \Comment{cluster assignment step}
        \Let{$c_i$}{$\operatorname{arg\, min}_j \| \mathbf d_i - \boldsymbol \mu_j \|^2$}
      \EndFor

      \For{$j \leftarrow 1 \ .. \ k$} \Comment{move centroids step}
        \Let{$\mathcal C_j$}{$\{\, \mathbf d_i \text{ s.t. } c_i = j \, \}$}
        \Let{$\boldsymbol \mu_j$}
            {$\cfrac{1}{| \mathcal C_j |} \sum_{\mathbf d_i \in \mathcal C_j} \mathbf d_i$}
      \EndFor
    \EndWhile

    \State \Return{$(c_1, c_2, \ ... \ , c_n)$}
  \EndFunction
\end{algorithmic}
\end{algorithm}

Usually, $K$-Means shows very good results for document clustering, and in
several studies it (or its variations) shows the best performance
\cite{steinbach2000comparison} \cite{hall2012evaluating} .

However for large document collections Lloyd's classical $K$-Means takes a lot
of time to converge. The problem is caused by the fact that it goes through
the entire collection many times. Mini-Batch $K$-Means \cite{sculley2010web}
uses Mini-Batch Gradient Descent method, which is a different optimization technique
that converges faster.

$K$-Means uses Euclidean distance, which does not always behave
well in high-dimensional sparse vector spaces like document vector
spaces. However, as discussed in section~\ref{sec:similarity-distance}, if
document vectors are normalized, the Euclidean distance and cosine distance
are related, and therefore Euclidean $K$-means is the same as
``Cosine Distance'' $K$-Means.



In cases when there are many documents, the centroids tend to contain a lot of words, 
which leads to a significant slowdown. To solve this problem, some terms of the 
centroid can be truncated. There are several possible ways of truncating the 
terms: for example, we can keep only the top $c$ terms, or remove the least 
frequent words such that at least 90\% (or 95\%) of the original vector norm is
retained \cite{schutze1997projections}.

\subsubsection{DBSCAN} \label{sec:dbscan} \ \\

DBSCAN is a density-based clustering algorithm that can discover
clusters of complex shapes based on the density of data points \cite{ester1996density}.

The \emph{density} associated with a data point is obtained by
counting the number of points in a region of radius $\varepsilon$
around the point, where $\varepsilon$  is defined by the user.
If a point has a density of at least some user defined
threshold \verb|MinPts|, then it is considered a \emph{core point}.
The clusters are formed around these core points, and if two core points
are within the radius $\varepsilon$, then they belong to the same cluster.
If a point is not a core point itself, but it belong to the neighborhood of some
core point, then it is a \emph{border point}. But if a point is not a core point
and it is not in the neighborhood of any other core point, then it does not
belong to any cluster and it is considered \emph{noise}.

DBSCAN works as follows: it selects an arbitrary data point $p$, and then
finds all other points in $\varepsilon$-neighborhood of $p$. If
there are more than  \verb|MinPts| points around $p$, then it is a core point,
and it is considered a cluster. Then the process is repeated for all points in
the neighborhood, and they all are assigned to the same cluster, as $p$.
If $p$ is not a core point, but it has a core point in its neighborhood, then
it's a border point and it is assigned to the same cluster and the core point.
But if it is a noise point, then it is marked as noise or discarded
(see listing~\ref{algo:dbscan}).

\begin{algorithm}
\caption{DBSCAN}
\label{algo:dbscan}

\begin{algorithmic}[0]
  \Statex
  \Function{DBSCAN}{database $\mathcal D$, radius $\varepsilon$, MinPts}
    \Let{$\text{result}$}{$\varnothing$}

    \ForAll{$p \in \mathcal D$}
      \If{$p$ is visited}
        \State{\textbf{continue}}
      \EndIf
      \State{mark $p$ as visited}
      \Let{$\mathcal N$}{\textsc{Region-Query}($p, \varepsilon$)}
          \Comment{$\mathcal N$ is the neighborhood of $p$}
      \If{$\mathcal N < \text{MinPts}$}
        \State{mark $p$ as \texttt{NOISE}}
      \Else
        \Let{$\mathcal C$}
            {\textsc{Expand-Cluster}$(p, \mathcal N, \varepsilon, \text{MinPts})$}
        \Let{result}{result $\cup \ \{ \mathcal C \}$}
      \EndIf
    \EndFor
    \State \Return{result}
  \EndFunction
\end{algorithmic}

\begin{algorithmic}[0]
  \Statex
  \Function{Expand-Cluster}{point $p$, neighborhood $\mathcal N$, radius $\varepsilon$, MinPts}
     \Let{$\mathcal C$}{$\{ p \}$}
     \ForAll{$x \in \mathcal N$}
        \If{$x$ is visited}
          \State{\textbf{continue}}
        \EndIf

        \State{mark $x$ as visited}
        \Let{$\mathcal N_x$}{\textsc{Region-Query}$(x, \varepsilon)$}
            \Comment{$\mathcal N_x$ is the neighborhood of $x$}
        \If{$| \mathcal N_x | \geqslant \text{MinPts}$}
          \Let{$\mathcal N$}{$\mathcal N \cup \mathcal N_x$}
        \EndIf

        \Let{$\mathcal C$}{$\mathcal C \cup \{ x \}$}
     \EndFor

     \State \Return{$\mathcal C$}
  \EndFunction
\end{algorithmic}

\begin{algorithmic}[0]
  \Statex
  \Function{Region-Query}{point $p$, radius $\varepsilon$}
     \State \Return{$\{ x \ : \ \| x - p \| \leqslant \varepsilon \}$} \Comment{all points within distance $\varepsilon$ from $p$}
  \EndFunction
\end{algorithmic}

\end{algorithm}

The details of implementation of \textsc{Region-Query} are not specified,
and it can be implemented differently. For example, it can use
Inverse Index to make the similarity search faster
\cite{manning2008introduction} \cite{ertoz2003finding}.

The DBSCAN algorithm uses the Euclidean distance, but can be adapted to
use any other distance or similarity function. For example, to modify the
algorithm to use the cosine similarity (or any other similarity function)
the \textsc{Region-Query} has to be modified to return
$\{ x \ : \ \text{similarity}(x, p) \geqslant \varepsilon \}$.

Shared Nearest Neighbors Similarity (SNN Similarity) \cite{ertoz2003finding}
is a special similarity function that is particularity useful for
high-dimensional spaces, it works well with DBSCAN, and it is
applicable to document clustering and topic discovery \cite{ertoz2004finding}.

SNN Similarity is specified in terms of the $K$ nearest neighbors.
Let $\text{NN}_{K, \, \text{sim}}(p)$ be a function that returns
top $K$ closest points of $p$ according to some similarity function
\texttt{sim}. Then the SNN similarity function is  defined as
$$\text{snn}(p, q) = \big| \text{NN}_{K, \, \text{sim}}(p) \cup \text{NN}_{K, \, \text{sim}}(q) \big|.$$

The extension of DBSCAN that uses the SNN Similarity is called
SSN Clustering algorithm. The user needs to specify the SSN similarity
function by setting parameter $K$ and choosing the base similarity
function $\text{\texttt{sim}}(\cdot, \cdot)$ (typically Cosine, Jaccard
or Euclidean). The algorithm itself has the same
parameters as DBSCAN: radius $\varepsilon$ (such that $\varepsilon < K$)
and the core points density threshold \verb|MinPts|. The
$\textsc{Region-Query}$ function is modified to return
$\{ q \ : \ \text{snn}(p, q) \geqslant \varepsilon \}$. For pseudocode,
see the listing~\ref{algo:snn-clustering}.

\begin{algorithm} \caption{SNN Clustering Algorithm} \label{algo:snn-clustering}

\begin{algorithmic}[0]
  \Statex
  \Function{SNN-Cluster}{database $\mathcal D$, $K$, similarity function \texttt{sim}, radius $\varepsilon$, MinPts}
    \ForAll{$p \in \mathcal D$} \Comment{Pre-compute the $K$NN lists}
      \Let{$\text{NN}[p]$}{$\text{NN}_{K, \, \text{sim}}(p)$}
    \EndFor

    \ForAll{$(p, q) \in (\mathcal D \times \mathcal D)$} \Comment{Pre-compute the SNN similarity matrix}
      \Let{$A[p, q]$}{$\big| \, \text{NN}[p] \ \cup \ \text{NN}[q] \, \big|$}
    \EndFor

    \State \Return{\textsc{DBSCAN}$(A, \varepsilon, \text{MinPts})$}
  \EndFunction
\end{algorithmic}

\end{algorithm}

The algorithm's running time complexity is $O(n^2)$ time, where $n = |\mathcal D|$,
but it can be sped up by using the Inverted Index \cite{ertoz2003finding}.

\subsection{Latent Semantic Analysis} \label{sec:lsa}

In section~\ref{sec:clusters-namespaces} we have discussed the
lexical variability and ambiguity problems in natural language: synonymy
and polysemy. We can treat these problems as ``statistical noise'' and
apply dimensionality reduction techniques to find the optimal dimensionality
for the data and thus reduce the amount of noise there.
This technique is called Latent Semantic Analysis (LSA) \cite{landauer1998introduction}
or Latent Semantic Indexing \cite{deerwester1990indexing}, and
it is often used for document clustering \cite{aggarwal2012survey} \cite{osinski2004lingo}.

There are three major steps in Latent Semantic Analysis  \cite{evangelopoulos2012latent}:
(1) preprocess documents;
(2) construct a term-document matrix $D$ using the Vector Space Model;
(3) de-noise $D$ by reducing its dimensionality with Singular Value Decomposition (SVD).

The first two steps are the same as for traditional Vector Space Models
and in the result we obtain a term-document matrix $D$.
If $D$ has rank $r$, then the SVD of $D$ is $D = U  \Sigma V^T$, where
$U$ is an $m \times r$ orthogonal matrix;
$\Sigma$ is a diagonal $r \times r$ matrix with singular values ordered by their magnitude;
and $V$ is an $n \times r$ orthogonal matrix.

The dimensionality reduction is done by finding the best $k$-rank approximation
of $D$, which is obtained by keeping only the first $k$ singular values of $\Sigma$
and setting the rest to 0.
Typically, not only $\Sigma$ is truncated, but also $U$ and $V$,
and therefore, the $k$-rank approximation of $D$ using SVD is written as
$D \approx D_k = U_k \Sigma_k V_k^T$ where $U_k$ is an $m \times k$
matrix with first $k$ columns of $U$, $\Sigma_k$ is an $k \times k$
diagonal matrix with singular values, and $V_k$ is an $n \times k$
matrix with first $k$ columns of $V$.  This decomposition
is called \emph{rank-reduced} SVD and when applied to text data
it reveals the ``true'' latent semantic space. The parameter $k$ corresponds
to the number of ``latent concepts'' in the data. The idea
of LSA is very nicely illustrated by examples  in
\cite{deerwester1990indexing} and \cite{landauer1998introduction}.

LSA can be used for clustering as well, and this is usually done
by first transforming the document space to the LSA space
and then doing applying transitional cluster analysis techniques
there \cite{schutze1997projections}.
Once $D$ is decomposed as $D \approx U_k \Sigma_k V_k^T$
it is enough to keep only the low dimensional representation $\hat D = V_k \Sigma_k$:
the calculation of inner product between two documents $i$ and $j$ in
the reduced semantic  space corresponds to computing the inner product
between $i$th and $j$th rows of $\hat D$ \cite{deerwester1990indexing}. Since the
Euclidean distance is defined in terms of inner product, it can also be used
directly on the rows of $\hat D$.

Therefore, a generic LSA-based clustering algorithm consists of the following steps:

\begin{enumerate}
\itemsep1pt\parskip0pt\parsep0pt
  \item Build a term-document matrix $D$ from the document collection;
  \item Select number of latent concepts $k$ and apply rank-reduced SVD on $D$
      to get $\hat D= V_k \Sigma_k$;
  \item Apply the cluster algorithm on the rows of $V_k \Sigma_k$.
\end{enumerate}

LSA has some drawbacks. Because SVD looks for an orthogonal basis for the new
reduced document space, there could be negative values that are harder
to interpret, and what is more, the cosine similarity can become negative as well.
However, it does not significantly affect the cosine distance: it still
will always give non-negative results.

Apart from SVD there are many other different matrix decomposition
techniques that can be applied for document clustering and for discovering
the latent structure of the term-document matrix \cite{osinski2006improving},
and one of them in Non-Negative Matrix Factorization (NMF) \cite{lee1999nnmf}.
Using NMF solves the problem of negative coefficients:
when it is applied to non-negative data such as term-document matrices,
NMF produces non-negative rank-reduced approximations.

\begin{figure}[h]
\centering\includegraphics[width=0.8\textwidth]{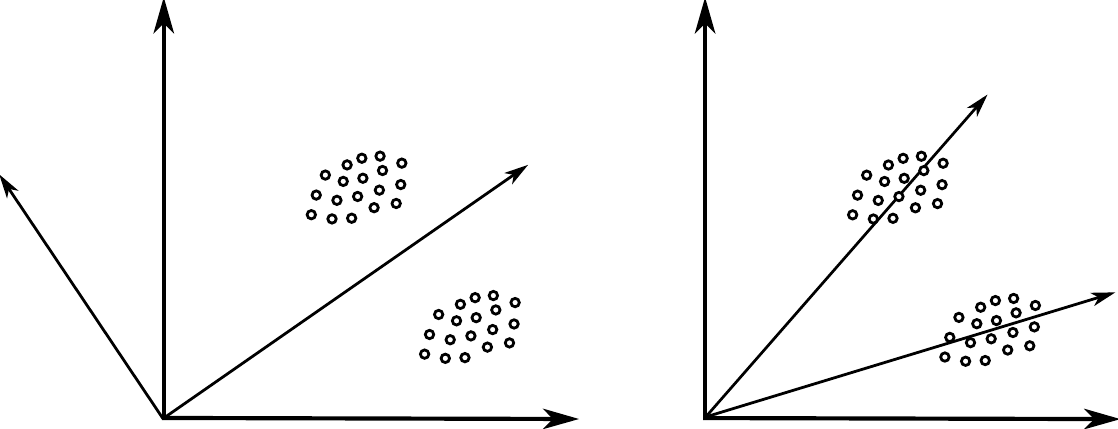}
\caption{Directions found by  SVD (on the left) vs directions by NMF (on the right)}
\label{fig:nmf-svd}
\end{figure}

The main conceptual difference between SVD and NMF is that SVD looks for
orthogonal directions to represent document space, while NMF does not
require orthogonality \cite{xu2003document} (see fig.~\ref{fig:nmf-svd}).

The NMF of an $m \times n$ term-document matrix $D$ is $D \approx D_k = U  V^T$
where $U$ is an $m \times k$ matrix, $V$ is an $n \times k$ matrix and
$k$ is the number of semantic concepts in $D$.
Non-negativity of elements in $D_k$ is very good for interpretability: it
ensures that documents can be seen as a non-negative combination of
the key concepts.

Additionally, NMF is useful for clustering: the results of NMF can
be directly interpreted as cluster assignment and there is no need
to use separate clustering algorithms \cite{xu2003document}. When $D$ is a
term-document matrix and $D \approx U V^T$, then elements $(V)_{ij}$
represent the degree to which document $i$ belongs to cluster $j$.

The document clustering using NMF consists of the following steps \cite{xu2003document}:

\begin{enumerate}
  \item Construct the term-document matrix $D$ and perform NMF on $D$ to get $U$ and $V$;
  \item Normalize rows $\mathbf v_i$ of $V$ by using the rule $\mathbf v_i \leftarrow \mathbf v_i \, \| \mathbf u_i \|$;
  \item Assign document $\mathbf d_i$ to cluster $x$ if $x = \operatorname{arg \, max}_j (V)_{ij}$.
\end{enumerate}

If the desired number of clusters $K$ is larger than the rank $k$ of the
reduced matrix $D_k$, the clustering can be performed directly on the rows
of $V$, for example, by using $K$-Means.


\section{Namespace Discovery} \label{sec:namespace-discovery-chap}

In this chapter, we introduce the problem of namespace discovery in 
mathematical notation and suggest how this problem can be approached. 

First, we extend the idea of namespaces to mathematics in section~\ref{sec:math-namespaces},
and discuss the problem of namespace discovery in section~\ref{sec:discovery-namespaces},
and then argue that it is possible to use document cluster analysis 
to solve the problem in section~\ref{sec:clusters-namespaces}. 
Finally, we propose a way of representing identifiers in a vector space 
in section~\ref{sec:ism}.

\subsection{Namespaces in Mathematical Notation} \label{sec:math-namespaces}

The idea of namespaces can be extended to identifiers in mathematical
formulae. 

In mathematics and other sciences, formulae are used to communicate the results 
to other scientists. An \emph{identifier} is a symbol used in a mathematical 
formula and it typically has some semantic meaning. For example, in a formula 
$E = m c^2$ there are three identifiers: ``$E$'', ``$m$'' and ''$c$''.
\emph{Mathematical notation} is a system of naming identifiers in
mathematical formulae, and for each identifier in the formula the notation assigns
a precise semantic meaning \cite{wikinotation}. For example, in the expression
``$E = mc^2$'' the notation assigns
unambiguous meaning to the symbols ``$E$'', ``$m$'' and ``$c$'', and the
meaning of these symbols is recognized among physicists.

However, notations may conflict. For example, while it is common to use
symbol ``$E$'' to denote ``Energy'' in Physics, it also is used in Probability and
Statistics to denote ``Expected Value'', or in Linear Algebra to denote
``Elimination Matrix''.
We can compare the conflict of notations with the name collision problem
in namespaces, and try to address this problem by extending the notion of
namespaces to mathematical notation.

Thus, let us define a \emph{notation} $\mathcal N$ as a set of pairs $\{ (i, s) \}$,
where $i$ is a symbol or \emph{identifier} and $s$ is its semantic meaning
or \emph{definition}, such that for any pair $(i, s) \in \mathcal N$ there
does not exist another pair $(i', s') \in \mathcal N$ with $i = i'$.
Two notations $\mathcal N_1$ and $\mathcal N_2$
\emph{conflict}, if there exists a pair $(i_1, s_1) \in \mathcal N_1$ and a pair
$(i_2, s_2) \in \mathcal N_2$ such that $i_1 = i_2$ and $s_1 \ne s_2$.

Then we can define \emph{namespace} as a named notation. For example,
$\mathcal N_\text{physics}$ can refer to the notation used in Physics.
For convenience, in this work we can use the Java syntax to refer to
specific entries of a namespace. If $\mathcal N$ is a namespace and $i$
is an identifier such that $(i, s) \in \mathcal N$ for some $s$, then
``$\mathcal N$.$i$'' is a \emph{fully qualified name} of the identifier $i$ that
relates $i$ to the definition $s$. For example,
given a namespace $\mathcal N_\text{physics} =$ $\{ (E, \text{``energy''}),$
$(m, \text{``mass''}),$ $(c, \text{``speed of light''}) \} \big)$,
``$\mathcal N_\text{physics}$.$E$'' refers to ``energy'' -- the definition of ``$E$'' in the
namespace ``Physics''.


Analogously to namespaces in Computer Science, formally a mathematical namespace
can contain any set of identifier-definition pairs that satisfies the definition of
the namespace, but typically namespaces of mathematical notation
exhibit the same properties as well-designed software packages: they
have low coupling and high cohesion, meaning that all definitions
in a namespace
come from the same area of mathematical knowledge and the definitions
from different namespace do not intersect heavily.

However, mathematical notation does not yet exist in isolation, and it is 
usually observed indirectly by its usage in documents. To account for this fact,
we need to introduce a document-centric view on mathematical
namespaces: suppose we have a collection of $n$ documents
$\mathcal D = \{ d_1, d_2, \ ... \ , d_n \}$ and a set of $K$ namespaces
$\{\mathcal N_1, \mathcal N_2, \ ... \ ,$ $\mathcal N_K \}$.
A document $d_j$ can use a namespace $\mathcal N_k$ by \emph{importing}
identifiers from it. To import an identifier, the document uses an import statement
where the identifier $i$ is referred by its fully qualified name.
For example, a document ``Energy-mass equivalence'' would import
``$\mathcal N_\text{physics}$.$E$'', ``$\mathcal N_\text{physics}$.$m$'',
and ``$\mathcal N_\text{physics}$.$c$'', and then these identifiers can be used in
formulae of this document unambiguously.

A namespace exhibits low coupling if it is used only in a small
subset of documents, and high cohesion if all the documents in this subset
are related to the same domain.

But in real-life scientific document there are no import statements
in the document preamble, and they contain only natural language
texts along with some mathematical formulae. Yet we may still assume
that these import exists, but they are implicit, i.e. they are latent
and cannot be observed directly. Additionally, the namespaces themselves are
also not observed.

Typically in mathematical texts, when an identifier is first introduced,
its definition is given in the natural language description that surrounds
the formula. This description can be extracted and used to assign the meaning to
the identifiers. Once identifier definitions are extracted, a document
can be represented as a set of identifier-definition pairs,
and these pairs can be used to discover the namespaces.

In the next section we discuss how this problem can be addressed.

\subsection{Discovery of Identifier Namespaces} \label{sec:discovery-namespaces}

There are several ways of constructing a set of namespaces given a collection 
of documents. 

It is possible to do it manually by pre-defining a set of namespaces and 
then by manually assigning each identifier-definition relation to some of these
namespace. It is not only time consuming, but also very difficult: one has 
to know where to put the identifiers, and set of namespaces needs to be exhaustive. 

Alternatively, it can be done automatically, and in this work we suggest a different 
approach: use Machine Learning techniques for discovering namespaces automatically.

We illustrate our idea by first drawing an analogy between identifier
namespaces and namespaces in programming languages. In a well-designed application, 
we can distinguish between two types of application packages \cite{evans2004domain}:

\begin{itemize}
  \item \emph{type 1}: domain-specific packages that deal with one particular
    concept or domain area, and
  \item \emph{type 2}: packages that use other packages of
    the first type
\end{itemize}


For example, for an application \verb|org.company.app|
there can be several domain-specific packages: \verb|org.company.app.domain.user|
with classes related to users, \verb|org.company.app.domain.account|
with classes related to user accounts, and a system-related package
\verb|org.company.app.tools.auth| that deals with authentication and
authorization. Then we also have a package \verb|org.company.app.web.manage|,
which belongs to the type 2: it handles web requests
while relying on classes from packages \verb|user| and \verb|account| to
implement the business logic and on \verb|auth| for making sure the
requests are authorized.

We can observe that the type 1 packages are mostly self-contained
and not highly coupled between each other, but type 2 packages mostly
use other packages of type 1: they depend on them.

This idea can be extended on the document-centric view on
identifier namespaces. Each document can be seen as a class that
imports identifiers defined in other documents.
Then the documents can be grouped together based on the identifiers
and the definitions they have, and then among these groups
there are some groups of documents that are of \emph{type 1}
and the rest are of \emph{type 2}. The type 1 document groups
contain information about closely related concepts, and they are
very homogenous (they have high cohesion), and they are also
not highly coupled with other document groups.
By using the import metaphor, we can say that the type 1 document groups
import only from few closely related namespaces.
Other documents are of \emph{type 2} and they do not have
low coupling: they are not very homogenous and they import from several namespaces

With this intuition we can refer to \emph{type 1} document groups
as \emph{namespace defining} groups. These groups can be seen as ``type 1''
packages: they define namespaces that are used by other \emph{type 2}
document groups. Once the namespace defining groups are found,
we can learn the namespace of these document.

Thus we need to find groups of homogenous documents given a
collection, and this is exactly what Cluster Analysis methods do.

In the next section we will argue why we can use traditional
document clustering techniques and what are the characteristics
that texts and identifiers have in common.

\subsection{Namespace Discovery by Cluster Analysis} \label{sec:clusters-namespaces}

We argue that cluster analysis techniques developed for text documents
should also work for cases when documents are represented by
identifers they contain.

The reason for this is that identifiers can be seen as ``words'' in 
the mathematical language, their senses are by described their definitions,
and the ``sentences'' of this language are formulae. 
Because identifiers are used like words, we can make the same assumptions 
about them. For example, words are distributed according to a power low distribution
\cite{manning2008introduction}, and therefore we can assume that identifiers 
also follow some power low. 

Additionally, natural languages suffer from lexical problems of variability
and ambiguity, and the two main problems are synonymy and polysemy
\cite{deerwester1990indexing} \cite{gliozzo2009semantic}:

\begin{itemize}
\itemsep1pt\parskip0pt\parsep0pt
  \item two words are \emph{synonymous} if they have the same meaning
        (for example, ``graph'' and ``chart'' are synonyms),
  \item a word is \emph{polysemous} is it can have multiple meanings
        (for example, ``trunk'' can refer to a part of elephant or a part of a car).
\end{itemize}

Note that identifiers have the same problems. For example,
``$E$'' can stand both for ``Energy'' and ``Expected value'',
so ``$E$'' is polysemous.

These problems have been studied in Information Retrieval and
Natural Language Processing literature.
One possible solution for the polysemy problem is \emph{Word Sense Disambiguation}
\cite{jurafsky2000speech}: either replace a word with its sense
\cite{stokoe2003word} or append the sense to the word. For example,
if the polysemous word is ``bank'' with meaning ``financial institution'',
then we replace it with ``bank\_finance''. The same idea can be used
for identifiers, for example if we have an identifier ``$E$'' which is
defined as ``energy'', then ``$E$'' can be replaced with ``$E$\_energy''.

Thus we see that text representation of documents and identifier representation
of documents have many similarities and therefore we can apply the set of
techniques developed for text representation for clustering documents based
on identifiers.

For document clustering, documents are usually represented using
Vector Space Models \cite{oikonomakou2005review} \cite{aggarwal2012survey}.
Likewise, we can introduce ``Identifier Vector Space Model'' analogously to
Vector Space Models for words, and then we can apply clustering algorithm
to documents represented in this space.

\subsection{Identifier Vector Space Model} \label{sec:ism}

The Vector Space Model discussed in section \ref{sec:vsm} can be adjusted to 
represent documents by identifers they contain instead of words. 
To do that we replace the vocabulary $\mathcal V$
with  a set of identifiers $\mathcal I = \{ i_1, i_2, \ ... \ , i_m \}$,
but documents are still represented as $m$-vectors $\mathbf d_j = (w_1, w_2, \ ... \ , w_m)$,
where $w_k$ is a weight of identifier $i_k$ in the document $\mathbf d_j$.
Likewise, we can define an identifier-document matrix $D$ as a matrix where
columns are document vectors and rows are indexed by the identifiers.

Identifiers, as terms, suffer from the problems of synonymy and polysemy,
and we solve this problem by extracting definitions for all the identifiers.
There are several ways of incorporating the extracted definitions into the
model:

\begin{itemize}
\itemsep1pt\parskip0pt\parsep0pt
  \item do not include definition information at all, use only identifiers;
  \item use ``weak'' identifier-definition association: include identifiers and
        definitions as separate dimensions;
  \item use ``strong'' association: append definition to identifier.
\end{itemize}

To illustrate how it is done, consider three relations ($E$, ``energy''),
($m$, ``mass'') and ($c$, ``speed of light''), and three documents
$d_1 = \{E, m, c\}, d_2 = \{ m, c\}, d_3 = \{ E \}$. Then

\begin{itemize}\itemsep1pt\parskip0pt\parsep0pt
  \item no definitions: dimensions are ($E$, $m$, $c$) and the identifier-document matrix is
  $$D = \left[
    \begin{array}{c|ccc}
       & d_1 & d_2 & d_3 \\
      \hline
      E & 1 & 0 & 1  \\
      m & 1 & 1 & 0 \\
      c & 1 & 1 & 0 \\
    \end{array}
  \right];$$
  \item ``weak'' association: dimensions are ($E$, $m$, $c$, energy, mass,
  speed of light), and the matrix is $$D = \left[
    \begin{array}{r|ccc}
       & d_1 & d_2 & d_3 \\
      \hline
      E                     & 1 & 0 & 1  \\
      m                     & 1 & 1 & 0 \\
      c                     & 1 & 1 & 0 \\
      \text{energy}         & 1 & 0 & 1  \\
      \text{mass}           & 1 & 1 & 0 \\
      \text{speed of light} & 1 & 1 & 0 \\
    \end{array}
  \right];$$
  \item ``strong'' association: dimensions are ($E$\_energy, $m$\_mass, $c$\_speed of light), and the matrix is $$D = \left[
    \begin{array}{r|ccc}
       & d_1 & d_2 & d_3 \\
      \hline
      E\text{\_energy} & 1 & 0 & 1  \\
      m\text{\_mass} & 1 & 1 & 0 \\
      c\text{\_speed of light} & 1 & 1 & 0 \\
    \end{array}
  \right].$$
\end{itemize}

Once a collection of documents is represented is some Identifier Vector Space, we 
can apply document clustering techniques discussed in the section~\ref{sec:doc-clustering}.

\section{Implementation} \label{sec:implementation}

In this chapter we give important implementation details.

First, we describe the data sets we use for our experiments and how they are
cleaned  in section~\ref{sec:dataset}.
Then we explain how definition extraction is implemented in
section~\ref{sec:defextraction-impl} and how we implement cluster analysis methods.
Finally, section~\ref{sec:building-namespaces} shows how we can build a namespace
from a cluster of documents.

\subsection{Data set} \label{sec:dataset}

Wikipedia is a big online encyclopedia where the content
are written and edited by the community. It contains a large
amount of articles on a variety of topics, including articles about
Mathematics and Mathematics-related fields such as Physics. It
is multilingual and available in several languages, including
English, German, French, Russian and others. The content of Wikipedia pages
are authored in a special markup language and the content of the entire
encyclopedia is freely available for download.


The techniques discussed in this work are mainly applied
to the English version of Wikipedia. At the moment of writing
(\today) the English Wikipedia contains about 4.9 million
articles\footnote{\url{https://en.wikipedia.org/wiki/Wikipedia:Statistics}}.
However, just a small portion of these articles are math related:
there are only 30\,000 pages that contain at least one \verb|<math>| tag.

Apart from the text data and formulas Wikipedia articles have information
about categories, and we can exploit this information as well.
The category information is encoded directly into each Wikipedia page
with a special markup tag. For example, the article
``Linear Regression''\footnote{\url{https://en.wikipedia.org/wiki/Linear_regression}}
belongs to the category ``Regression analysis'' and \verb|[[Category:Regression analysis]]|
tag encodes this information.


Wikipedia is available in other languages, not only English.
While the most of the analysis is performed on the English Wikipedia,
we also apply some of the techniques to the Russian version \cite{ruwikidump}
to compare it with the results obtained on the English Wikipedia.
The Russian Wikipedia is smaller that the English Wikipedia and contains
1.9 million
articles\footnote{\url{https://en.wikipedia.org/wiki/Russian_Wikipedia}},
among which only 15\,000 pages are math-related (i.e. contain at
least one \verb|<math>| tag).

\subsection{Definition Extraction} \label{sec:defextraction-impl}

Before we can proceed to discovering identifier namespaces, we
need to extract identifier-definition relations. For this we use the probabilistic
approach, discussed in the section~\ref{sec:mlp}.
The extraction process is implemented using Apache Flink \cite{flink}
and it is based on the open source implementation provided by Pagel and
Schubotz in \cite{pagael2014mlp}\footnote{\url{https://github.com/rbzn/project-mlp}}.

The first step is to keep only mathematical articles and discard the rest.
This is done by retaining only those articles that contain
at least one \verb|<math>| tag with a simple python script
\verb|wikiFilter|\footnote{\url{https://github.com/physikerwelt/wikiFilter}}.
Once the data set is filtered, then
all the \LaTeX\ formulas form the \verb|<math>| tags are converted
to MathML, an XML-based representation of mathematical
formulae \cite{mathml}.

The dataset is stored in a big XML file in the Wiki XML
format. It makes it easy to extract the title and the content
of each document, and then process the documents separately.
The formulas are extracted by looking for the \verb|<math>| tags.
However some formulas for some reasons are typed without the tags
using the unicode symbols, and such formulas are very hard to
detect and therefore we choose not to process them.
Once all \verb|<math>| tags are found, they (along with the content)
are replaced with a special placeholder \verb|FORMULA_%HASH%|, where
\verb|%HASH%| is MD5 hash \cite{rivest1992md5} of the tag's content represented as
a hexadecimal string. After that
the content of the tags is kept separately from the document content.

The next step is to find the definitions for identifiers in formulas.
We are not interested in the semantics of a formula, only in the identifiers
it contains. In MathML \verb|<mi>| corresponds to identifiers, and
hence extracting identifiers from MathML formulas amounts to finding
all \verb|<mi>| tags and retrieving their content. It is enough to
extract simple identifiers such as ``$t$'', ``$C$'', ``$\mu$'', but
there also are complex identifiers with subscripts, such as
``$x_1$'', ``$\xi_i$'' or even ``$\beta_{\text{slope}}$''.
To extract them we need to look for tags \verb|<msub>|. We do not
process superscripts because they are usually powers (for example, ``$x^2$''),
and therefore they are not interesting for this work.
There are exceptions to this, for example, ``$\sigma^2$'' is an identifier,
but these cases are rare and can be ignored.

Since MathML is XML, the identifiers are extracted with XPath
queries \cite{moller2006introduction}:

\begin{itemize}
  \item \verb|//m:mi[not(ancestor::m:msub)]/text()| for all \verb|<mi>| tags
that are not subscript identifers;
  \item \verb|//m:msub| for subscript identifiers.
\end{itemize}

Once the identifiers are extracted, the rest of the formula is discarded.
As the result, we have a ``Bag of Formulae'': analogously to the Bag of Words
approach (see section~\ref{sec:vsm}) we keep only the counts of occurrences
of different identifiers and we do not preserve any other structure.

The content of Wikipedia document is authored with Wiki markup --
a special markup language for specifying document layout elements
such as headers, lists, text formatting and tables.
Thus the next step is to process the Wiki markup and extract the textual
content of an article, and this is done using a Java library
``Mylyn Wikitext'' \cite{mylynwikitext}.
Almost all annotations are discarded at this stage,
and only inner-wiki links are kept: they can be useful as candidate definitions.
The implementation of this step is taken entirely from \cite{pagael2014mlp}
with only a few minor changes.

Once the markup annotations are removed and the text content of an article is
extracted, we then apply Natural Language Processing (NLP) techniques.
Thus, the next step is the NLP step, and for NLP we use the
Stanford Core NLP library (StanfordNLP) \cite{manning2014stanford}.
The first part of this stage is to tokenize the text and also split it by sentences.
Once it is done, we then apply Math-aware POS tagging
(see section~\ref{sec:postagging}). For English documents from the English Wikipedia
we use StanfordNLP's Maximal Entropy POS Tagger \cite{toutanova2003feature}.
Unfortunately, there are no trained models available for POS tagging the Russian
language for the StanfordNLP library and we were not able to find a
suitable implementation of any other POS taggers in Java. Therefore we
implemented a simple rule-based POS tagger ourselves. The implementation is based on
a PHP function from \cite{habr2012postag}: it is translated into Java
and seamlessly integrated into the StanfordNLP pipeline.
The English tagger uses the Penn Treebank POS Scheme \cite{santorini1990part},
and hence we follow the same convention for the Russian tagger.

For handling mathematics we introduce two new POS classes:
``\verb|ID|'' for identifiers and ``\verb|MATH|'' for formulas.
These classes are not a part of the Penn Treebank POS Scheme,
and therefore we need to label all the instances of these tags ourselves
during the additional post-processing step. If a token starts
with ``\verb|FORMULA_|'', then we recognize that it is a placeholder
for a math formula, and therefore we annotate it with the ``\verb|MATH|''
tag. Additionally, if this formula contains only one identifier, this
placeholder token is replaced by the identifier and it is tagged with
``\verb|ID|''. We also keep track of all identifiers found in
the document and then for each token we check if this token is in the list.
If it is, then it is re-annotated with the ``\verb|ID|'' tag.

At the Wikipedia markup processing step we discard almost all markup
annotations, but we do keep inner Wikipedia links, because these links
are good definition candidates. To use them, we introduce
another POS Tag: ``\verb|LINK|''. To detect all inner-wiki links,
we first find all token subsequences that start with \verb|[[|
and end with \verb|]]|, and then these subsequences are
concatenated and tagged as ``\verb|LINK|''.

Successive nouns (both singular and plurals), possible modified
by an adjective, are also candidates for definitions. Therefore
we find all such sequences on the text and then concatenate each
into one single token tagged with ``\verb|NOUN_PHRASE|''.

The next stage is selecting the most probable identifier-definition
pairs, and this is done by ranking definition candidates.
The definition candidates are tokens annotated with ``\verb|NN|'' (noun singular),
``\verb|NNS|'' (noun plural), ``\verb|LINK|'' and ``\verb|NOUN_PHRASE|''.
We rank these tokens by a score that depends how far it is from the identifer
of interest and how far is the closest formula that contains this
identifier (see section~\ref{sec:mlp}).
The output of this step is a list of identifier-definition pairs along
with the score, and only the pairs with scores above
the user specified threshold are retained. The implementation of
this step is also taken entirely from \cite{pagael2014mlp} with very minor
modifications.

\subsubsection{Data Cleaning} \label{sec:datacleaning} \ \\

The Natural Language data is famous for being noisy and hard to
clean \cite{sonntag2004assessing}. The same is true for
mathematical identifiers and  scientific texts with formulas.
In this section we describe how the data was preprocessed and
cleaned at different stages of Definition Extraction.

Often identifiers contain additional semantic information visually conveyed
by special diacritical marks or font features. For example, the diacritics can be
hats to denote ``estimates'' (e.g., ``$\hat w$''), bars to denote the average
(e.g., ``$\bar X$''), arrows to denote vectors (e.g., ``$\vec x\, $'') and others.
As for the font features, bold lower case single characters are often used to
denote vectors (e.g., ``$\mathbf w$'') and bold upper case single characters denote 
matrices (e.g., ``$\mathbf X$''), calligraphic fonts are
used for sets (e.g., ``$\mathcal H$''), double-struck fonts often denote spaces
(e.g., ``$\mathbb R$''), and so on.

Unfortunately, there is no common notation established across all fields of
mathematics and there is a lot of variance. For example,
a vector can be denoted by ``$\vec x\, $'', ``$\boldsymbol x$'' or ``$\mathbf x$'',
and a real line by ``$\mathbb R$'', ``$\mathbf R$'' or ``$\mathfrak R$''.
In natural languages there are related problems of lexical ambiguity such as
synonymy, when different words refer to the same concept, and it can be solved
by replacing the ambiguous words with some token, representative of the concept.
Therefore, this problem with identifiers can be solved similarly
by reducing identifiers to their ``root'' form. This can be done
by discarding all additional visual information, such that
``$\bar X$'' becomes ``$X$'', ``$\mathbf w$'' becomes ``$w$'' and ``$\mathfrak R$''
becomes ``$R$''.

The disadvantage of this approach is that we lose
the additional semantic information about the identifier that overwise
could be useful. Additionally, in some cases we will treat different
identifiers like they are the same. For example, in Statistics, ``$\bar X$''
usually denotes the mean value of a random variable $X$, but when we remove
the bar, we lose this semantic information, and it becomes impossible to
distinguish between different usages of $X$.

The diacritic marks can easily be discarded because they are represented
by special MathML instructions that can be ignored when
the identifiers are retrieved. But, on the other hand,
the visual features are encoded directly on the character level:
the identifiers use special unicode symbols to convey font features such
as bold type or Fraktur, so it needs to be normalized by converting characters
from special ``Mathematical Alphanumeric Symbols'' unicode block \cite{allen2007unicode}
back to the standard ASCII positions (``Basic Latin'' block).
Some identifiers (such as ``$\hbar$'' or ``$\ell$'') are expressed using
characters from a special ``Letterlike Symbols'' table, and these characters
are normalized as well.

Additionally, there is a lot of noise on the annotation level in MathML formulas:
many non-identifiers are captured as identifiers inside \verb|<mi>| tags. Among
them there are many mathematical symbols
like ``\textasciicircum'', ``\#'',``$\forall$'', ``$\int$'';
miscellaneous symbols like ``$\diamond$'' or
``$\circ$'', arrows like ``$\to$'' and ``$\Rightarrow$'', and special characters like
``$\lceil$''. Ideally, these symbols should be represented inside \verb|<mo>| tags.
However, there are many cases when they are not.

To filter out these one-symbol false identifiers we fully exclude all characters from
the following unicode blocks: ``Spacing Modifier Letters'', ``Miscellaneous Symbols'',
``Geometric Shapes'', ``Arrows'', ``Miscellaneous Technical'', ``Box Drawing'',
``Mathematical Operators'' (except ``$\nabla$'' which is sometimes used as an identifier)
and ``Supplemental Mathematical Operators'' \cite{allen2007unicode}.
Some symbols (like ``='', ``+'', ``\verb|~|'', ``\%'', ``?'', ``!'')
belong to commonly used unicode blocks which we cannot exclude altogether.
For these symbols we manually prepare a stop list for filtering them.

It also captures multiple-symbol false positives: operator and function names
like ``\texttt{sin}'', ``\texttt{cos}'', ``\texttt{exp}'', ``\texttt{max}'', ``\texttt{trace}'';
words commonly used in formulas like ``\texttt{const}'', ``\texttt{true}'', ``\texttt{false}'',
``\texttt{vs}'', ``\texttt{iff}''; auxiliary words like ``\texttt{where}'',
``\texttt{else}'', ``\texttt{on}'', ``\texttt{of}'', ``\texttt{as}'', ``\texttt{is}'';
units like ``\texttt{mol}'', ``\texttt{dB}'', ``\texttt{mm}''.
These false identifiers are excluded by a stop list as well: if a
candidate identifier is in the list, it is filtered out.
The stop list of false positives is quite similar for
both English and Russian: for the Russian wikipedia we only need
to handle the auxiliary words such as ``\texttt{где}'' (``\texttt{where}''),
``\texttt{иначе}'' (``\texttt{else}'') and so on. The names for operators and functions
are more or less consistent across both data sources.

Then, at the next stage, the definitions are extracted. However many
shortlisted definitions are either not valid definitions or too general.
For example, some identifiers become associated with ``\texttt{if and only if}'',
``\texttt{alpha}'', ``\texttt{beta}'', ``\texttt{gamma}'', which are not valid definitions.

Other definitions like ``\texttt{element}'' (``\texttt{элемент}''),
``\texttt{number}'' (``\texttt{число}'') or \\ ``\texttt{variable}'' (``\texttt{переменная}'' )
are valid, but they are too general and not descriptive. We maintain a stop list of such
false definitions and filter them out from the result. The elements
of the stop list are also consistent across both data data sets,
in the sense that the false definition candidates are the same but expressed
in different languages.

The Russian language is highly inflected, and due to this extracted
definitions have many different forms, depending on grammatical gender,
form (singular or plural) and declensions. This highly increases the
variability of the definitions, and to reduce it lemmatize the definitions:
they are reduced to the same common form (nominative, singular, and masculine).
This is done using Pymorphy2: a Python library for Russian and
Ukrainian morphology \cite{korobov2015morphological}.

At the next stage the retrieved identifier/definition pairs
are used for document clustering. Some definitions are used only
once and we can note that they are not very useful because
they do not have any discriminative power. Therefore, all such
definitions are excluded.

\subsubsection{Dataset Statistics}  \ \\

At the identifier extraction step when the data set is cleaned,
some identifiers are discarded, and after that some documents
become empty: they no longer contain any identifiers, which is why
these documents are not considered for further analysis.
Additionally, we discard all the documents that have only one identifier.
This leaves only 22\,515 documents out of 30\,000, and they
contain 12\,771 distinct identifiers, which occur about 2 million times.

\begin{figure}[h]
\centering
\hfill
\begin{subfigure}[b]{0.47\textwidth}
  \centering
  \includegraphics[width=\textwidth]{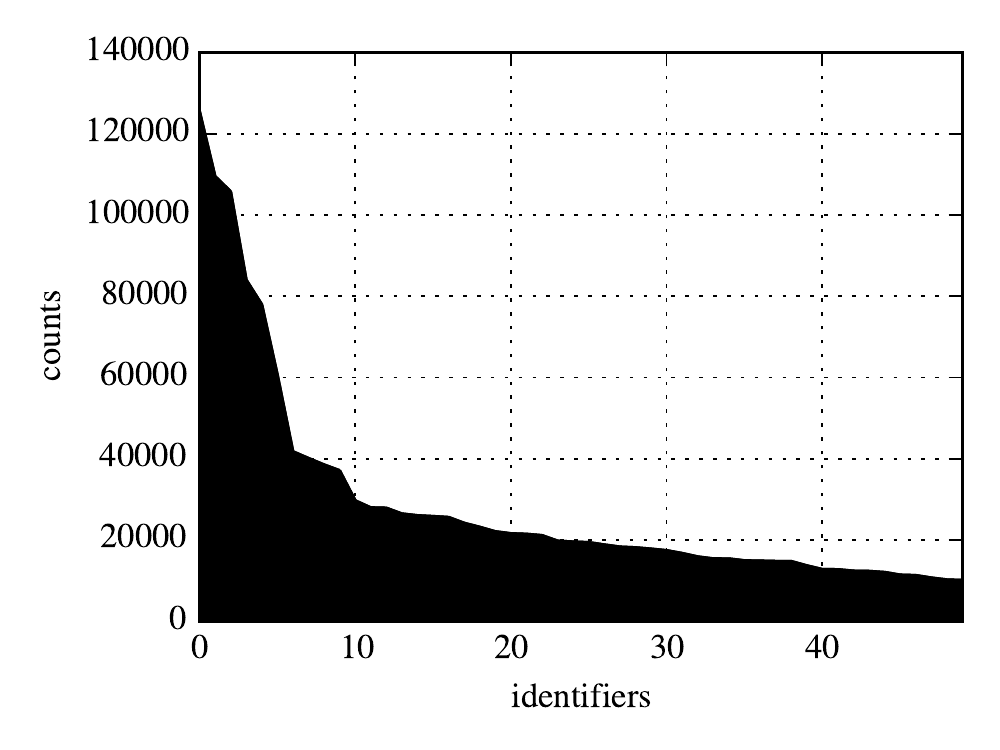}
  \caption{Frequencies of the first 50 identifiers}
  \label{fig:en-wiki-ids-1}
\end{subfigure}
\hfill
\begin{subfigure}[b]{0.47\textwidth}
  \centering
  \includegraphics[width=\textwidth]{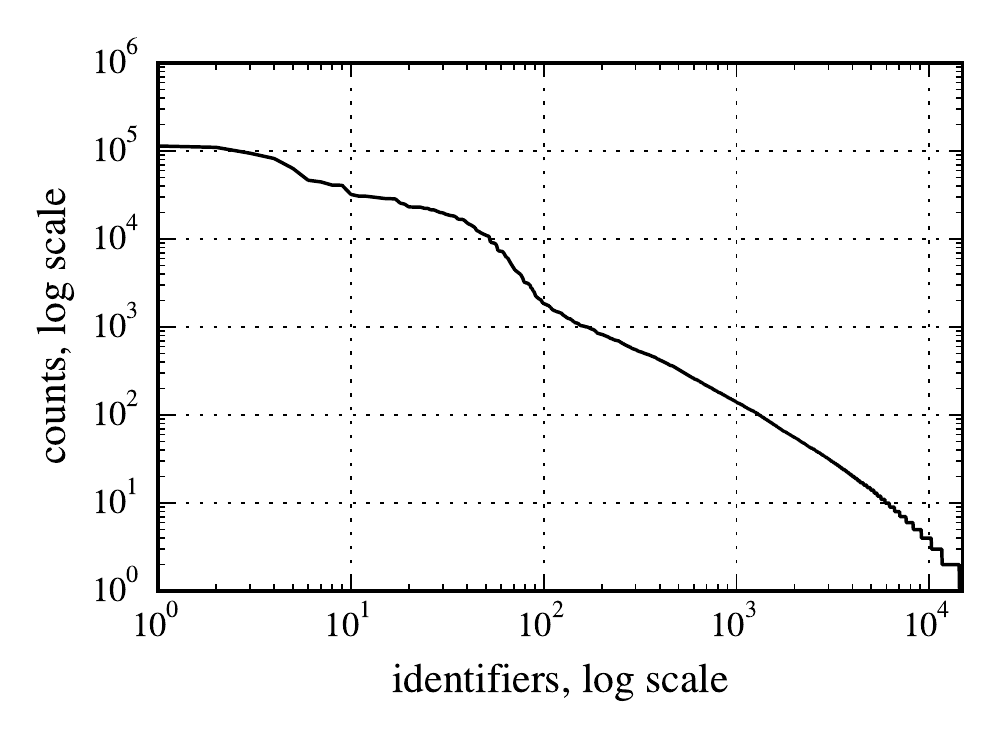}
  \caption{Frequencies, log-log scale}
  \label{fig:en-wiki-ids-2-log.pdf}
\end{subfigure}
\hfill
\caption{Distribution of frequencies of identifiers}
\label{fig:ed-wiki-ids}
\end{figure}

The most frequent identifiers are ``$x$'' (125\,500 times), ``$p$'' (110\,000),
``$m$'' (105\,000 times) and ``$n$'' (83\,000 times), but about 3\,700 identifiers occur
only once and 1\,950 just twice. Clearly, the distribution of
identifiers follows some power law distribution (see fig.~\ref{fig:ed-wiki-ids}).

\begin{figure}[h]
\centering

\begin{subfigure}[b]{\textwidth}
  \centering
  \includegraphics[width=0.9\textwidth]{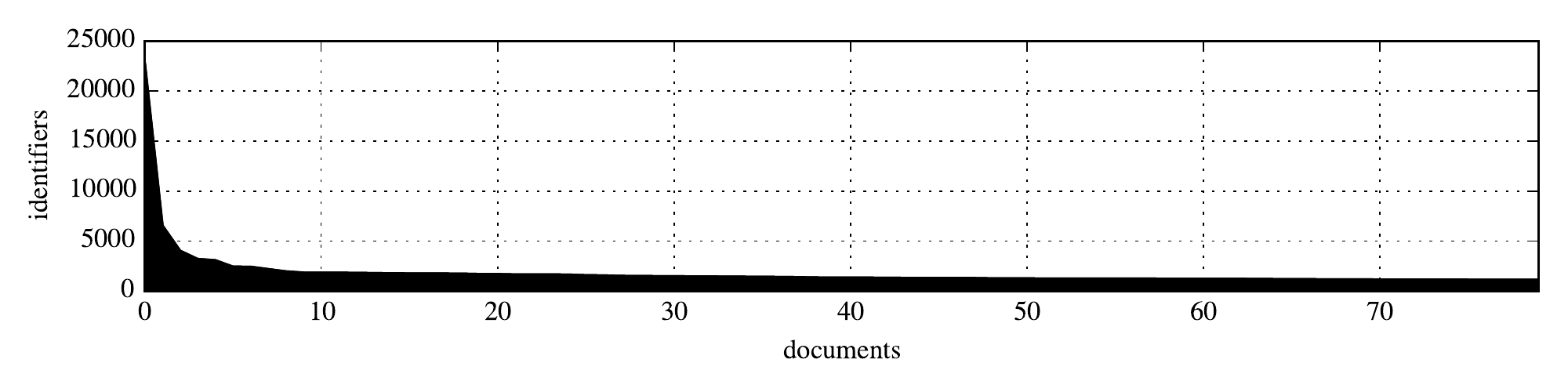}
  \caption{Identifier frequencies per document for first 80 most largest documents}
  \label{fig:en-wiki-doc-ids-1.pdf}
\end{subfigure}

\begin{subfigure}[b]{\textwidth}
  \centering
  \includegraphics[width=0.9\textwidth]{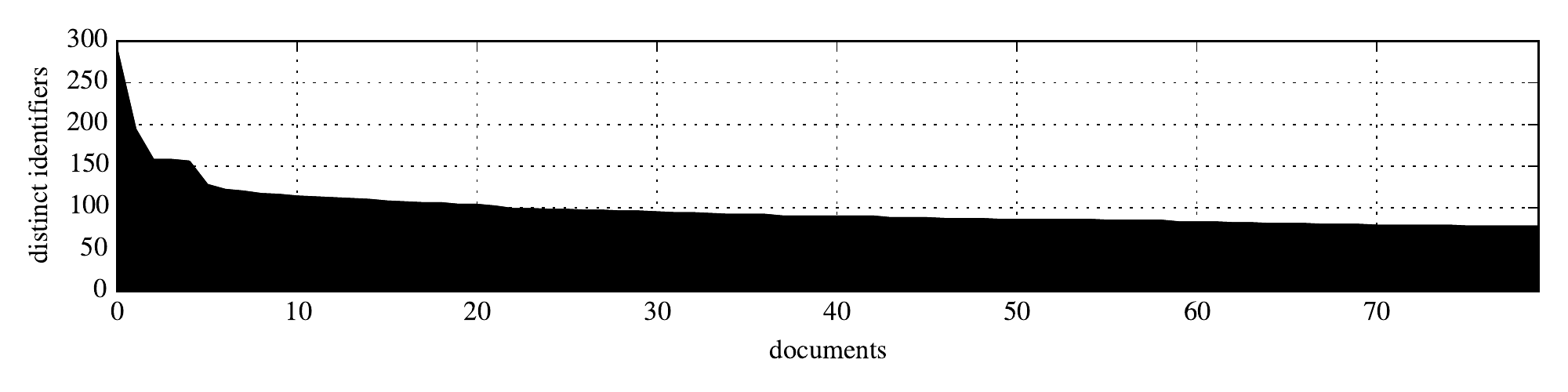}
  \caption{No. of distinct identifiers per document for first 80 most largest documents}
  \label{fig:en-wiki-doc-ids-2.pdf}
\end{subfigure}

\begin{subfigure}[b]{\textwidth}
  \centering
  \includegraphics[width=0.9\textwidth]{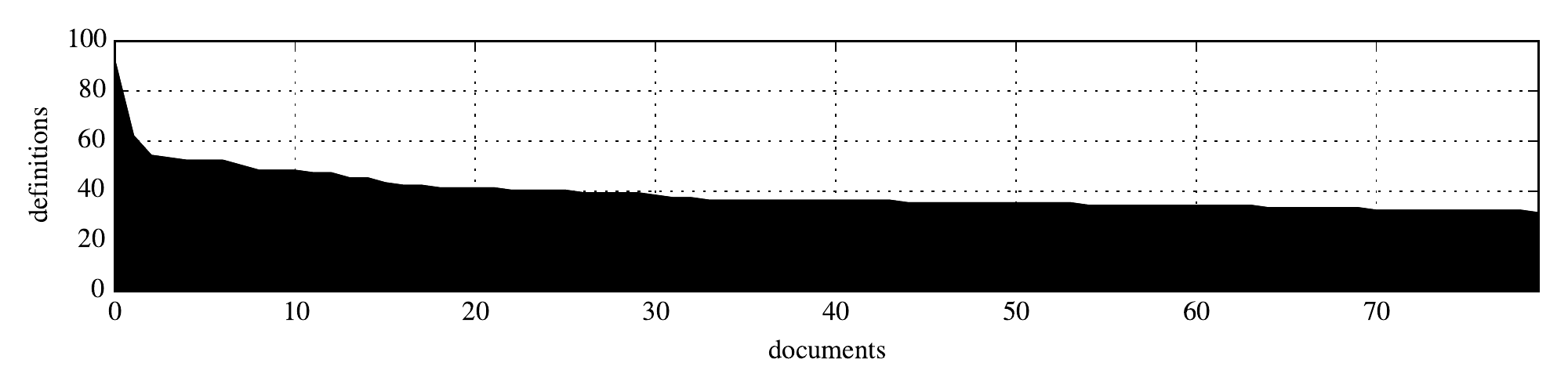}
  \caption{Definitions per document}
  \label{fig:en-wiki-doc-def.pdf}
\end{subfigure}

\caption{Frequencies per documents}
\label{fig:freqs-doc}

\end{figure}

The distribution of counts for identifiers inside the documents also
appears to follow a long tail power law distribution: there are few articles
that contain many identifiers, while most of the articles do not
(see fig.~\ref{fig:en-wiki-doc-ids-1.pdf}).
The biggest article (``Euclidean algorithm'') has 22\,766 identifiers,
and the second largest (``Lambda lifting'') has only 6\,500 identifiers.
The mean number of identifiers per document is 33.
The distribution of the number of distinct identifiers per document
is less skewed (see fig.~\ref{fig:en-wiki-doc-ids-2.pdf}).
The largest number of distinct identifiers is 287 (in the article
``Hooke's law''), and it is followed by 194 (in ``Dimensionless quantity'').
The median number of identifiers per document is 10.

For 12\,771 identifiers the algorithm extracted 115\,300 definitions, and the
number of found definitions follows a long tail distribution as well
(see fig.~\ref{fig:en-wiki-doc-def.pdf}), with the median number of
definitions per page being 4.

\begin{table}[h!]
\centering
\begin{tabular}{|c|c|c|c|c|}
  \hline
  ID & Definition & Count & Definition Rus & Count \\
  \hline
  $t$ & time & 1086 & время & 130 \\
$m$ & mass & 424 & масса & 103 \\
$\theta$ & angle & 421 &    &  \\
$T$ & temperature & 400 & температура & 69 \\
$r$ & radius & 395 &  &    \\
$v$ & velocity & 292 & скорость & 50 \\
$\rho$ & density & 290 & плотность & 57 \\
$G$ & group & 287 & группа & 87 \\
$V$ & volume & 284 &  &    \\
$\lambda$ & wavelength & 263 &   &  \\
$R$ & radius & 257 & радиус & 38 \\
$n$ & degree & 233 &  &   \\
$r$ & distance & 220 &  &   \\
$c$ & speed of light & 219 & скорость свет & 89 \\
$L$ & length & 216 &  &   \\
$n$ & length & 189 &  &    \\
$n$ & order & 188 &  &    \\
$n$ & dimension & 185 &    &  \\
$n$ & size & 178 &  &    \\
$M$ & mass & 171 &  &    \\
$d$ & distance & 163 &    &  \\
$X$ & topological space & 159 & топ. пространство & 46 \\
\hline
\end{tabular}
\caption{Most frequent relations from the English Wikipedia and corresponding relations
from the Russian Wikipedia.}
\label{tab:top-def}
\end{table}

Table~\ref{tab:top-def} shows the list of the most common
identifier-definition relations extracted from the English Wikipedia.

In the Russian Wikipedia only 5\,300 articles contain enough identifiers,
and the remaining 9\,500 are discarded.

The identifiers and definitions extracted from the Russian version of
Wikipedia exhibit the similar properties. The most frequently
occurring identifier is ``$x$'' with 13\,248 occurrences,
but the median frequency of an identifer is only 3 times.
The article with the largest number of identifiers is ``Уравнения Максвелла''
(``Maxwell's equations''), which contains 1\,831 identifiers, while
the median number of identifiers is just 3;
the article with the largest number of distinct identifiers
is also  ``Уравнения Максвелла'' with 112 unique identifiers, and
the median number of distinct identifiers in the data set is 5.
Finally, the largest number of extracted definitions is 44
(again, for ``Уравнения Максвелла'') with 2 being the median number of
definitions per page.

We can compare the most frequent identifier-definition relations extracted
from the Russian Wikipedia (see table~\ref{tab:top-def}): some of the top relations
appear in both datasets. Other frequent identifier-definition relations
extracted from the Russian Wikipedia include:

\begin{itemize}
\item $f$: ``функция'' (``function'') (215)
\item $X$: ``множество'' (``set'') (113)
\item $h$: ``постоянный планка'' (``Planck constant'') (68)
\item $M$: ``многообразие'' (``manifold'') (53)
\item $K$: ``поль'' (``field'') (53)
\item $X$: ``пространство'' (``space'') (50)
\item $G$: ``граф'' (``graph'') (44)
\item $R$: ``радиус'' (``radius'') (38)
\item $R$: ``кольцо'' (``ring'') (36)
\item $G$: ``гравитационный постоянный'' (``gravitational constant'') (34)
\item $E$: ``энергия'' (``energy'') (34)
\item $m$: ``модуль'' (``modulo'') (33)
\item $S$: ``площадь'' (``area'') (32)
\item $k$: ``постоянный больцмана'' (``Boltzmann constant'') (30)
\end{itemize}

\subsection{Document Clustering} \label{sec:clustering-impl}

At the document clustering stage we want to find cluster of documents
that are good namespace candidates.

Before we can do this, we need to vectorize our dataset: i.e., build the
identifier space (see section~\ref{sec:ism}) and represent each document
in this space.

There are three choices for dimensions of the identifier space:

\begin{itemize}
  \item identifiers alone,
  \item ``weak'' identifier-definition association,
  \item ``strong'' association: using identifier-definition pairs.
\end{itemize}

In the first case we are only interested in identifier information and
discard the definitions altogether.

In the second and third cases we keep the definitions and use them to
index the dimensions of the identifier space. But there is some
variability in the definitions: for example, the same identifier
``$\sigma$'' in one document can be assigned to ``Cauchy stress tensor'' and
in other it can be assigned to ``stress tensor'', which are almost the same thing.
To reduce this variability we perform some preprocessing: we tokenize
the definitions and use individual tokens to index dimensions of the space.
For example, suppose we have two pairs ($\sigma$, ``Cauchy stress tensor'')
and ($\sigma$, ``stress tensor''). In the ``weak'' association case
we have will dimensions $(\sigma, \text{Cauchy}, \text{stress}, \text{tensor})$,
while for the ``strong'' association case we will have
$(\sigma\text{\_Cauchy}, \sigma\text{\_stress}, \sigma\text{\_tensor})$.

Additionally, the effect of variability can be decreased further
by applying a stemming technique for each definition token.
In this work we use Snowball stemmer for English \cite{porter2001snowball}
implemented in NLTK \cite{bird2006nltk}: a Python library for
Natural Language Processing. For Russian we use Pymorphy2 \cite{korobov2015morphological}.

Using \verb|TfidfVectorizer|  from  scikit-learn \cite{scikit-learn} we vectorize
each document. The experiments are performed with $(\log \text{TF}) \times \text{IDF}$ weighting,
and therefore we use \verb|use_idf=False, sublinear_tf=True| parameters
for the vectorizer. Additionally, we use \verb|min_df=2| to discard identifiers
that occurs only once.

The output is a document-identifier matrix (analogous to ``document-term''):
documents are rows and identifiers/definitions are columns.
The output of \verb|TfidfVectorizer| is row-normalized, i.e. all rows have unit length.

Once we the documents are vectorized, we can apply clustering techniques
to them. We use $K$-Means (see section~\ref{sec:kmeans}) implemented as a
class \verb|KMeans| in scikit-learn and Mini-Batch $K$-Means (class \verb|MiniBatchKMeans|) \cite{scikit-learn}. Note that if rows are unit-normalized, then running $K$-Means with
Euclidean distance is equivalent to cosine distance
(see section~\ref{sec:cosine}).


DBSCAN and SNN Clustering (see section~\ref{sec:dbscan}) algorithms were
implemented manually: available DBSCAN implementations usually employ a distance
measure rather than a similarity measure. The similarity matrix created by similarity measures
are typically very sparse, usually because only a small fraction of the documents
are similar to some given document. Similarity measures
can be converted to distance measures, but in this case
the matrix will no longer be sparse, and we would like to avoid that.
Additionally, available implementations are usually general purpose
implementations and do not take advantage of the structure of the data:
in text-like data clustering algorithms can be sped up significantly
by using an inverted index.

Dimensionality reduction techniques are also important: they
not only reduce the dimensionality, but also help reveal the latent
structure of data. In this work we use Latent Semantic Analysis (LSA) (section~\ref{sec:lsa})
which is implemented using randomized Singular Value Decomposition (SVD)
\cite{tropp2009finding}, The implementation of randomized SVD is taken from scikit-learn
\cite{scikit-learn} -- method \verb|randomized_svd|. Non-negative Matrix Factorization
is an alternative technique for dimensionality reduction (section~\ref{sec:lsa}).
Its implementation is also taken from scikit-learn \cite{scikit-learn},
class \verb|NMF|.

To assess the quality of produced clusters we use wikipedia categories. It is
quite difficult to extract category information from raw wikipedia text,
therefore we use DBPedia \cite{bizer2009dbpedia} for that: it provides
machine-readable information about categories for each wikipedia article.
Additionally, categories in wikipedia form a hierarchy, and this hierarchy
is available as a SKOS ontology.

Unfortunately, there is no information about articles from the Russian Wiki\-pedia on
DBPedia. However the number of documents is not very large, and therefore
this information can be retrieved via MediaWiki API\footnote{\url{http://ru.wikipedia.org/w/api.php}} individually for each
document. This information can be retrieved in chunks for a group of several
documents at once, and therefore it is quite fast.

\subsection{Building Namespaces}  \label{sec:building-namespaces}

Once a cluster analysis algorithm assigns documents in our collection to
some clusters, we need to find namespaces among these clusters. We assume that
some clusters are namespace-defining: they are not only homogenous in the cluster
analysis sense (for example, in case of $K$-Means it means that within-cluster sum
of squares is minimal), but also ``pure'': they are about the same topic.

A cluster is \emph{pure} if all documents belong to the same category.
Using categories information we can find the most frequent category of the
cluster, and then we can define purity of a cluster $C$ as
$$\operatorname{purity}(C) = \cfrac{\max_i \operatorname{count}(c_i)}{|C|},$$
where $c_i$'s are cluster categories.
Thus we can select all clusters with purity above some pre-defined threshold
and refer to them as namespace-defining clusters.

Then we convert these clusters into namespaces by collecting all the identifiers
and their definitions in the documents of each cluster. To do this, we first
collect all the identifier-definition pairs, and then group them by identifier.
When extracting, each definition candidate is scored, and this score is used
to determine, which definition an identifier will be assigned in the namespace.

For example, consider three documents with the following extracted relations:

\begin{itemize}
  \item Document A:
  \begin{itemize}
\item $n$: (predictions, 0.95), (size, 0.92), (random sample, 0.82), (population, 0.82)
\item $\theta$: (estimator, 0.98), (unknown parameter, 0.98), (unknown parameter, 0.94)
\item $\mu$: (true mean, 0.96), (population, 0.89)
\item $\mu_4$: (central moment, 0.83)
\item $\sigma$: (population variance, 0.86), (square error, 0.83), (estimators, 0.82)
  \end{itemize}

  \item Document B:
    \begin{itemize}
\item $P_\theta$: (family, 0.87)
\item $X$: (measurable space, 0.95), (Poisson, 0.82)
\item $\theta$: (sufficient statistic, 0.93)
\item $\mu$: (mean, 0.99), (variance, 0.95), (random variables, 0.89), (normal, 0.83)
\item $\sigma$: (variance, 0.99), (mean, 0.83)
  \end{itemize}

  \item Document C:
    \begin{itemize}
\item $n$: (tickets, 0.96), (maximum-likelihood estimator, 0.89)
\item $x$: (data, 0.99), (observations, 0.93)
\item $\theta$: (statistic, 0.95), (estimator, 0.93), (estimator, 0.93), (rise, 0.91), (statistical model, 0.85), (fixed constant, 0.82)
\item $\mu$: (expectation, 0.96), (variance, 0.93), (population, 0.89)
\item $\sigma$: (variance, 0.94), (population variance, 0.91), (estimator, 0.87)
  \end{itemize}
\end{itemize}

We take all these relations, and combine together. If an identifer
has two or more definitions that are exactly the same, them we
merge them into one and its score is the sum of scores:

\begin{itemize}
\item $P_\theta$: (family, 0.87)
\item $X$: (measurable space, 0.95), (Poisson, 0.82)
\item $n$: (tickets, 0.96), (predictions, 0.95), (size, 0.92), (maximum-likelihood estimator, 0.89), (random sample, 0.82), (population, 0.82)
\item $x$: (data, 0.99), (observations, 0.93)
\item $\theta$: (estimator, 0.98+0.93+0.93), (unknown parameter, 0.98+0.94), (statistic, 0.95), (sufficient statistic, 0.93), (rise, 0.91), (statistical model, 0.85), (fixed constant, 0.82)
\item $\mu$: (random variables, 0.89+0.89+0.89), (variance, 0.95+0.93), (mean, 0.99), (true mean, 0.96), (expectation, 0.96), (normal, 0.83)
\item $\mu_4$: (central moment, 0.83)
\item $\sigma$: (variance, 0.99+0.94), (population variance, 0.91+0.86), (estimator, 0.87), (square error, 0.83), (mean, 0.83), (estimators, 0.82)
\end{itemize}

There is some lexical variance in the definitions. For example, ``variance'' and ``population
variance'' or ``mean'' and ``true mean'' are very related definitions, and
it makes sense to group them together to form one definition.
It can be done by fuzzy string matching (or approximate matching)
\cite{navarro2001guided}. To implement it, we use a Python library FuzzyWuzzy
\cite{fuzzywuzzy}, and using fuzzy matching we group related identifiers and then
sum over their scores.

Then we have the following:

\begin{itemize}
\item $P_\theta$: (family, 0.87)
\item $X$: (measurable space, 0.95), (Poisson, 0.82)
\item $n$: (tickets, 0.96), (predictions, 0.95), (size, 0.92), (maximum-likelihood estimator, 0.89), (random sample, 0.82), (population, 0.82)
\item $x$: (data, 0.99), (observations, 0.93)
\item $\theta$: (estimator, 2.84), (unknown parameter, 1.92), (\{statistic, sufficient statistic\}, 1.88), (rise, 0.91), (statistical model, 0.85), (fixed constant, 0.82)
\item $\mu$: (random variables, 2.67), (\{mean, true mean\}, 1.95), (variance, 1.88),  (expectation, 0.96), (normal, 0.83)
\item $\mu_4$: (central moment, 0.83)
\item $\sigma$: (\{variance, population variance\}, 3.7), (\{estimator, estimators\}, 1.69), (square error, 0.83), (mean, 0.83)
\end{itemize}

In a namespace an identifier can have at most one definition, and therefore the next step
is selecting the definition with the highest score. This gives us the following namespace:

\begin{itemize}
\item ($P_\theta$, family, 0.87)
\item ($X$, measurable space, 0.95)
\item ($n$, tickets, 0.96)
\item ($x$, data, 0.99)
\item ($\theta$, estimator, 2.84)
\item ($\mu$, random variables, 2.67)
\item ($\mu_4$, central moment, 0.83)
\item ($\sigma$: variance, 3.7)
\end{itemize}

Intuitively, the more a relation occurs,  the higher the score, and it
gives us more confidence that the definition is indeed correct.

\begin{figure}[h!]
\centering\includegraphics[width=0.6\textwidth]{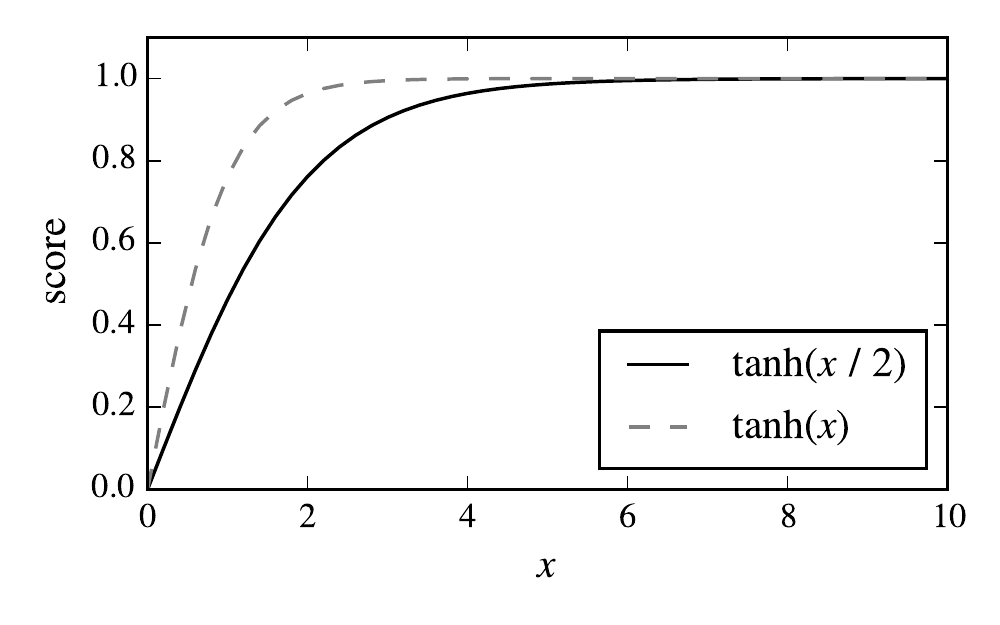}
\caption{Non-linear transformation of scores with $\tanh$ function.}
\label{fig:tanh}
\end{figure}

However, it is more convenient to compare scores when they are on the $[0, 1]$
scale, and therefore we may apply additional transformation to convert the scores.
Hyperbolic tangent function is a good choice for such transformation:
it is near zero for small values and it never exceeds one (see fig.~\ref{fig:tanh})
But it is quite steep and converges to 1 rather quickly: for a relation
with score of 3 it would produce a score of 0.99, which is quite high.
Instead, we can use a less steep $\tanh (x/2)$ (see fig.~\ref{fig:tanh}):
for 3,  $\tanh (3/2)$ produces a score of 0.90, which is better.

Because the hyperbolic tangent is a monotonously increasing function, larger
values of the original score correspond to larger values of the transformed
score, and therefore we still can use the output of $\tanh (x/2)$ to
rank the relations.

Thus, after applying this transformation, we obtain the following namespace:

\begin{itemize}
\item ($P_\theta$, family, 0.41)
\item ($X$, measurable space, 0.44)
\item ($n$, tickets, 0.45)
\item ($x$, data, 0.46)
\item ($\theta$, estimator, 0.89)
\item ($\mu$, random variables, 0.87)
\item ($\mu_4$, central moment, 0.39)
\item ($\sigma$: variance, 0.95)
\end{itemize}

The name for this namespace is selected as the category that the majority of the documents
in the namespace-defining cluster share.

\section{Evaluation} \label{sec:evaluation}

In this chapter we describe the experimental setup and the obtained
results.

The following software is used for the experiments: 
Apache Flink 0.8.1 \cite{flink}
numpy 1.9.2 \cite{walt2011numpy}, scipy 0.15.1 \cite{scipy}, 
scikit-learn 0.16.1 \cite{scikit-learn} and IPython notebook 3.1.0. 
The experiments are run on a 64-bit Windows machine with two CPUs of 2.10 GHz 
and 8 GB of RAM. 

First, section~\ref{sec:jlp} verifies that the namespace discovery is possible
by applying the proposed technique to Java source code.
Next, section~\ref{sec:param-tuning} describes parameter tuning: there
are many possible choices of parameters, and we find the best.
Once the best algorithm and its parameters are selected, we analyze
the obtained results in section~\ref{sec:result-analysis}. Next,
we describe how the discovered clusters can be mapped to a hierarchy
in section~\ref{sec:hierarchy}, and finish by summarizing our findings
in section~\ref{sec:evaluation-summary}.

\subsection{Java Language Processing} \label{sec:jlp}

There is no ``gold standard'' data set for the namespace discovery problem
because this problem has not been studied, and it is hard to verify if our
approach works or not.

Previously we have illustrated the idea of identifier namespaces by comparing
it with namespaces in Computer Science, and it allowed us to develop an intuition
behind the namespaces in mathematics and also propose a method to discover them:
we motivated the assumption that there exist ``namespace defining''
groups of documents by arguing that these groups also exist in
programming languages.

Therefore we try to use source code as a dataset, and see whether our method is able to
recover namespace information or not.

If a programming language is statically typed, like Java or Pascal,
usually it is possible to know the type of a variable from the declaration
of this variable. Therefore we can see variable names as ``identifiers''
and variable types as ``definitions''. Clearly, there is a difference
between variable types and identifier definitions, but we believe
that this comparison is valid because the type carries additional semantic
information about the variable and in what context it can be used --
like the definition of an identifier.

The information about variables and their types can be extracted from a
source code repository, and each source file can be processed to
obtain its Abstract Syntax Tree (AST). By processing the ASTs,
we can extract the variable declaration information. Thus, each
source file can be seen as a document, which is represented
by all its variable declarations.

In this work we process Java source code, and for parsing it
and building ASTs we use a library JavaParser \cite{javaparser}.
The Java programming language was chosen because it requires the programmer
to always specify the type information when declaring a variable.
It is different for other languages when the type information is
usually inferred by the compilers at compilation time.

In Java a variable can be declared in three places:
as an inner class variable (or a ``field''), as a method (constructor)
parameter or as a local variable inside a method or a constructor.
We need to process all three types of variable declarations
and then apply additional preprocessing, such as converting the name
of the type from short to fully qualified using the information from the
import statements. For example, \verb|String| is converted to
\verb|java.lang.String| and \verb|List<Integer>| to \verb|java.util.List<Integer>|,
but primitive types like \verb|byte| or \verb|int| are left unchanged.

Consider an example in the listing~\ref{code:javaclass}. There is a
class variable \texttt{threshold}, a method parameter \texttt{in} and
two local variables \texttt{word} and \texttt{posTag}. The following
relations will be extracted from this class: (``threshold'', \verb|double|),
(``in'', \verb|domain.Word|), (``word'', \verb|java.lang.String|),
(``posTag'', \verb|java.lang.String|).
Since all primitives and classes from packages that start with
\verb|java| are discarded, at the end the class \verb|WordProcesser|
is represented with only one relation (``in'', \verb|domain.Word|).

\begin{lstlisting}[language=Java,caption={A Java class},label={code:javaclass}]
package process;

import domain.Word;

public class WordProcesser  {

    private double threshold;

    public boolean isGood(Word in) {
        String word = in.getWord();
        String posTag = in.getPosTag();
        return isWordGood(word) && isPosTagGood(posTag);
    }

    // ...

}
\end{lstlisting}

In the experiments we applied this source code analysis to
the source code of Apache Mahout 0.10 \cite{mahout}, which is an open-source
library for scalable Machine Learning and Data Mining.
As on \today, this dataset consists of 1\,560 java classes with 45\,878
variable declarations. After discarding declarations from the standard Java API,
primitives and types with generic parameters, only 15\,869 declarations were
retained.

The following is top-15 variable/type declarations extracted from the Mahout
source code:

\begin{itemize}
\item ``conf'', \verb|org.apache.hadoop.conf.Configuration| (491 times)
\item ``v'', \verb|org.apache.mahout.math.Vector| (224 times)
\item ``dataModel'', \verb|org.apache.mahout.cf.taste.model.DataModel| (207 times)
\item ``fs'', \verb|org.apache.hadoop.fs.FileSystem| (207 times)
\item ``log'', \verb|org.slf4j.Logger| (171 times)
\item ``output'', \verb|org.apache.hadoop.fs.Path| (152 times)
\item ``vector'', \verb|org.apache.mahout.math.Vector| (145 times)
\item ``x'', \verb|org.apache.mahout.math.Vector| (120 times)
\item ``path'', \verb|org.apache.hadoop.fs.Path| (113 times)
\item ``measure'', \verb|org.apache.mahout.common.distance.DistanceMeasure| (102 times)
\item ``input'', \verb|org.apache.hadoop.fs.Path| (101 times)
\item ``y'', \verb|org.apache.mahout.math.Vector| (87 times)
\item ``comp'', \verb|org.apache.mahout.math.function.IntComparator| (74 times)
\item ``job'', \verb|org.apache.hadoop.mapreduce.Job| (71 times)
\item ``m'', \verb|org.apache.mahout.math.Matrix| (70 times)
\end{itemize}

We use the ``soft'' association method to incorporate ``definition''
(i.e. types), and considering each source code file as a document,
we build an identifier-document matrix of dimensionality
$1436 \times 1560$. Only identifiers that occur at least twice are used
to build the matrix.

\begin{figure}[h!]
\centering
\hfill
\begin{subfigure}[b]{0.47\textwidth}
  \centering
  \includegraphics[width=\textwidth]{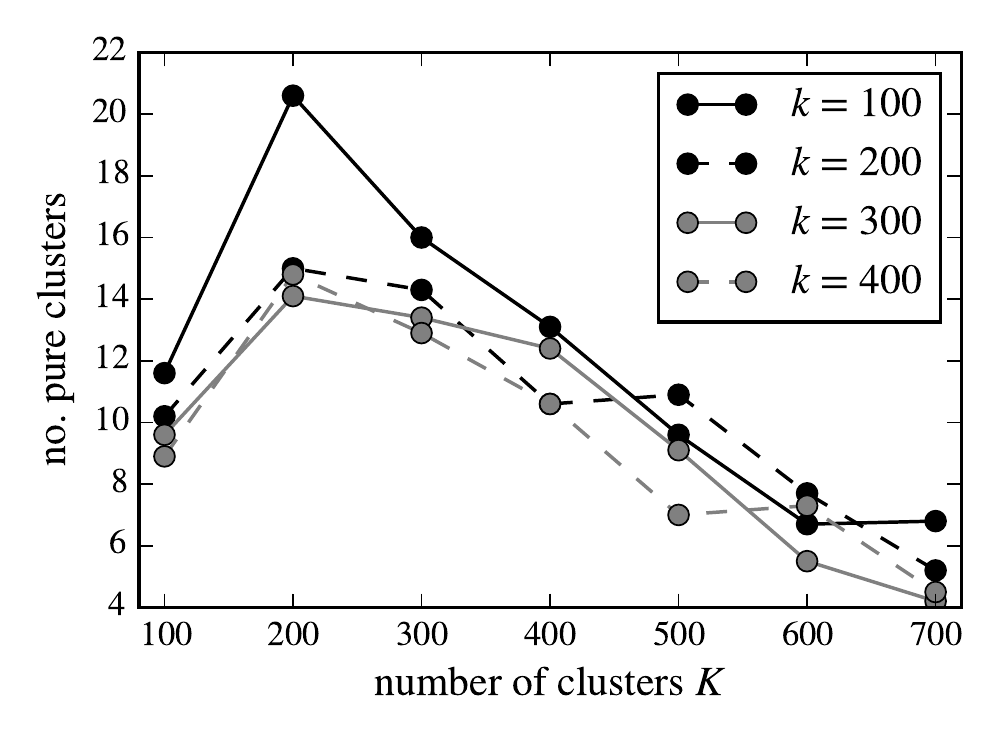}
  \caption{Usual TF $\times$ IDF weighting}
  \label{fig:jlp-perf-tf}
\end{subfigure}
~
\begin{subfigure}[b]{0.47\textwidth}
  \centering
  \includegraphics[width=\textwidth]{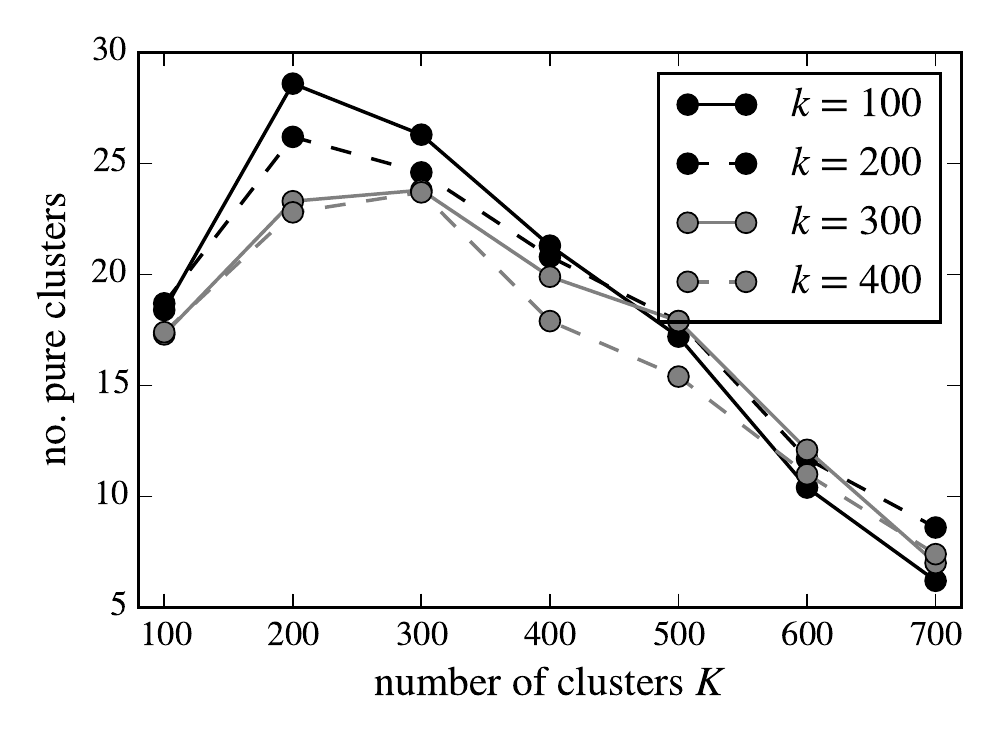}
  \caption{$(\log \text{TF}) \times \text{IDF}$ weighting}
  \label{fig:jlp-perf-subtf}
\end{subfigure}
\caption{The performance MiniBatch $K$-Means on the Mahout dataset.}
\label{fig:jlp-perf}
\end{figure}

To discover namespace-defining cluster, we use LSA with SVD and MiniBatch $K$-Means.
For weight assignment, we try two weighting systems: usual TF component and sublinear TF.
The best performance is achieved with the rank $k=200$ of SVD and
number of clusters $K=200$ using sublinear weighting (see fig.~\ref{fig:jlp-perf}).

With these parameters the best result is 33 clusters (see table~\ref{tab:mahout-namespaces}).
One of such clusters is a cluster about SVD: there are 5 classes
from the \verb|svd|\footnote{full name: \texttt{org.apache.mahout.cf.taste.impl.recommender.svd}} package
(\texttt{Factorization}, \texttt{FilePersistenceStrategy}, \texttt{NoPersistenceStrategy}, \texttt{PersistenceStrategy}, \texttt{FilePersistenceStrategyTest})
and one from \\ \verb|kddcup.track1.svd|\footnote{full name: \texttt{org.apache.mahout.cf.taste.example.kddcup.track1.svd}} package
(\texttt{Track1SVDRunner}).
Although this cluster is not 100\% pure, in the sense that not all of
these classes belong to the same package, these classes are
clearly related: they are all about SVD.
The top dimensions with the most influence in this cluster are
\verb|svd.Factorization|\footnote{full name: \texttt{org.apache.mahout.cf.taste.impl.recommender.svd.Factorization}}
and \texttt{factorization}.

Another interesting namespace is the distance namespace, with classes mostly
from \verb|common.distance|\footnote{full name: \texttt{org.apache.mahout.common.distance}}
(see table~\ref{tab:mahout-dist-classes}).
What is interesting, there are mostly tests in this cluster, but because they use
classes that they test, they roughly can be seen as documents that refer to
some concepts defined outside of the cluster (see table~\ref{tab:mahout-dist-ids}).

\begin{table}[h!]
\centering
\begin{subtable}{\textwidth}
\centering
\begin{tabular}{|c|c|c|}
\hline
Size & Namespace & Purity \\
\hline
16 & \verb|org.apache.mahout.h2obindings.ops| & 0.88 \\
12 & \verb|org.apache.mahout.math.map| & 0.92 \\
11 & \verb|org.apache.mahout.text| & 0.91 \\
10 & \verb|org.apache.mahout.vectorizer.collocations.llr| & 1.00 \\
10 & \verb|org.apache.mahout.classifier.sequencelearning.hmm| & 1.00 \\
10 & \verb|org.apache.mahout.math.list| & 1.00 \\
9 & \verb|org.apache.mahout.math.map| & 0.89 \\
9 & \verb|org.apache.mahout.cf.taste.hadoop.item| & 1.00 \\
8 & \verb|org.apache.mahout.cf.taste.hadoop.als| & 1.00 \\
8 & \verb|org.apache.mahout.classifier.sgd| & 0.88 \\
\hline
\end{tabular}
\caption{Top namespace-defining clusters discovered from Mahout.}
\label{tab:mahout-namespaces}
\end{subtable}
\begin{subtable}{\textwidth}
\centering
\begin{tabular}{|c|}
  \hline
Class name \\
\hline
\verb|org.apache.mahout.math.neighborhood.UpdatableSearcher| \\
\verb|org.apache.mahout.common.distance.CosineDistanceMeasureTest| \\
\verb|org.apache.mahout.common.distance.DefaultDistanceMeasureTest| \\
\verb|org.apache.mahout.common.distance.DefaultWeightedDistanceMeasureTest| \\
\verb|org.apache.mahout.common.distance.TestChebyshevMeasure| \\
\verb|org.apache.mahout.common.distance.TestMinkowskiMeasure| \\
\hline
\end{tabular}
\caption{A namespace-defining cluster about Distances.}
\label{tab:mahout-dist-classes}
\end{subtable}
\begin{subtable}{\textwidth}
\centering
\begin{tabular}{|c|c|}
  \hline
  ID & Class  \\
  \hline
\verb|chebyshevDistanceMeasure| &  \verb|org.apache.mahout.common.distance.DistanceMeasure| \\
\verb|distanceMeasure| &  \verb|org.apache.mahout.common.distance.DistanceMeasure|  \\
\verb|euclideanDistanceMeasure| &  \verb|org.apache.mahout.common.distance.DistanceMeasure|  \\
\verb|manhattanDistanceMeasure| &  \verb|org.apache.mahout.common.distance.DistanceMeasure|  \\
\verb|minkowskiDistanceMeasure| &  \verb|org.apache.mahout.common.distance.DistanceMeasure|  \\
\verb|v| &  \verb|org.apache.mahout.math.Vector|  \\
\verb|vector| &  \verb|org.apache.mahout.math.Vector| \\
\hline
\end{tabular}
\caption{Definitions in the v namespace.}
\label{tab:mahout-dist-ids}
\end{subtable}
\caption{Namespaces extracted from Apache Mahout source code.}
\label{tab:mahout-namespaces-dist}
\end{table}


We are able to discover 33 namespaces using the Mahout source code as a dataset.
There are 150 packages in total. This means that using only identifier information
it is enough to recover namespace information at least partially (in our case,
we can recover 22\% of the namespaces).

Therefore, in principle, the approach works and it can be used to discover
namespaces in mathematics.

\subsection{Parameter Tuning} \label{sec:param-tuning}

There are many different clustering algorithms, each with its own set
of parameter. In this section we describe how we find the settings that
find the best namespaces.

The following things can be changed:

\begin{itemize}
  \item Ways to incorporate definition information (no definitions, soft association, hard association);
  \item Weighting schemes for the identifier-document matrix $D$: TF, sublinear TF, TF-IDF;
  \item There are different clustering algorithms: agglomerative clustering, DBSCAN, SNN
  clustering, $K$-Means, 
  and each algorithm has its own set of parameters;
  \item Dimensionality of $D$ can be reduced via SVD or NMF, parameter $k$ controls the
      rank of output.
\end{itemize}

To find the best parameters set we use the grid search approach: we try
different combinations of parameters and keep track on the number of
pure clusters and the purity.

The overall purity of cluster assignment is calculated as a weighed sum
of individual  cluster purities, where the weight is chosen proportionally to
the size of a cluster.

\begin{figure}[h!]
\centering
\begin{subfigure}[b]{0.75\textwidth}
  \centering\includegraphics[width=\textwidth]{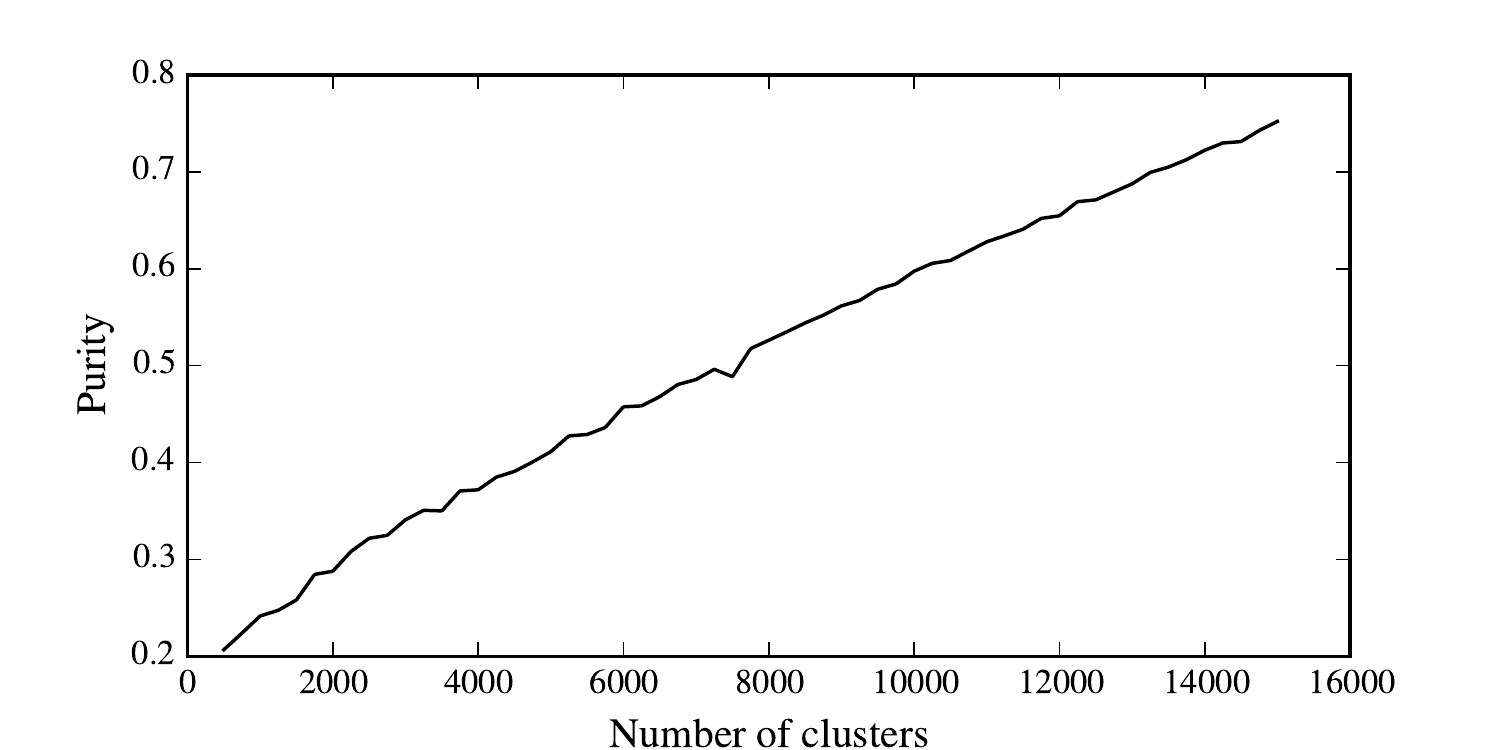}
  \caption{Number of clusters $K$ vs overall purity of clustering:
      the purity increases linearly with $K$ ($R^2 = 0.99$).}
  \label{fig:k-vs-purity}
\end{subfigure}
\begin{subfigure}[b]{0.75\textwidth}
  \centering\includegraphics[width=\textwidth]{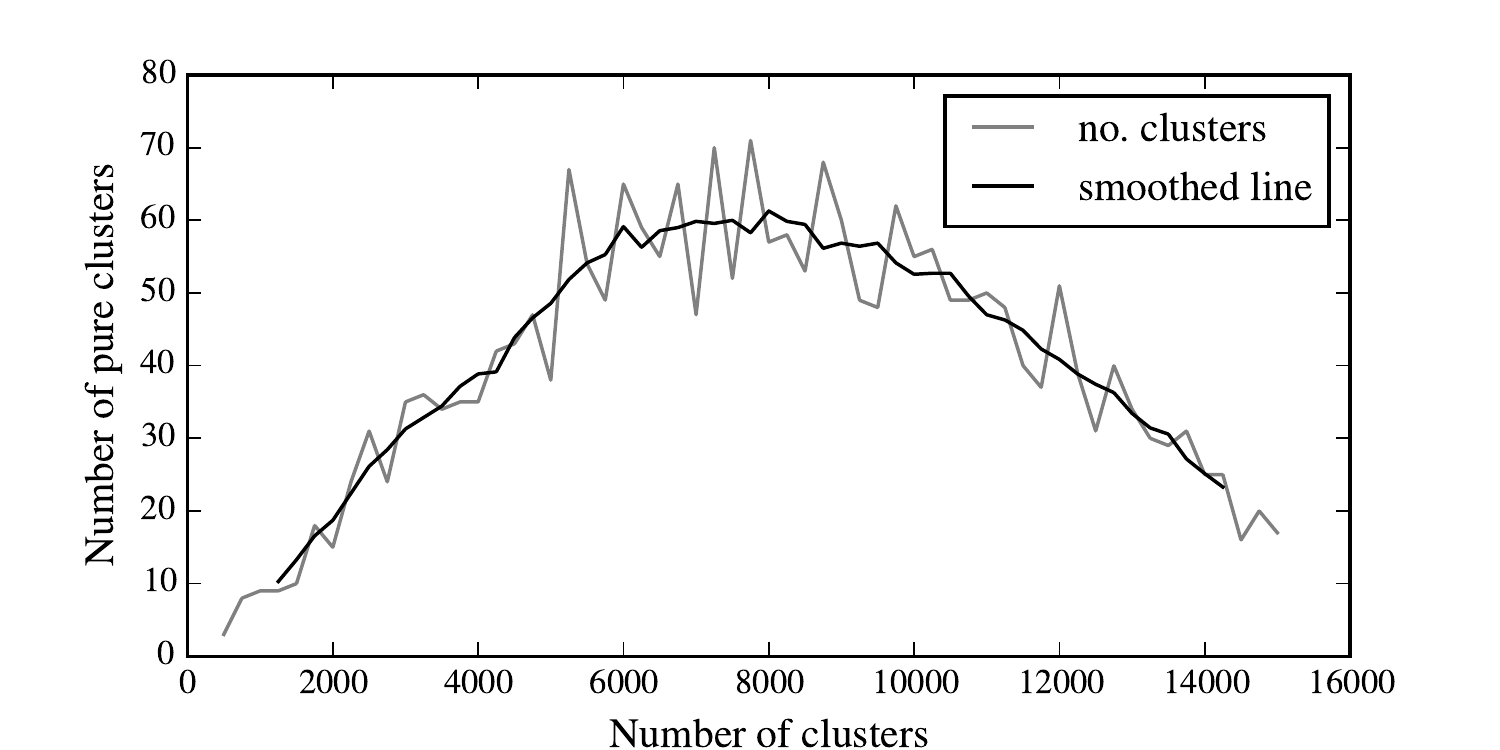}
  \caption{Number of clusters $K$ vs the number of pure clusters: it grows initially, but after $K\approx 8\,000$ starts to decrease.}
  \label{fig:k-vs-pureclusters}
\end{subfigure}
\caption{Purity and number of pure clusters as measures of algorithm performance.}
\label{fig:performace}
\end{figure}

However it is not enough just to find the most pure cluster assignment: because
as the number of clusters increases the overall purity also grows.
Thus we can also optimize for the number of clusters with purity $p$ of
size at least $n$.
When the number of clusters increase, the purity always grows
(see fig.~\ref{fig:k-vs-purity}), but at some point the number of pure clusters
will start decreasing (see fig.~\ref{fig:k-vs-pureclusters}).

\subsubsection{Baseline} \ \\

We compare the performance of clustering algorithms against a random
categorizer. The simplest version of such a categorizer is the random
cluster assignment categorizer, which assigns each document to some random
cluster.
In this case, we constrain the categorizer to include 3 documents in each
cluster, and once a document belongs to some cluster, it cannot be re-assigned.
It is done by first creating a vector of assignments and shuffling it.

Then we record how many pure clusters (at least 80\% pure) are in the cluster
assignment.

\begin{figure}[h!]
\centering\includegraphics[width=0.7\textwidth]{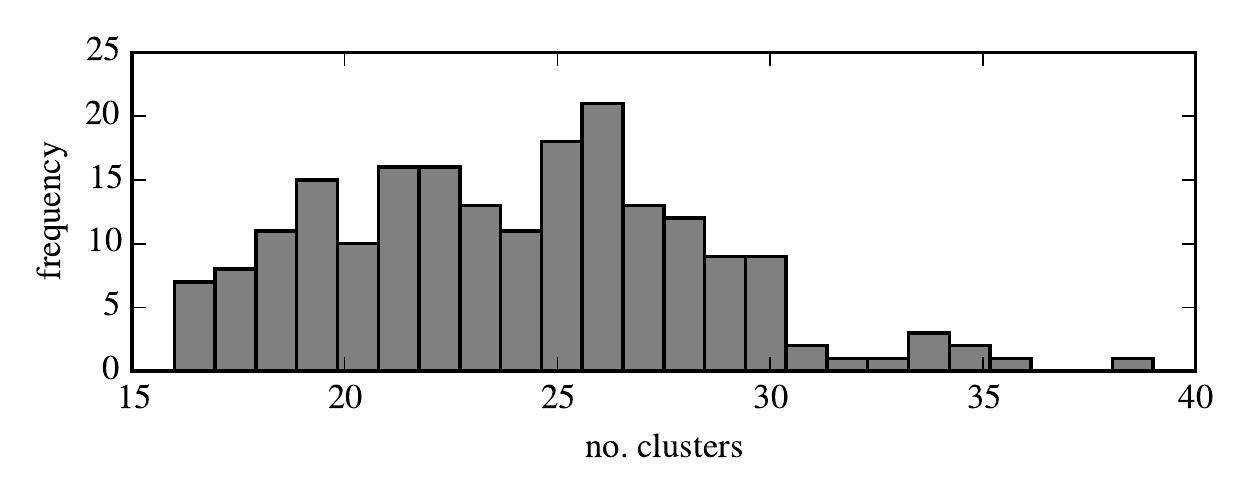}
\caption{Distribution of the number of pure clusters across 200 trials.}
\label{fig:baseline}
\end{figure}

To establish the baseline, we repeated this experiment for 200
times (see fig.~\ref{fig:baseline}), and the maximal achieved value is 39 pure
clusters, while the mean value is 23.85.


\subsubsection{Only Identifiers} \ \\

The first way of building the identifier space is to use only identifiers
andW not use definitions at all.
If we do this, the identifier-document matrix is $6075 \times 22512$
(we keep only identifiers that occur at least twice), and it contains  302\, 541
records, so the density of this matrix is just 0.002.

First, we try to apply agglomerative clustering, then DBSCAN with SNN similarity
based on Jaccard coefficient and cosine similarity, then we
do $K$-Means and finally we apply LSA using SVD and NMF and apply
$K$-Means on the reduced space.

\begin{figure}[h!]
\centering\includegraphics[width=0.75\textwidth]{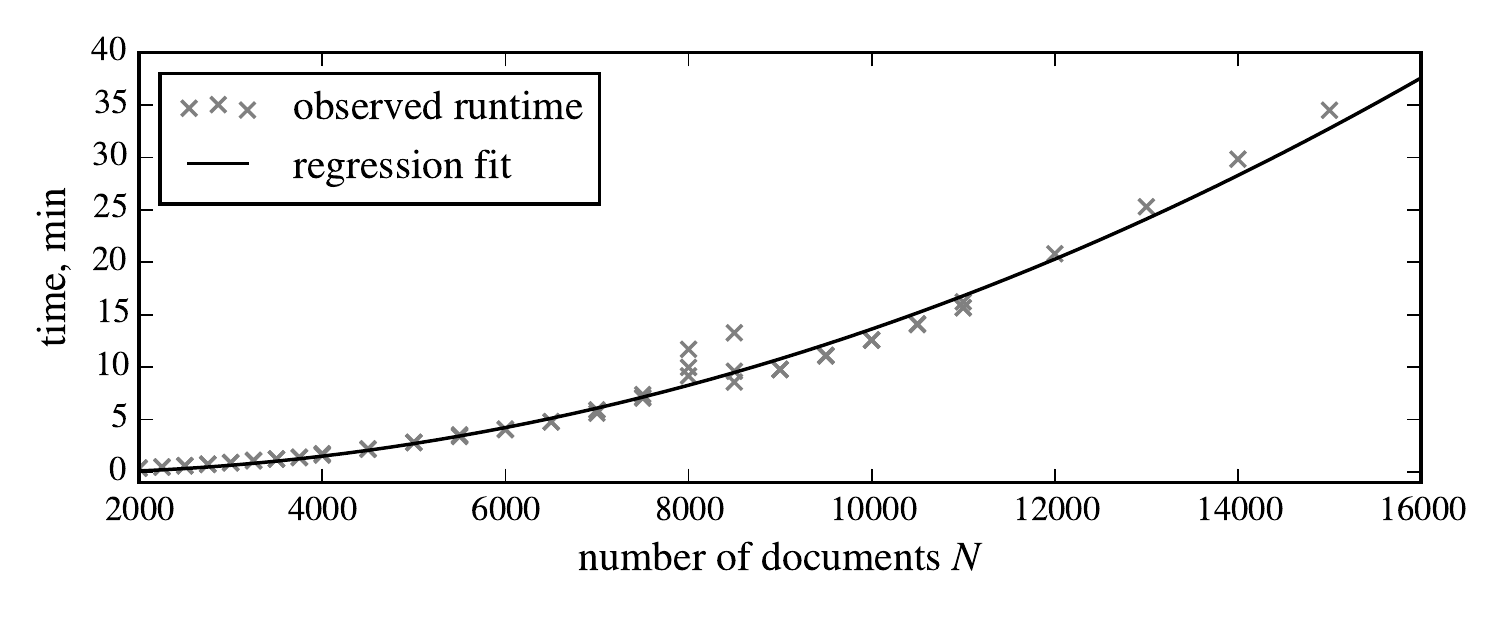}
\caption{Runtime of agglomerative clustering is quadratic with the number of documents
in the collection ($R^2 = 0.99$).}
\label{fig:agglo-time}
\end{figure}

\textbf{Agglomerative clustering} algorithms are quite fast for small datasets,
but they become more computationally expensive as the dataset size grows.
We run a series of experiments on subsamples of our dataset and we can observe that the run
time is quadratic with the number of documents to cluster
(see fig.~\ref{fig:agglo-time}). The polynomial regression model that we built predicts
that it should process the entire dataset of 22\,512 documents in 90 minutes,
but it was not able to finish in several hours, so we stopped the computation.
Additionally, the implementation we use from scikit-learn
requires a dense matrix. But when densified, the identifier-document matrix
occupies a lot of space: if each element of the matrix is represented with a double
precision number, then this matrix occupies 1.01 GB of RAM in total. While
it is small enough to fit into memory, matrices of larger dimensionality might not.
Therefore, we exclude these clustering algorithms from further analysis.

\begin{figure}[h!]
\centering
\begin{subfigure}[b]{0.5\textwidth}
  \centering
  \includegraphics[width=\textwidth]{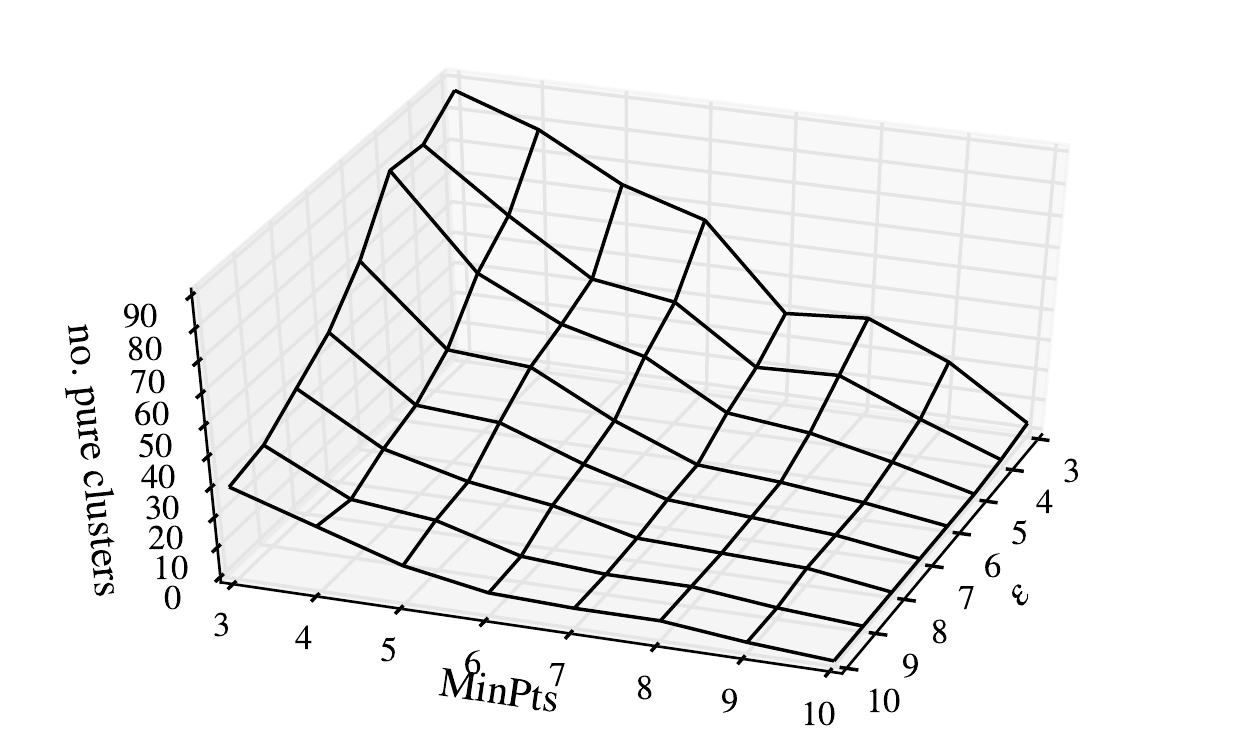}
  \caption{Number of clusters when 10 nearest \mbox{neighbors} are considered}
  \label{fig:nodef-dbscan-jac10}
\end{subfigure}%
\begin{subfigure}[b]{0.5\textwidth}
  \centering
  \includegraphics[width=\textwidth]{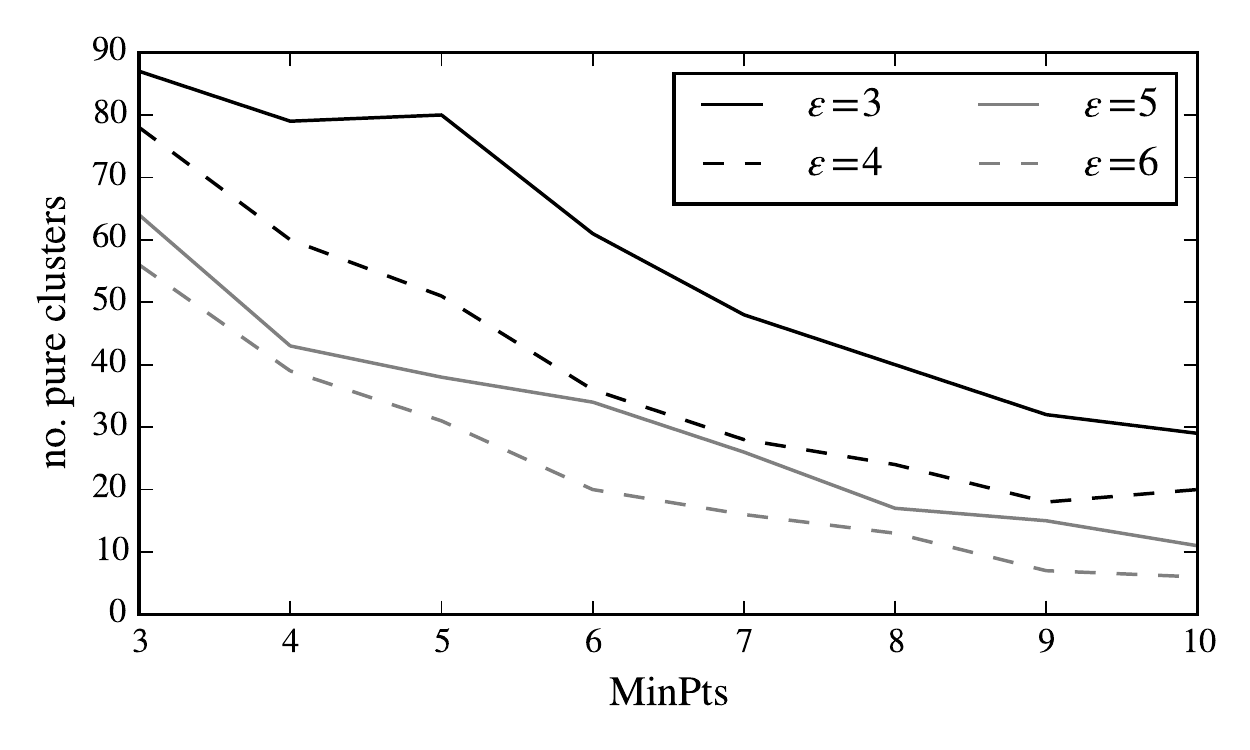}
  \caption{Performance of selected $\varepsilon$ with 10 nearest neighbors}
  \label{fig:nodef-dbscan-jac10-2}
\end{subfigure}
\begin{subfigure}[b]{0.5\textwidth}
  \centering
  \includegraphics[width=\textwidth]{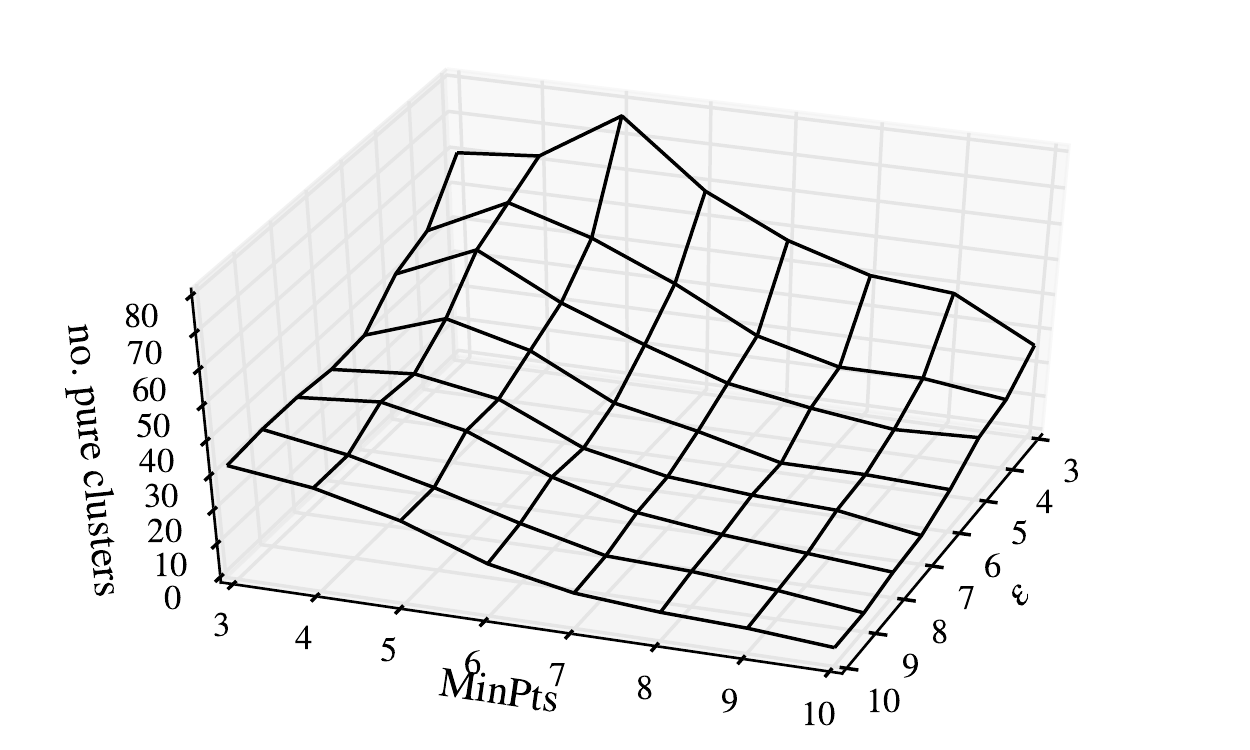}
  \caption{Number of clusters when 15 nearest \mbox{neighbors} are considered}
  \label{fig:nodef-dbscan-jac15}
\end{subfigure}%
\begin{subfigure}[b]{0.5\textwidth}
  \centering
  \includegraphics[width=\textwidth]{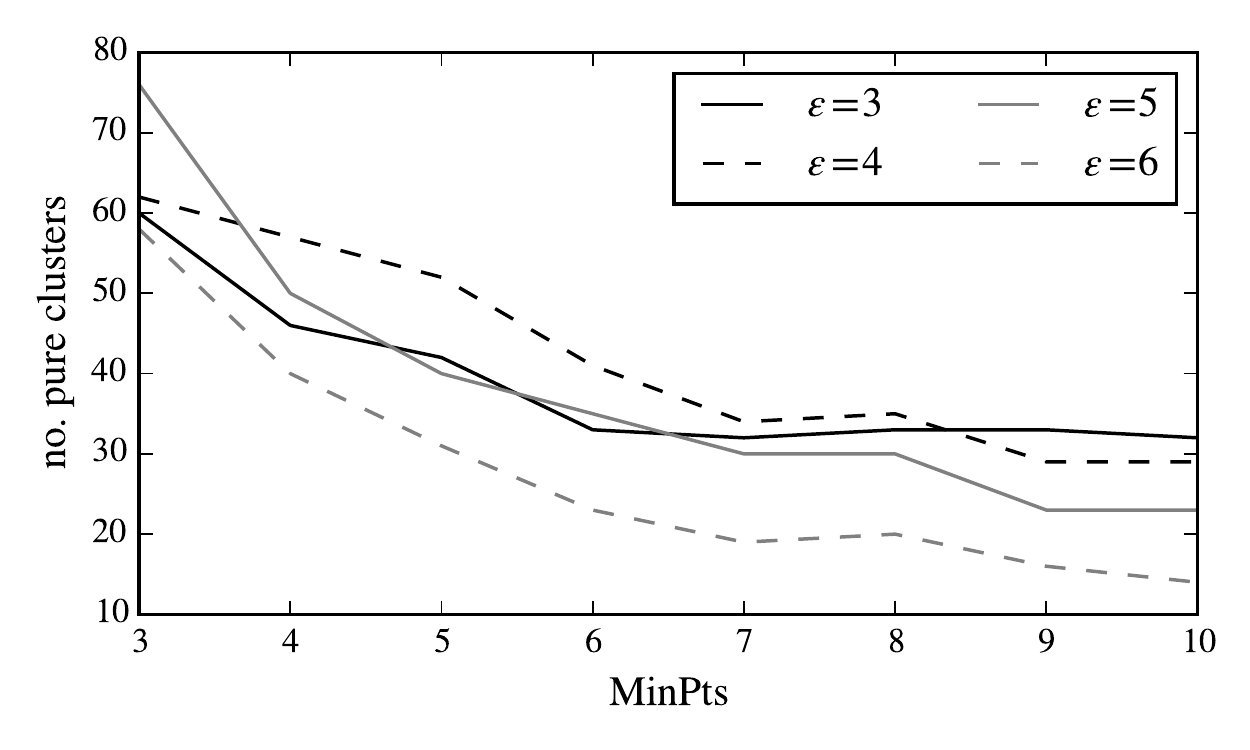}
  \caption{Performance of selected $\varepsilon$ with 15 nearest neighbors}
  \label{fig:nodef-dbscan-jac15-2}
\end{subfigure}
\caption{Effect of parameters $\varepsilon$, \texttt{MinPts} and number of nearest
 neighbors on performance of SNN DBSCAN when Jaccard coefficient is used.}
\label{fig:nodef-dbscan-jac}
\end{figure}

The second method is \textbf{DBSCAN} with \textbf{SNN Similarity}.
To compute SSN similarity we need to use some other base similarity measure.
We start with Jaccard coefficient, and use a binarized identifier-document
matrix: a matrix with only ones and zeros.
For example, the closest article to ``Linear Regression'' is
``Linear predictor function'' with Jaccard coefficient of 0.59
and ``Low-rank approximation'' is the closest to ``Singular value decomposition''
with coefficient of 0.25. With Jaccard, we were able to discover 87
clusters, which is two times better than the baseline (see fig.~\ref{fig:nodef-dbscan-jac})
and the best parameters are 10 nearest neighbors,
$\varepsilon=3$ and \texttt{MinPts} $=4$ (see fig.~\ref{fig:nodef-dbscan-jac10-2}).

Then we run the same algorithm, but with cosine similarity, using an
identifier-document matrix with $(\log \text{TF}) \times \text{IDF}$
weights, and calculate pair-wise similarity between each document.
For example, let us take an article ``Linear regression''
and calculate the cosine with the rest of the corpus. The closest document
is ``Linear predictor function''.
They have 23 identifiers in common, and they indeed look related.
However cosine is not always giving the best closets neighbors. For example,
the nearest neighbor of ``Singular value decomposition'' is ``Rule of Sarrus'',
and although their cosine score is 0.92, they have only 3 identifiers in common.

\begin{figure}[h!]
\centering
\begin{subfigure}[b]{0.5\textwidth}
  \centering
  \includegraphics[width=\textwidth]{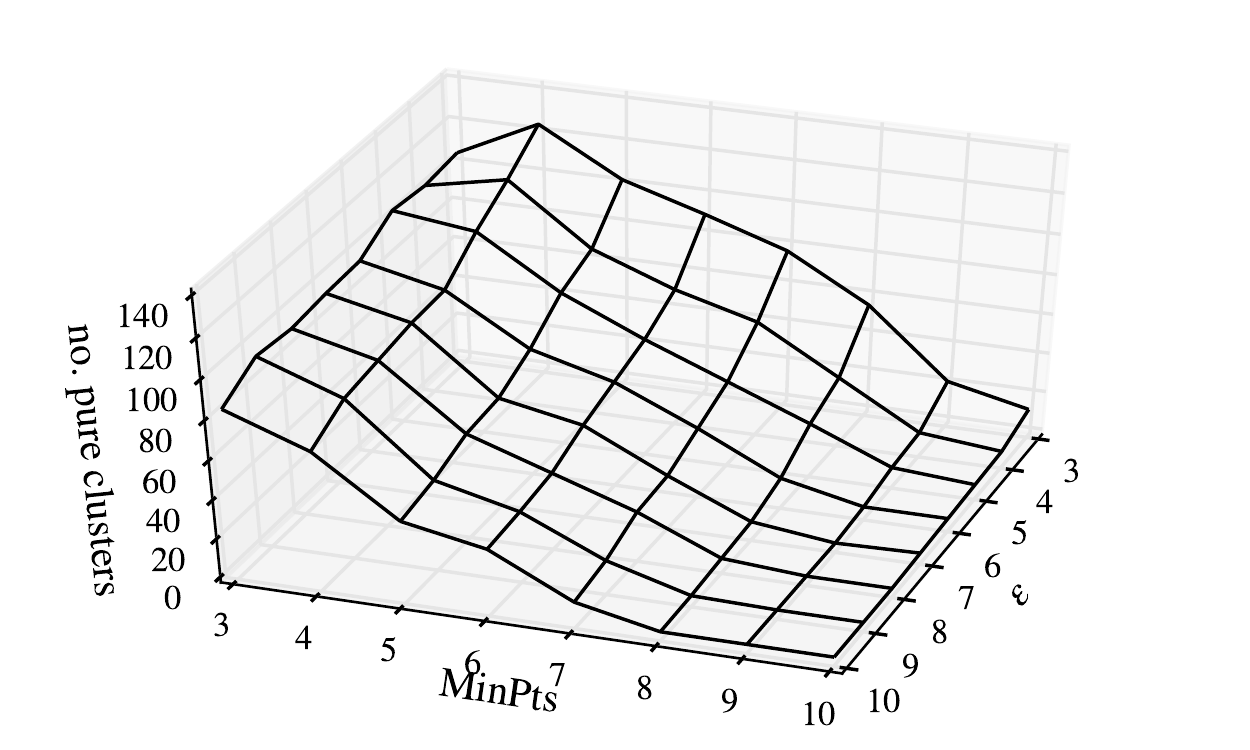}
  \caption{Number of clusters when 10 nearest \mbox{neighbors} are considered}
  \label{fig:nodef-dbscan-cos10}
\end{subfigure}%
\begin{subfigure}[b]{0.5\textwidth}
  \centering
  \includegraphics[width=\textwidth]{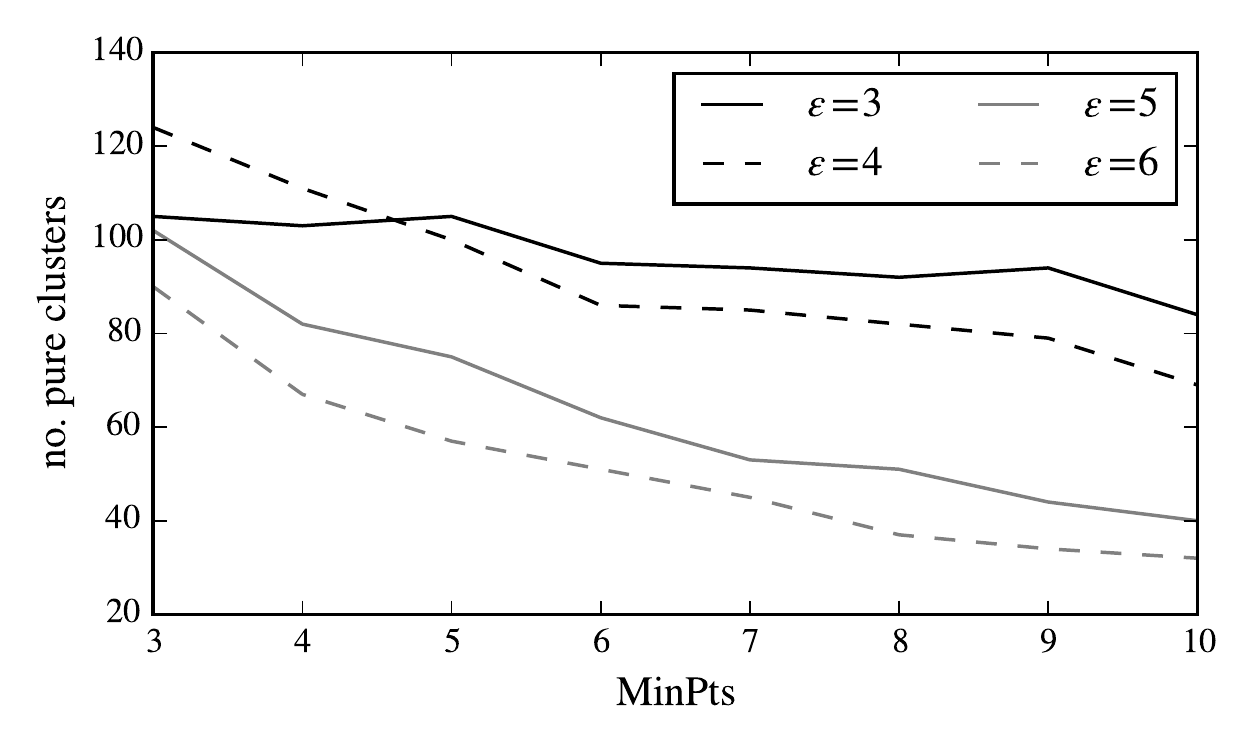}
  \caption{Performance of selected $\varepsilon$ with 10 nearest neighbors}
  \label{fig:nodef-dbscan-cos10-2}
\end{subfigure}
\begin{subfigure}[b]{0.5\textwidth}
  \centering
  \includegraphics[width=\textwidth]{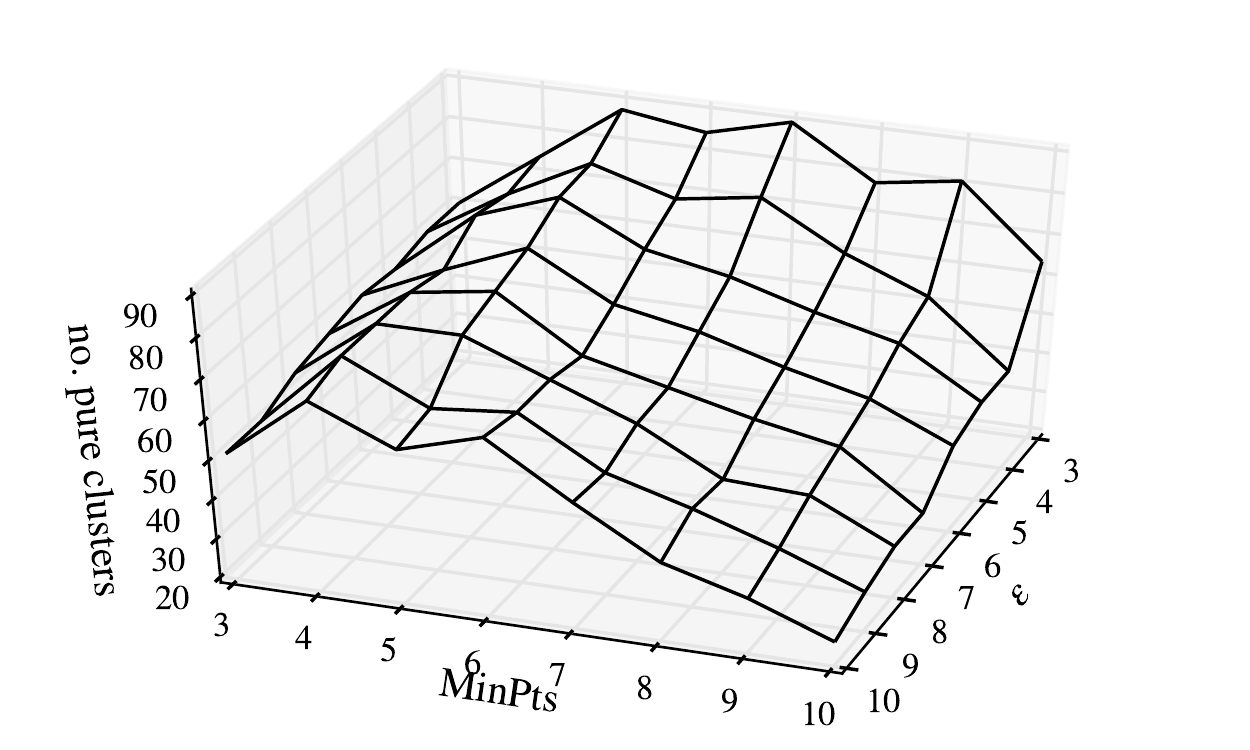}
  \caption{Number of clusters when 15 nearest \mbox{neighbors} are considered}
  \label{fig:nodef-dbscan-cos15}
\end{subfigure}%
\begin{subfigure}[b]{0.5\textwidth}
  \centering
  \includegraphics[width=\textwidth]{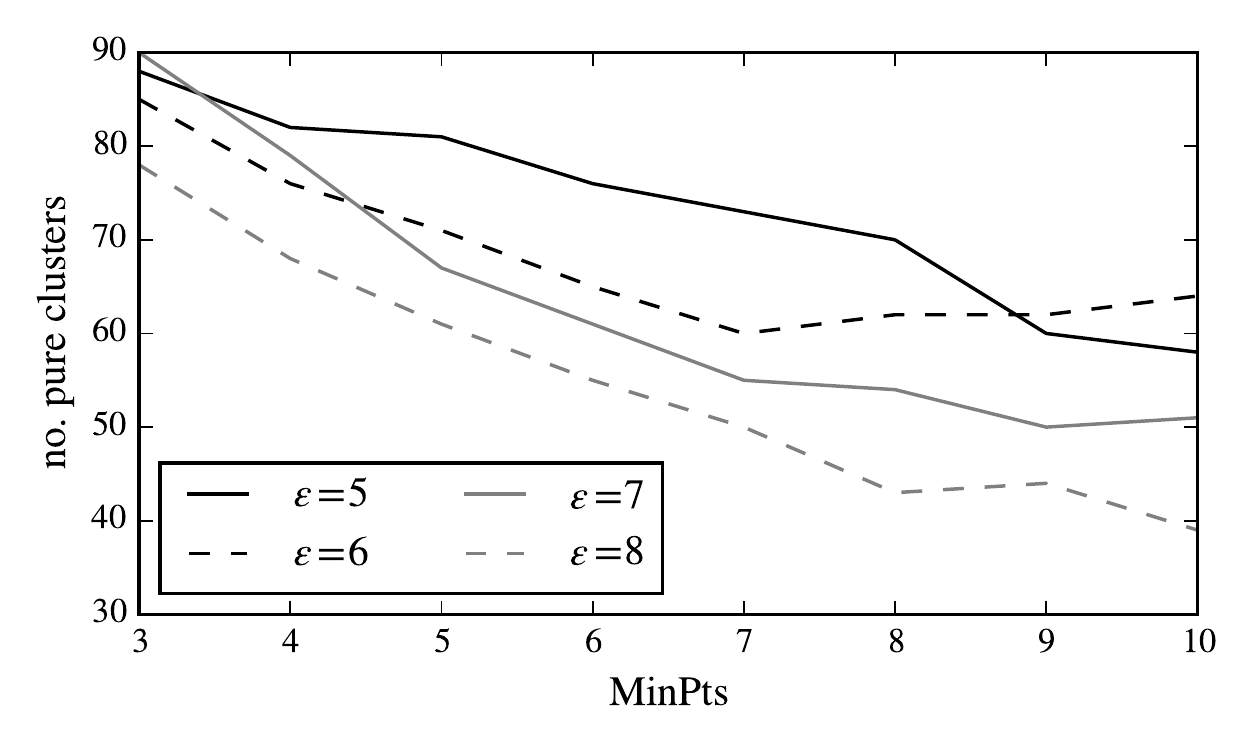}
  \caption{Performance of selected $\varepsilon$ with 15 nearest neighbors}
  \label{fig:nodef-dbscan-cos15-2}
\end{subfigure}
\caption{Effect of parameters $\varepsilon$, \texttt{MinPts} and number of nearest
 neighbors on performance of SNN DBSCAN when cosine is used.}
\label{fig:nodef-dbscan-cos}
\end{figure}

With cosine as the base similarity function for SNN DBSCAN we
were able to discover 124 namespace-defining clusters (see fig.~\ref{fig:nodef-dbscan-cos}),
which is significantly better than the baseline. The best parameters
are 10 nearest neighbors and $\varepsilon=4$, \texttt{MinPts} $=3$
(see fig.~\ref{fig:nodef-dbscan-cos10-2}).

\begin{figure}[h!]
\centering
\begin{subfigure}[b]{0.75\textwidth}
  \centering\includegraphics[width=\textwidth]{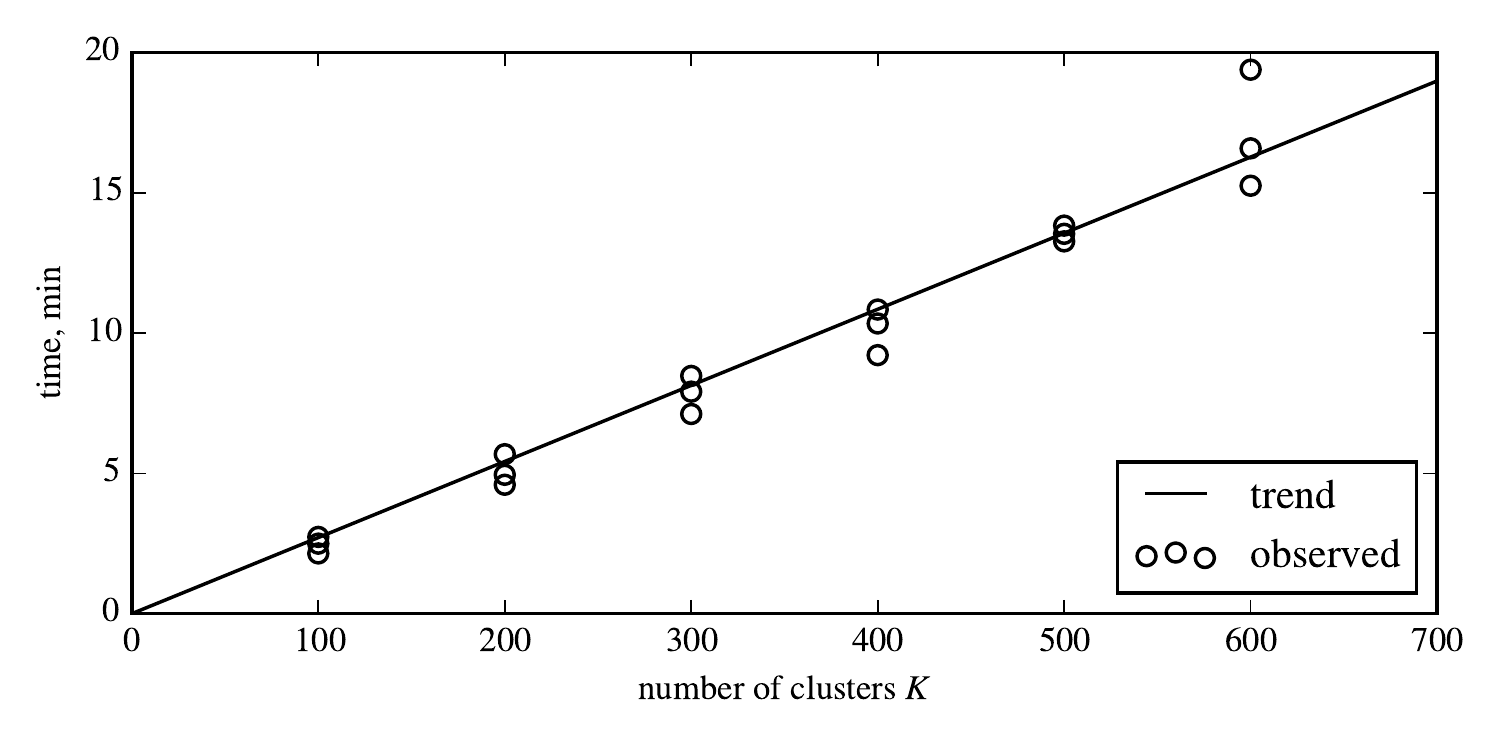}
  \caption{$K$ in $K$-Means vs time in minutes ($R^2 = 0.99$).}
  \label{fig:k-vs-time}
\end{subfigure}

\begin{subfigure}[b]{0.75\textwidth}
  \centering\includegraphics[width=\textwidth]{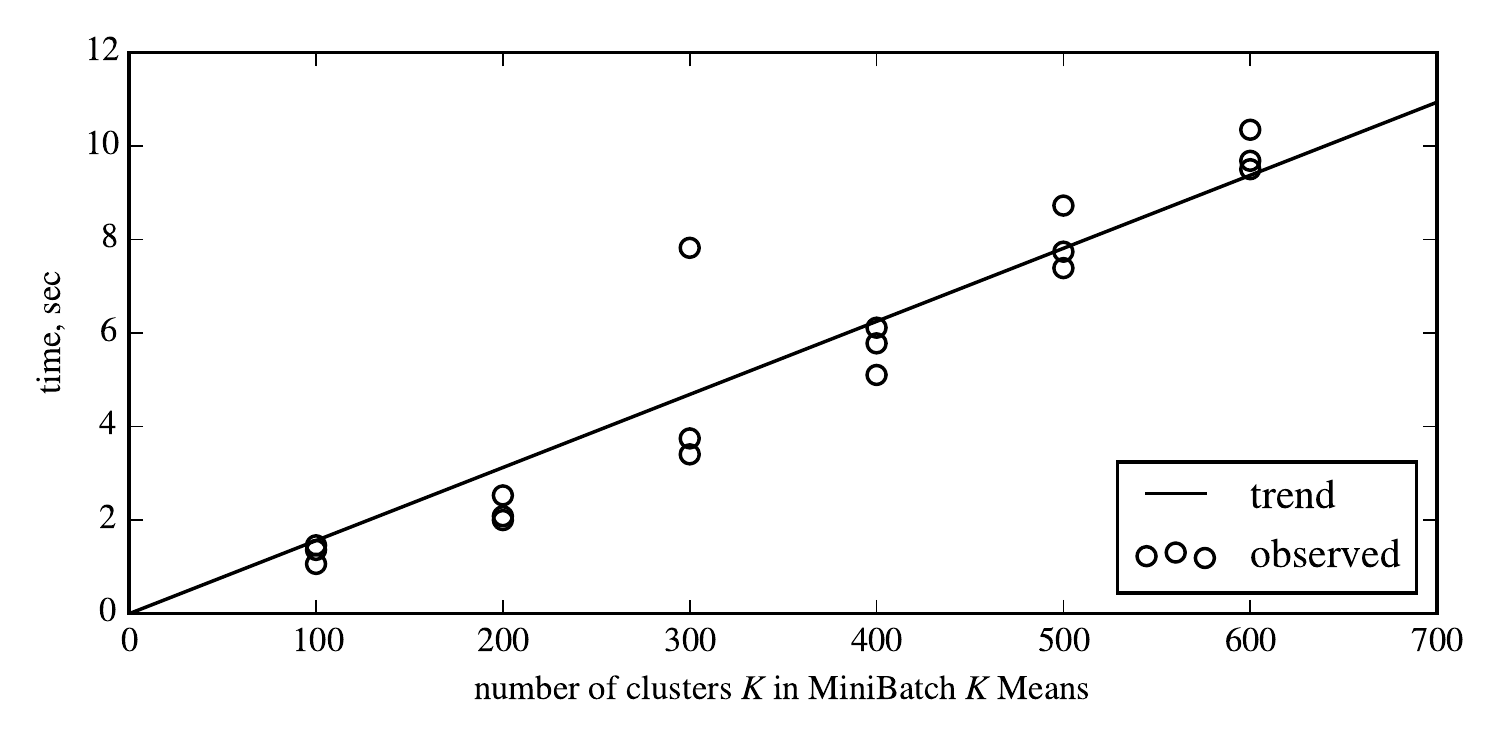}
  \caption{$K$ in MiniBatch $K$-Means vs time in seconds ($R^2 = 0.97$).}
  \label{fig:k-vs-time-minibatch}
\end{subfigure}
\caption{Runtime of $K$-Means and MiniBatch $K$-Means}
\label{fig:kmeans-vs-minibatch1}
\end{figure}

Next, we apply \textbf{$K$-Means}. We observe that increasing~$K$
leads to linear increase in time (see fig.~\ref{fig:k-vs-time}),
which means that for bigger values of~$K$, it takes longer, so it is not
feasible to run: for example, we estimate the runtime of $K$-Means with $K = 10\, 000$
to be about 4.5 hours. As \textbf{MiniBatch $K$-Means} is expected to be significantly
faster than usual $K$-Means, we use it as well. Although we observe that the run
time of  MiniBatch $K$-Means also increases linearly with~$K$
(see fig.~\ref{fig:k-vs-time-minibatch}), it indeed runs considerably faster.
For example, MiniBatch $K$-Means takes 15 seconds with $K=700$ while
usual $K$-Means takes about 15 minutes (see fig.~\ref{fig:k-vs-time}
and fig.~\ref{fig:k-vs-time-minibatch}).

\begin{figure}[h!]
\centering
\hfill
\begin{subfigure}[b]{0.5\textwidth}
  \centering
  \includegraphics[width=\textwidth]{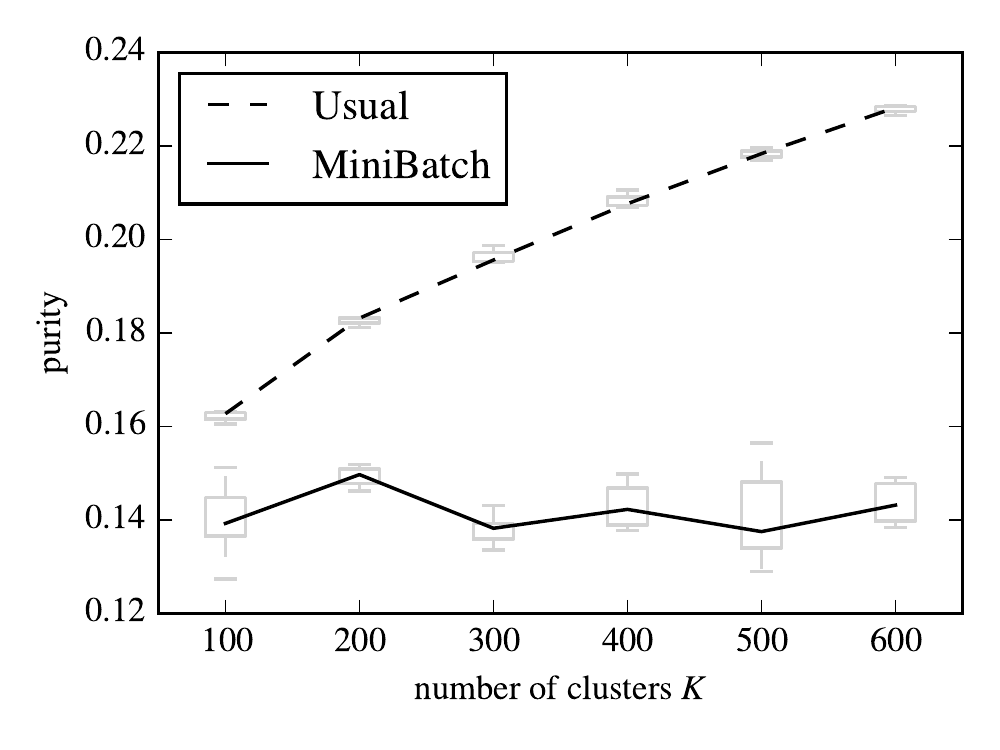}
  \caption{Purity vs number of clusters $K$ in \mbox{$K$-Means} and MiniBatch $K$-Means}
  \label{fig:k-vs-purity2}
\end{subfigure}%
\begin{subfigure}[b]{0.5\textwidth}
  \centering
  \includegraphics[width=\textwidth]{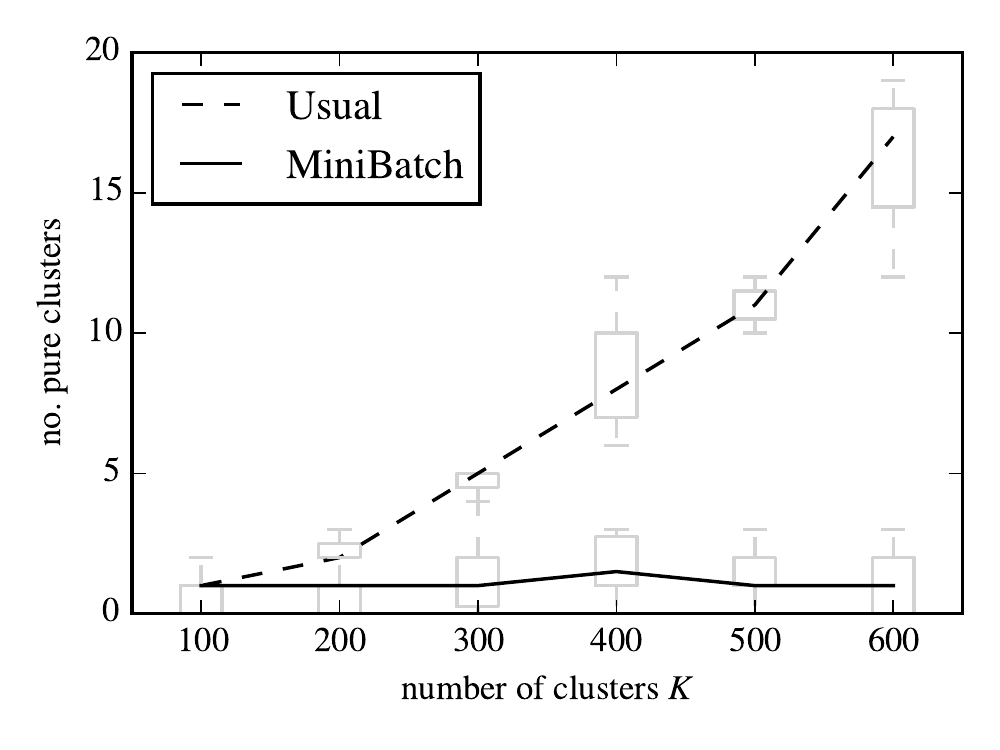}
  \caption{Number of pure clusters vs $K$ in $K$-Means and MiniBatch $K$-Means}
  \label{fig:k-vs-len2}
\end{subfigure}
\hfill
\begin{subfigure}[b]{\textwidth}
  \centering
  \includegraphics[width=0.85\textwidth]{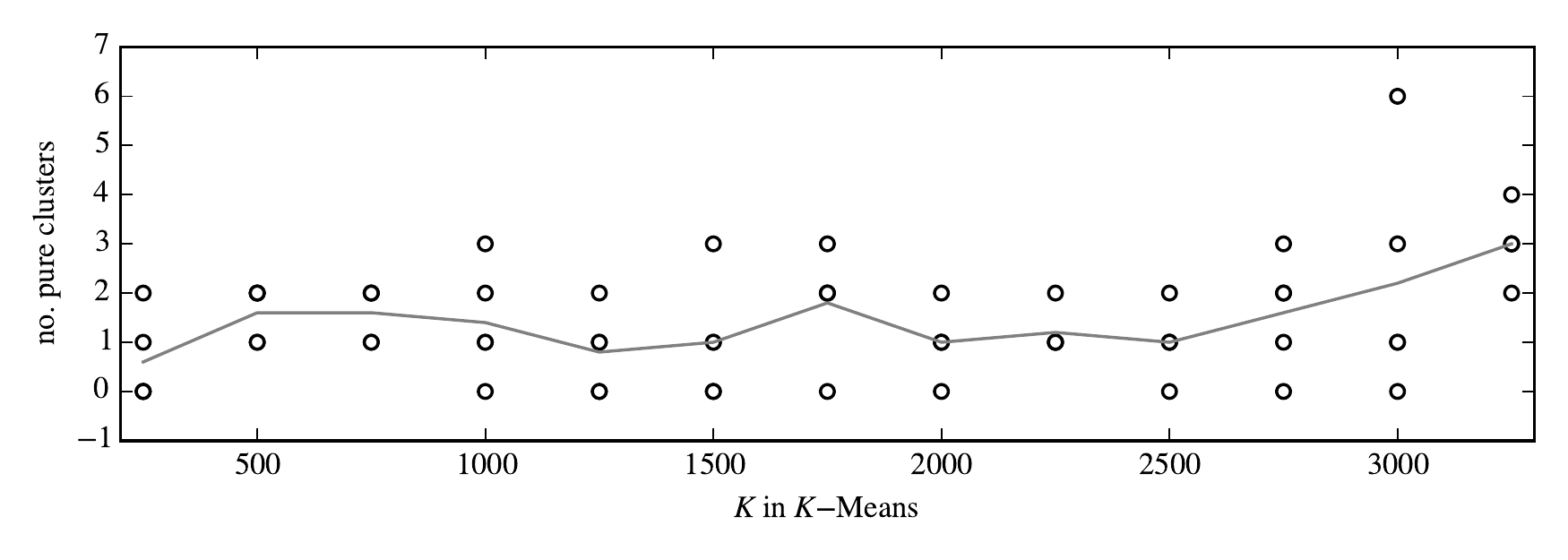}
  \caption{Number of pure clusters vs $K$ in MiniBatch $K$-Means for larger $K$}
  \label{fig:k-vs-len-mb}
\end{subfigure}
\hfill
\caption{Effect of $K$ on performance in $K$-Means}
\label{fig:kmeans-vs-minibatch}
\end{figure}

Usual $K$-Means with small $K$ does not find many pure clusters,
and MiniBatch $K$-Means does even worse: independently of the choice of $K$,
the number of pure clusters and purity does not change significantly
(see fig.~\ref{fig:k-vs-purity2} and fig.~\ref{fig:k-vs-len2}). This is also true
for larger values of $K$ (see fig.~\ref{fig:k-vs-len-mb}).


The best result was found by usual $K$-Means with $K=600$: it was able
to discover 19 clusters with purity at least 0.8 (note that this is worse
than the baseline of 39 pure clusters).

Next, we use \textbf{Latent Semantic Analysis} with \textbf{SVD}
to reduce the dimensionality of the identifier-document
matrix $D$, and then apply $K$-Means on the reduced space. As discussed in the
LSA section (see section~\ref{sec:lsa}), it should reveal the latent structure of data.
Hence, we expect that it should improve the results achieved by usual $K$-Means.

\begin{figure}[h!]
\centering\includegraphics[width=0.75\textwidth]{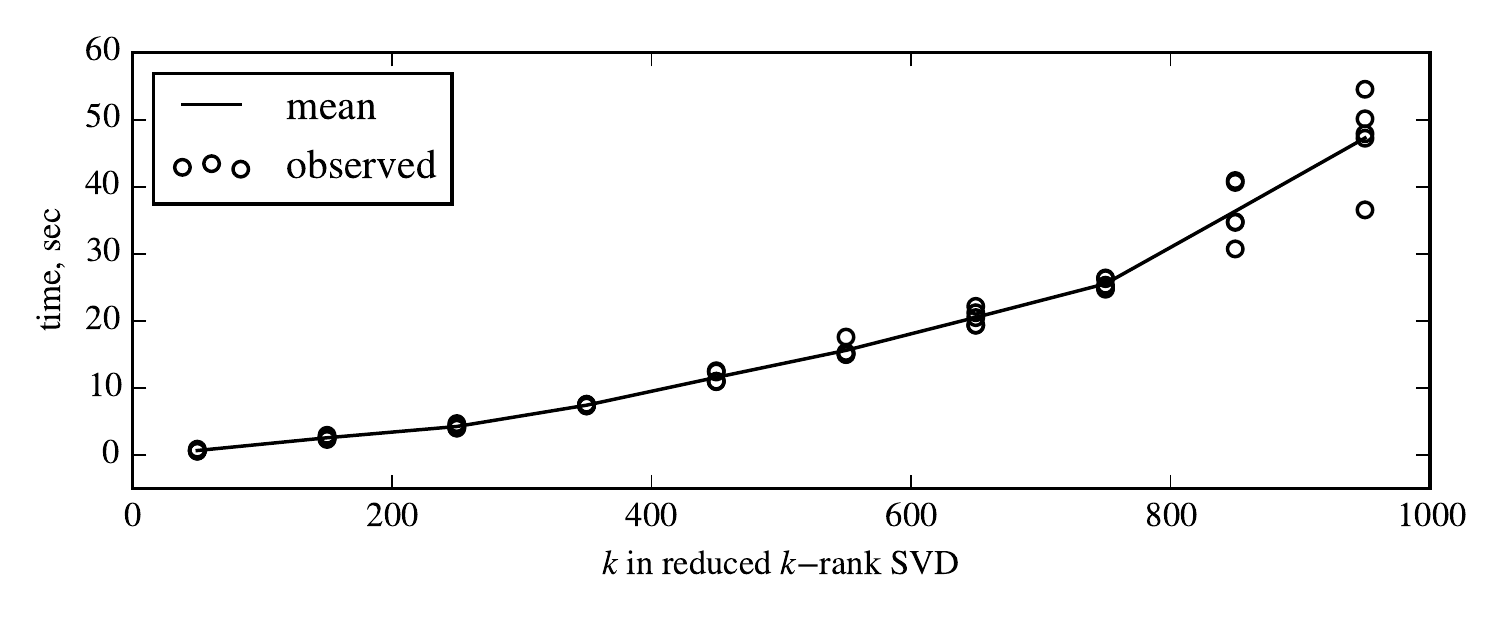}
\caption{Effect of $k$ in $k$-rank-reduced randomized SVD on the runtime in seconds.}
\label{fig:k-svd-vs-time}
\end{figure}

Randomized SVD is very fast, but the runtime does not grow linearly with $k$,
it looks quadratic (see fig.~\ref{fig:k-svd-vs-time}).
However, the typical values of $k$ for SVD used in latent semantic analysis is
150-250 \cite{aggarwal2012survey} \cite{evangelopoulos2012latent},
therefore the run time is not prohibitive, and we do not need to rut it
with very large $k$.

\begin{figure}[h!]
\centering
\hfill
\begin{subfigure}[b]{0.5\textwidth}
  \centering
  \includegraphics[width=\textwidth]{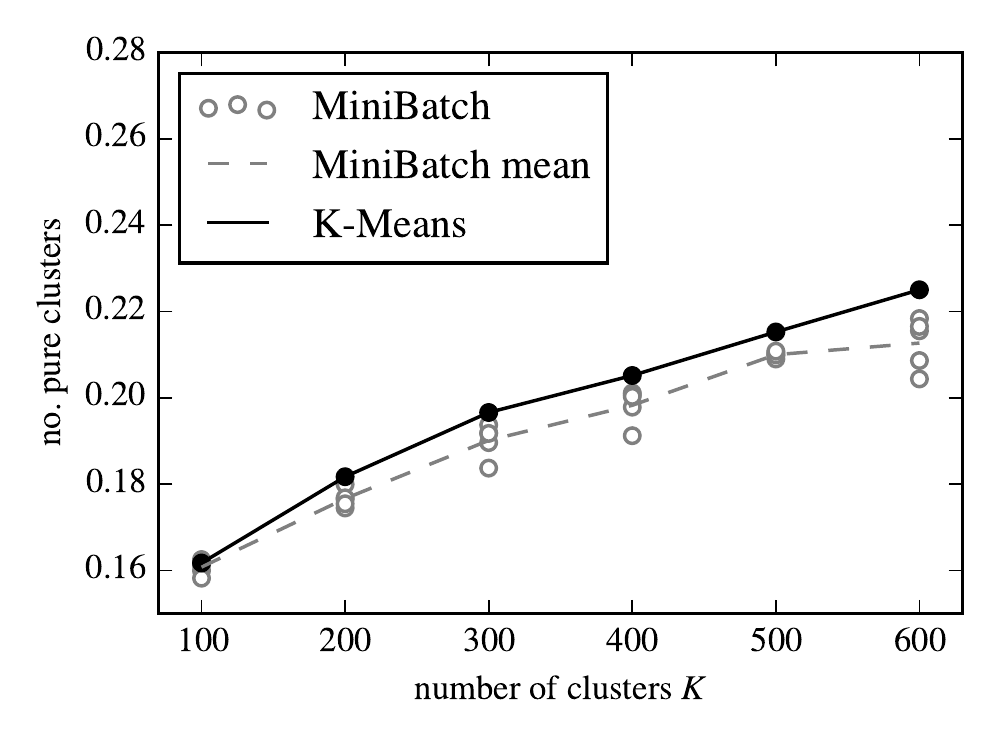}
  \caption{Purity vs number of clusters $K$ in \mbox{$K$-Means} and MiniBatch $K$-Means.}
  \label{fig:k-vs-mb-svd-purity}
\end{subfigure}%
\begin{subfigure}[b]{0.5\textwidth}
  \centering
  \includegraphics[width=\textwidth]{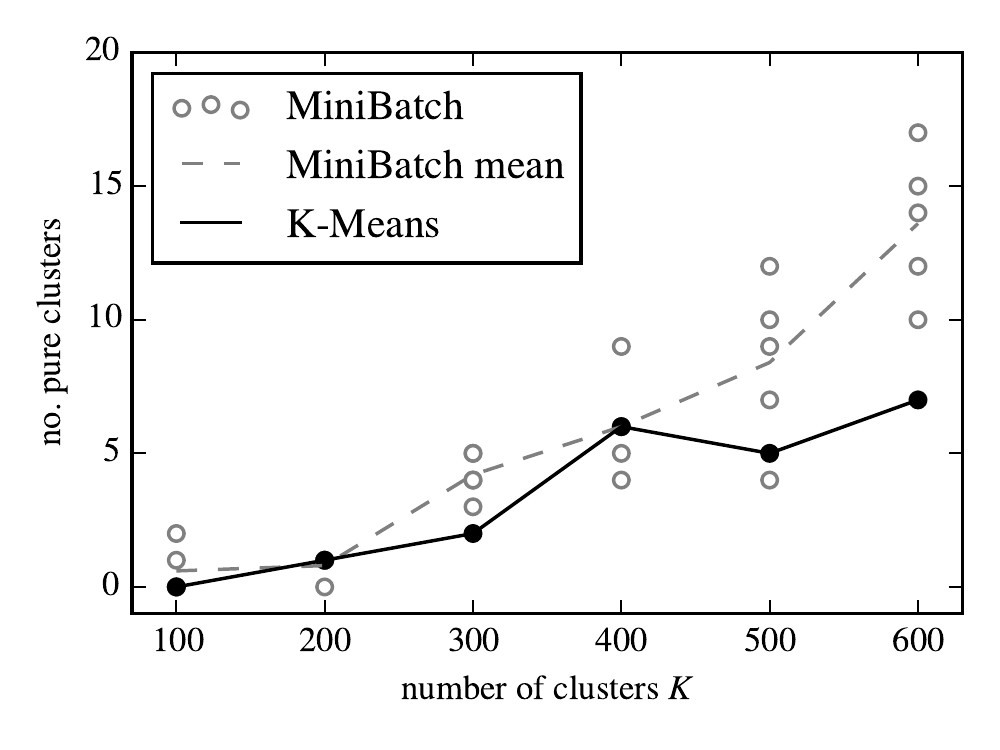}
  \caption{Number of pure clusters vs $K$ in $K$-Means and MiniBatch $K$-Means.}
  \label{fig:k-vs-mb-svd-len}
\end{subfigure}
\caption{The performance of $K$-Means and MiniBatch $K$-Means on the reduced document space with $k=150$.}
\label{fig:kmeans-vs-minibatch-svd}
\end{figure}

When the dimensionality is reduced, the performance of $K$-Means and
MiniBatch $K$-Means is similar (see fig.~\ref{fig:k-vs-mb-svd-purity}),
but with MiniBatch $K$-Means we were able to discover more interesting pure
clusters (see fig.~\ref{fig:k-vs-mb-svd-len}). The reason for this may be the fact that in the reduced space there is less noise and both methods find equally good clusters,
but because MiniBatch $K$-Means works faster, we are able to run it multiple
times thus increasing its chances to find a good local optimum where there
are many pure document clusters. Note that the obtained result is below
the baseline.

\begin{figure}[h!]
\centering\includegraphics[width=0.9\textwidth]{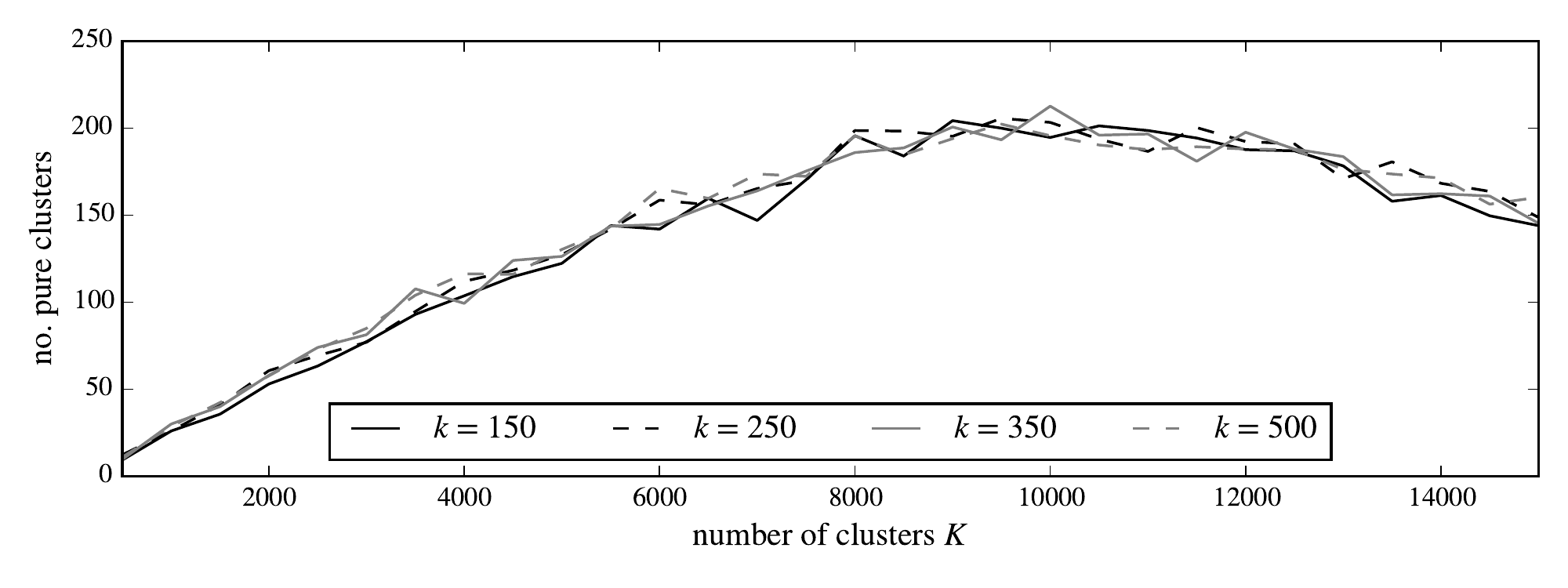}
\caption{Number of discovered pure clusters in $K$-Means for different number of clusters $K$ and rank $k$.}
\label{fig:k-vs-kmeans-len-nodef}
\end{figure}

We can observe that as $K$ increases, the number of interesting clusters increases
(see fig.~\ref{fig:k-vs-mb-svd-len}).
Therefore, we try a wide range of larger $K$ for different $k \in \{150, 250, 350, 500\}$.
The performance in terms of discovered pure clusters does not depend much on the
rank $k$ of the reduced space (see fig.~\ref{fig:k-vs-kmeans-len-nodef}). In fact,
it is very hard to distinguish different lines because they are quite perplexed.
The maximum for is achieved at $K \approx 10\,000$ for all $k$.

\begin{figure}[h!]
\centering
\includegraphics[width=0.6\textwidth]{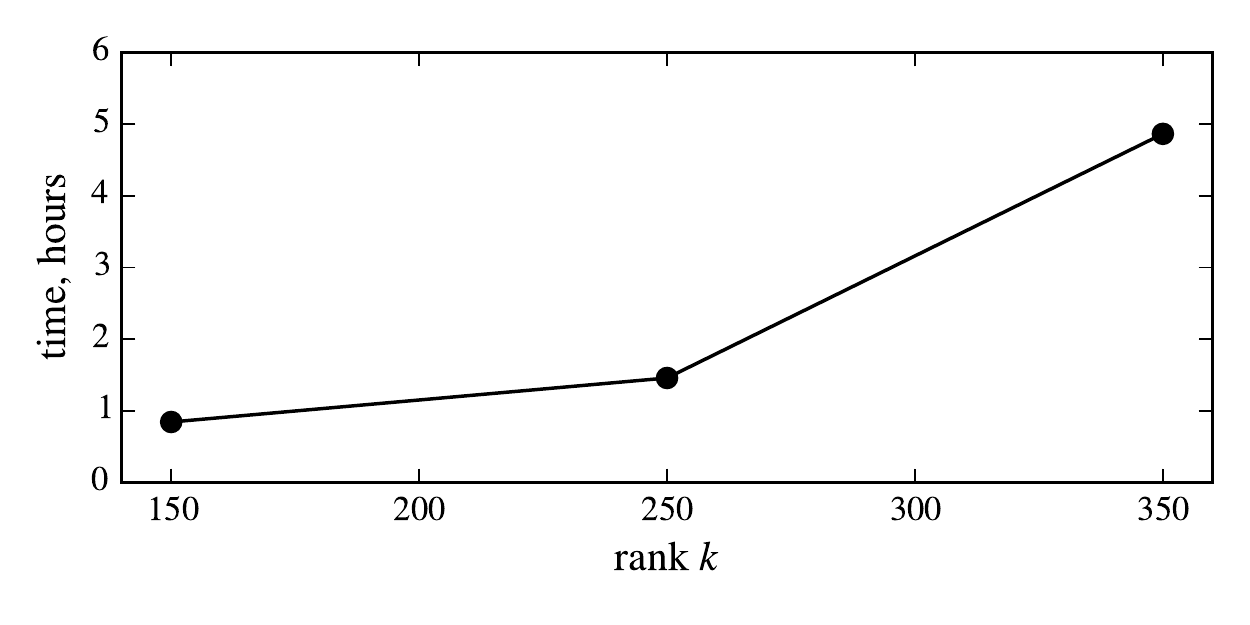}
\caption{Runtime of NMF for different $k$.}
\label{fig:nmf-runtime}
\end{figure}

\begin{figure}[h!]
\centering
\includegraphics[width=0.9\textwidth]{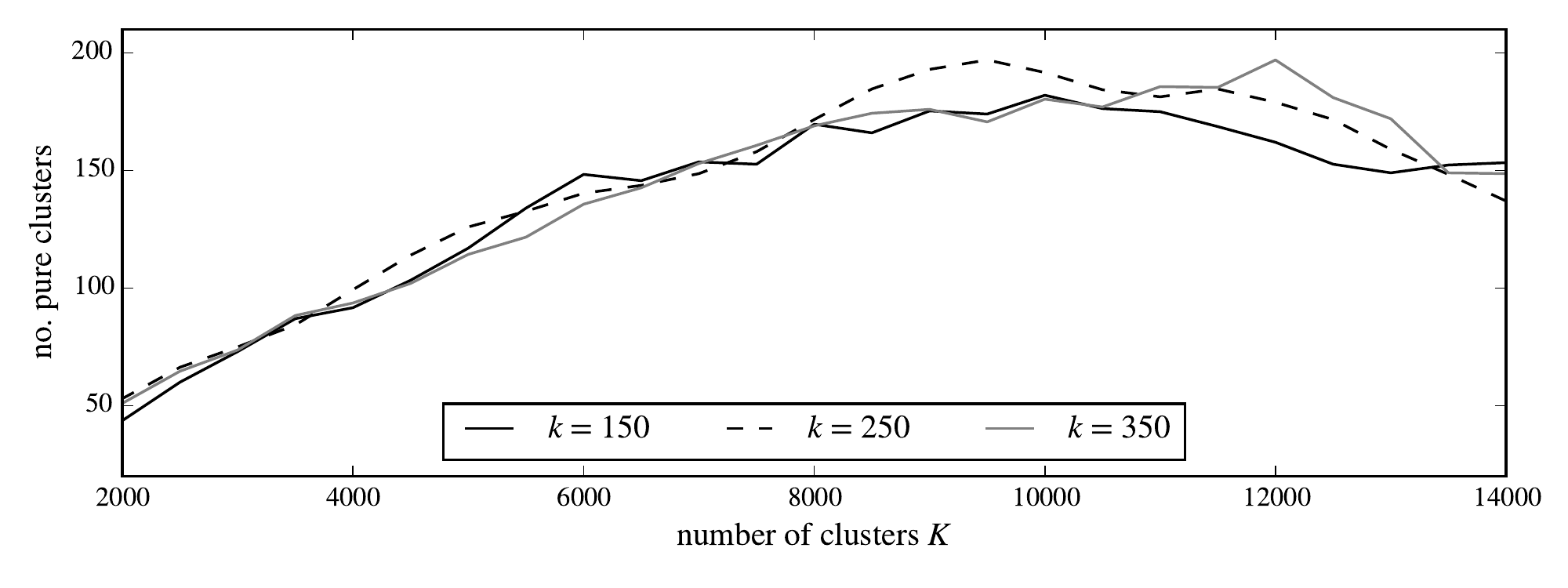}
\caption{Number of discovered pure clusters in $K$-Means and NMF for different number of clusters $K$ and rank $k$.}
\label{fig:k-vs-kmeans-len-nmf-nodef}
\end{figure}

We also can apply \textbf{Non-Negative Matrix Factorization} for LSA.
NMF takes significantly more time than randomized SVD (see fig.~\ref{fig:nmf-runtime}).
In addition, although the runtime should be $O(nk)$ \cite{xu2003document}, we do not
observe that it grows linearly with $k$. On the contrary, it appears that there is rather
quadratic relationship. We expected that the results produced by NMF will
be better than SVD because of non-negativity of produced results,
but the performance is quite similar (see fig.~\ref{fig:k-vs-kmeans-len-nmf-nodef}).
For NMF, however, it is easier to see the difference in performance when different rank $k$ is used, and the curves are not as perplexed as for SVD. We see that $k=250$ does better on
$K=[8000; 12000]$ than $k=150$ and $k=350$. For example, $K$-Means with
$K=9500$ and $k=250$ discovered a clustering with 200 namespace-defining clusters.

We can also observe that generally clustering works better on reduced spaces.

\begin{figure}[h!]
\centering\includegraphics[width=0.9\textwidth]{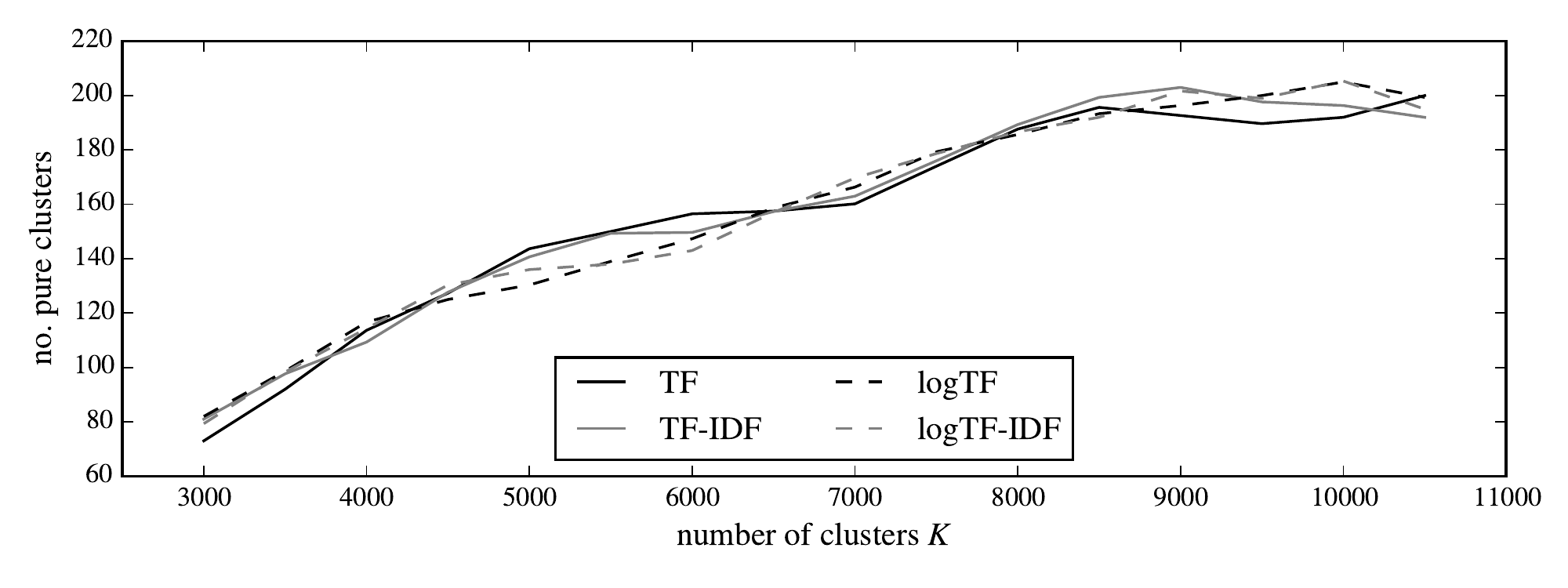}
\caption{The effect of using different weighting systems on $K$-Means with SVD.}
\label{fig:nodef-weighing}
\end{figure}

In the experiments above we used the $(\log \text{TF}) \times \text{IDF}$ weighting scheme.
Let us compare the effect of different weighting on the resulting clusters.
To do that, we apply SVD  with $k=150$ and run MiniBatch $K$-Means for a set of smaller $K$'s
because it is computationally faster. We can observe that performance of $K$-Means
does not depend significantly on the weighting system when no definitions
are used (see fig.~\ref{fig:nodef-weighing}).

\subsubsection{Weak Association} \ \\

The identifier-document matrix has the dimensionality $10419 \times 22512$,
and there are 485\,337 elements in the matrix, so the density is about 0.002.

\begin{figure}[h!]
\centering
\begin{subfigure}[b]{0.5\textwidth}
  \centering
  \includegraphics[width=\textwidth]{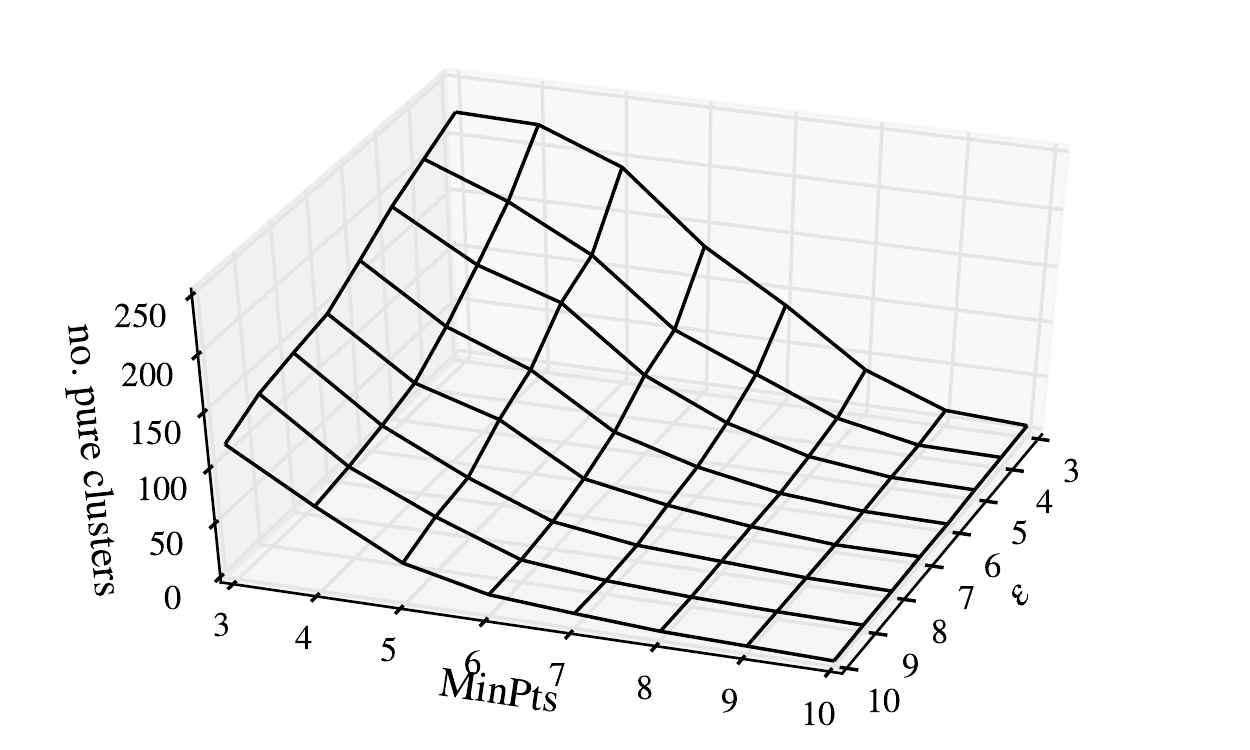}
  \caption{Number of clusters when 10 nearest neighbors are considered}
  \label{fig:soft-dbscan-cos10}
\end{subfigure}%
\begin{subfigure}[b]{0.5\textwidth}
  \centering
  \includegraphics[width=\textwidth]{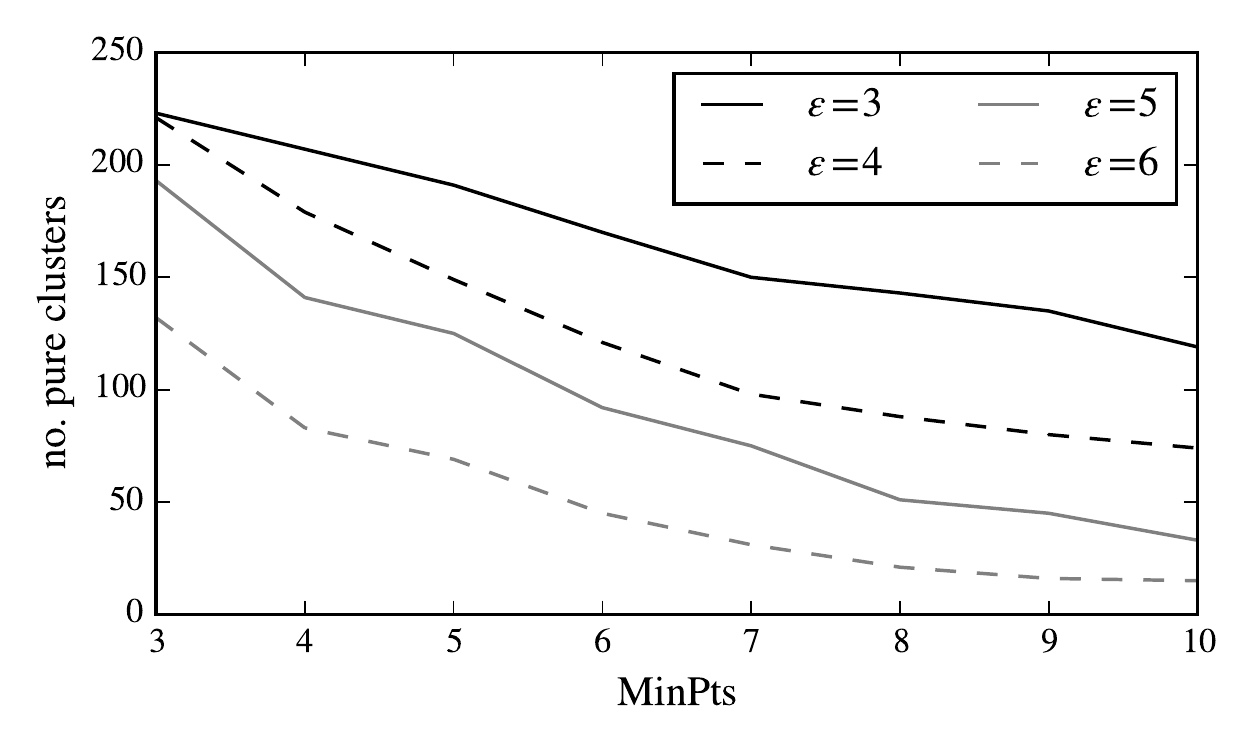}
  \caption{Performance of selected $\varepsilon$ with 10 nearest neighbors}
  \label{fig:soft-dbscan-cos10-2}
\end{subfigure}
\begin{subfigure}[b]{0.5\textwidth}
  \centering
  \includegraphics[width=\textwidth]{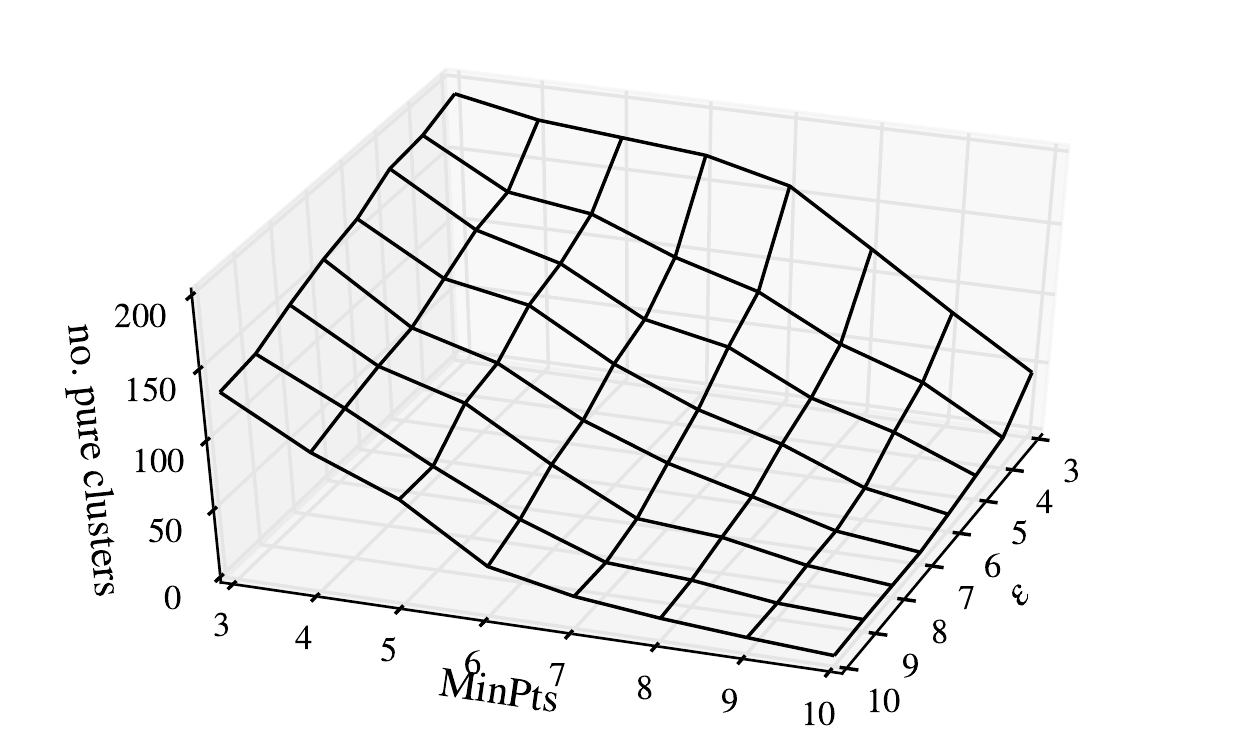}
  \caption{Number of clusters when 15 nearest neighbors are considered}
  \label{fig:soft-dbscan-cos15}
\end{subfigure}%
\begin{subfigure}[b]{0.5\textwidth}
  \centering
  \includegraphics[width=\textwidth]{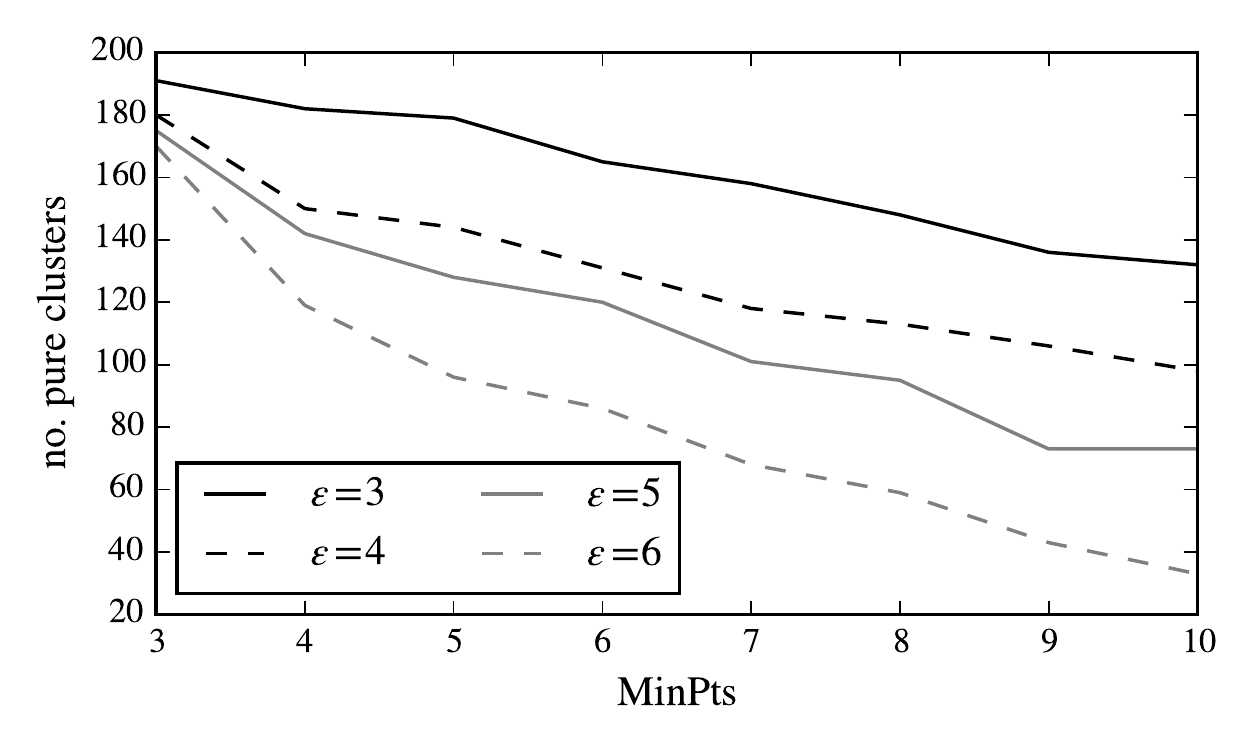}
  \caption{Performance of selected $\varepsilon$ with 15 nearest neighbors}
  \label{fig:soft-dbscan-cos15-2}
\end{subfigure}
\caption{Effect of parameters $\varepsilon$, \texttt{MinPts} and number of nearest
 neighbors on performance of SNN DBSCAN when cosine is used.}
\label{fig:soft-dbscan-cos}
\end{figure}

We do not attempt to use hierarchical methods and start with DBSCAN. Previously we have
observed that Jaccard is inferior to cosine similarity, and therefore we start directly
with cosine.

Like in no-definition case, we calculate the cosine similarity on document vectors
where elements are weighed with $(\log \text{TF}) \times \text{IDF}$. Using definitions
it gives better results, than just identifiers. For example,
for ``Linear regression'' the closest document is ``Linear predictor function'' with
cosine of 0.62, which is the same result, obtained when no definitions are used.
However, for ``Singular value decomposition'' the most similar document is
``Moore–Penrose pseudoinverse'' with cosine score of 0.386, and this is more meaningful
than the most similar document when no definitions are used.
As previously, we applied \textbf{SNN DBSCAN} with 10 and 15 nearest neighbors, and the best
result was obtained with 10 nearest neighbors, $\varepsilon=3$ and \texttt{MinPts}$=3$
(see fig.~\ref{fig:soft-dbscan-cos}). It was able to discover 223 namespace-defining
clusters, which is slightly better than the best case when no definitions are
used.

\begin{figure}[h!]
\centering\includegraphics[width=0.9\textwidth]{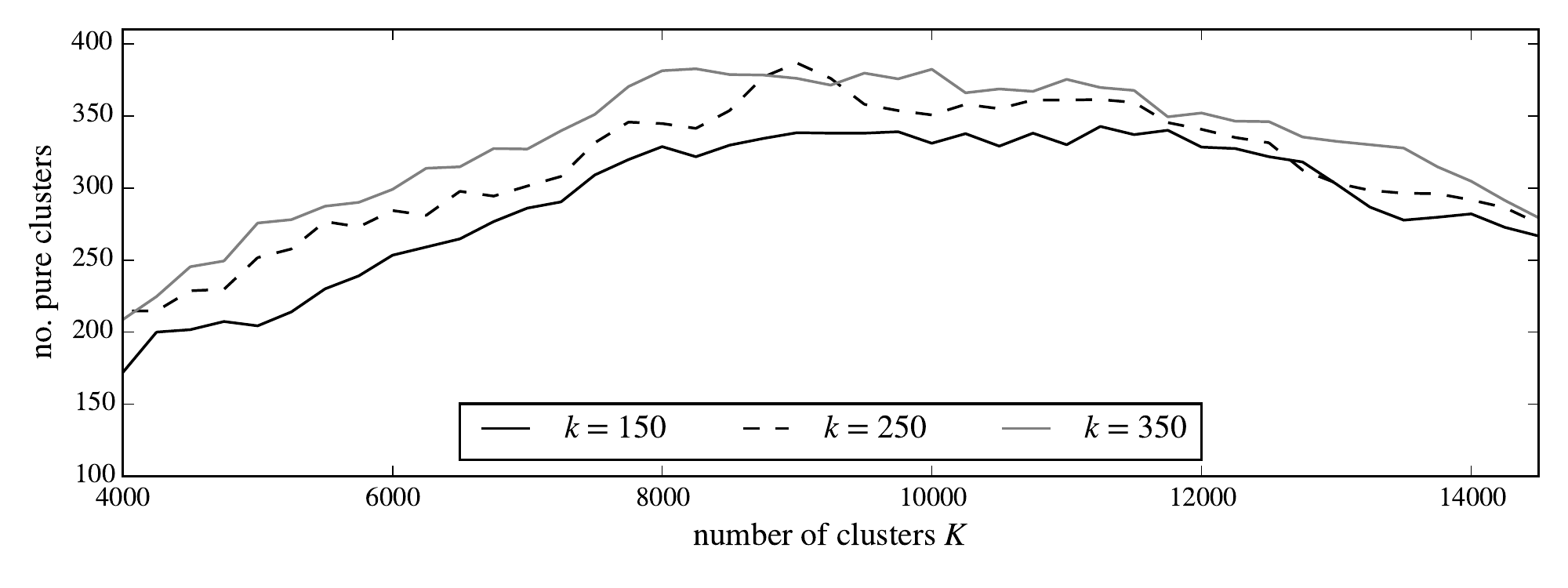}
\caption{Number of discovered pure clusters in $K$-Means and SVD for different number of clusters $K$ and rank $k$.}
\label{fig:k-vs-kmeans-len-svd-soft}
\end{figure}

With  \textbf{MiniBatch $K$-Means} applied on the plain untransformed document
space we are able to find some interesting clusters, but it general, similarity to the no-definition case, the does not show good results overall. Therefore we
apply it to the LSA space reduced by \textbf{SVD}, when identifier-document matrix
is reduced to rank $k$. We search for
the best combination trying $K = [500; 15000]$ and $k \in \{150, 250, 350\}$.
Unlike the case where no definitions are used, the space produced by the
soft definition association is affected by $k$ (see fig.~\ref{fig:k-vs-kmeans-len-svd-soft})
and the results produced by $k = 350$ are almost always better.
The weighing scheme used for this experiment is $(\log \text{TF}) \times \text{IDF}$.

\begin{figure}[h!]
\centering\includegraphics[width=0.9\textwidth]{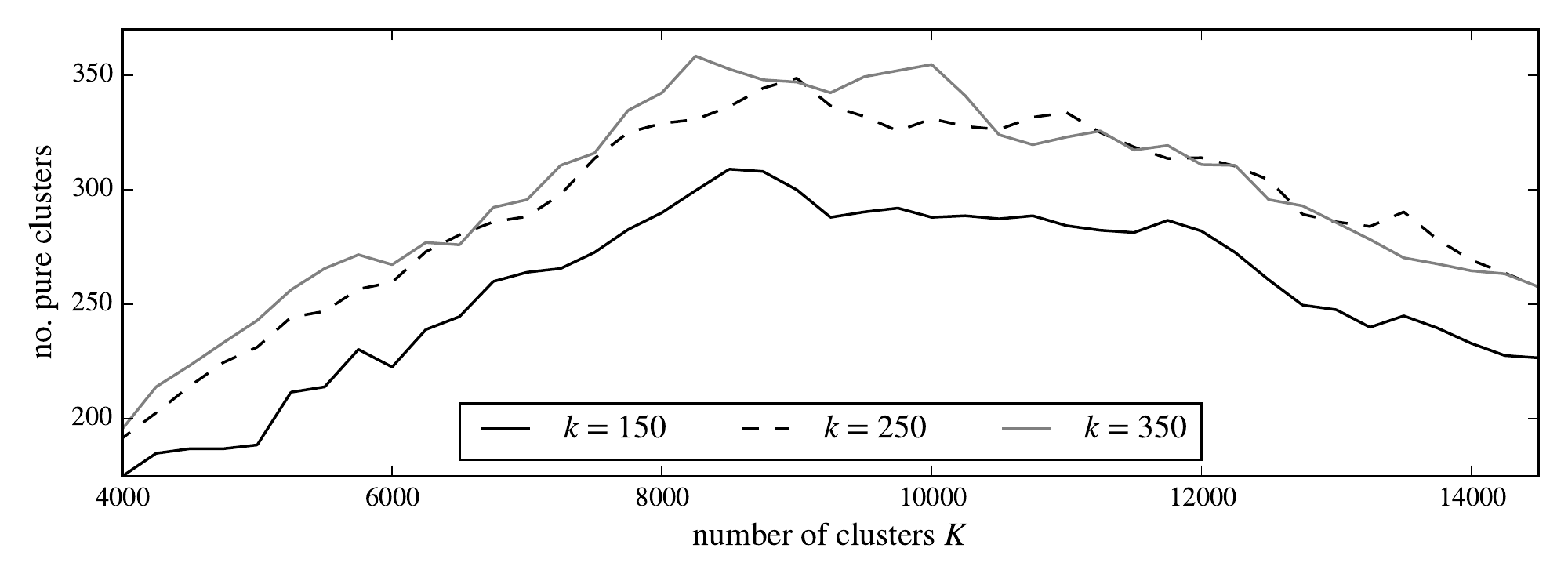}
\caption{The effect of rank $k$ used in NMF on $K$-Means.}
\label{fig:nnmf-soft}
\end{figure}

\textbf{Non-Negative Matrix Factorization} gives good results, but does not improve on
the best result obtained with SVD (see fig.~\ref{fig:nnmf-soft}).
The largest number of namespace-defining clusters is 370 and it is achieved
with $K=10000$ and $k=350$.

\begin{figure}[h!]
\centering\includegraphics[width=0.9\textwidth]{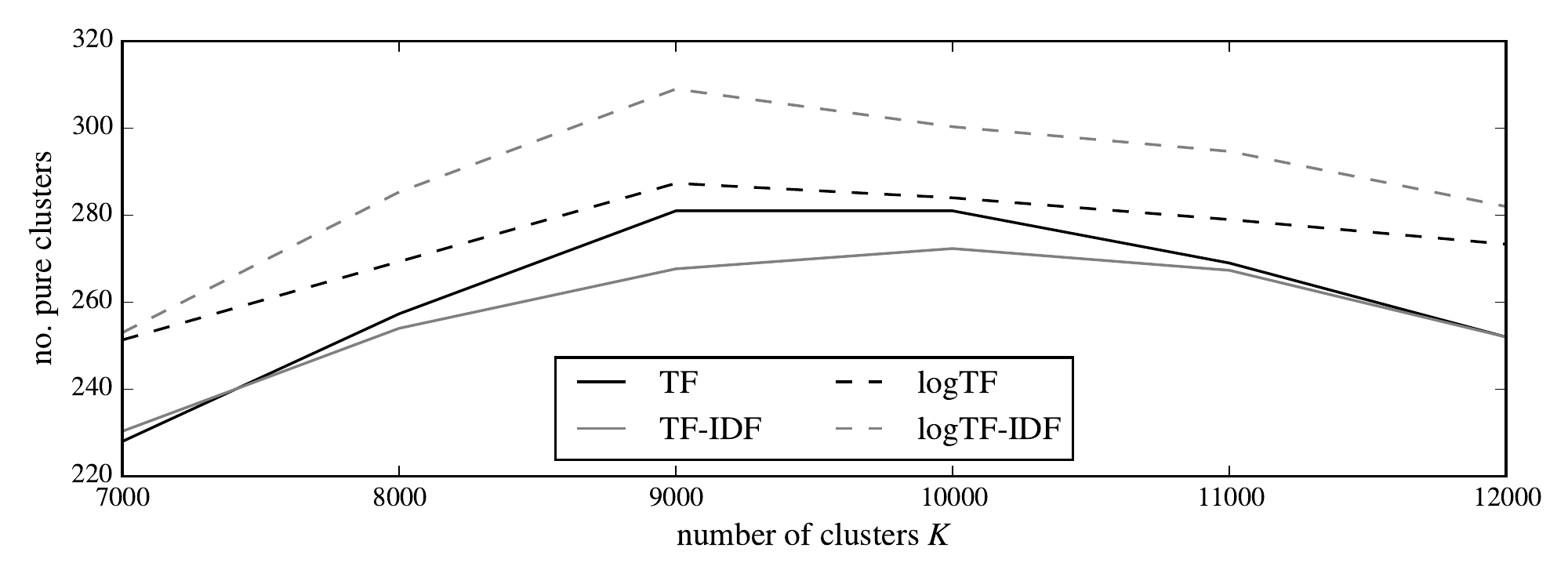}
\caption{The effect of using different weighting systems on $K$-Means with SVD
($k=150$).}
\label{fig:soft-weighing}
\end{figure}

We also experiment with different weighting schemes, and, unlike the no-definition
case, it has a significant effect on the results: we can observe that sublinear
TF is better that untransformed TF, and $(\log \text{TF}) \times \text{IDF}$
achieves the best performance (see fig.~\ref{fig:soft-weighing}).

\subsubsection{Strong Association} \ \\

In the case when we use the strong association, the identifier-document
matrix has the dimensionality of $37879 \times 22512$ identifier-document matrix.
It has 499070 entries, so the density of this matrix is just 0.00058.

\begin{figure}[h!]
\centering
\includegraphics[width=0.9\textwidth]{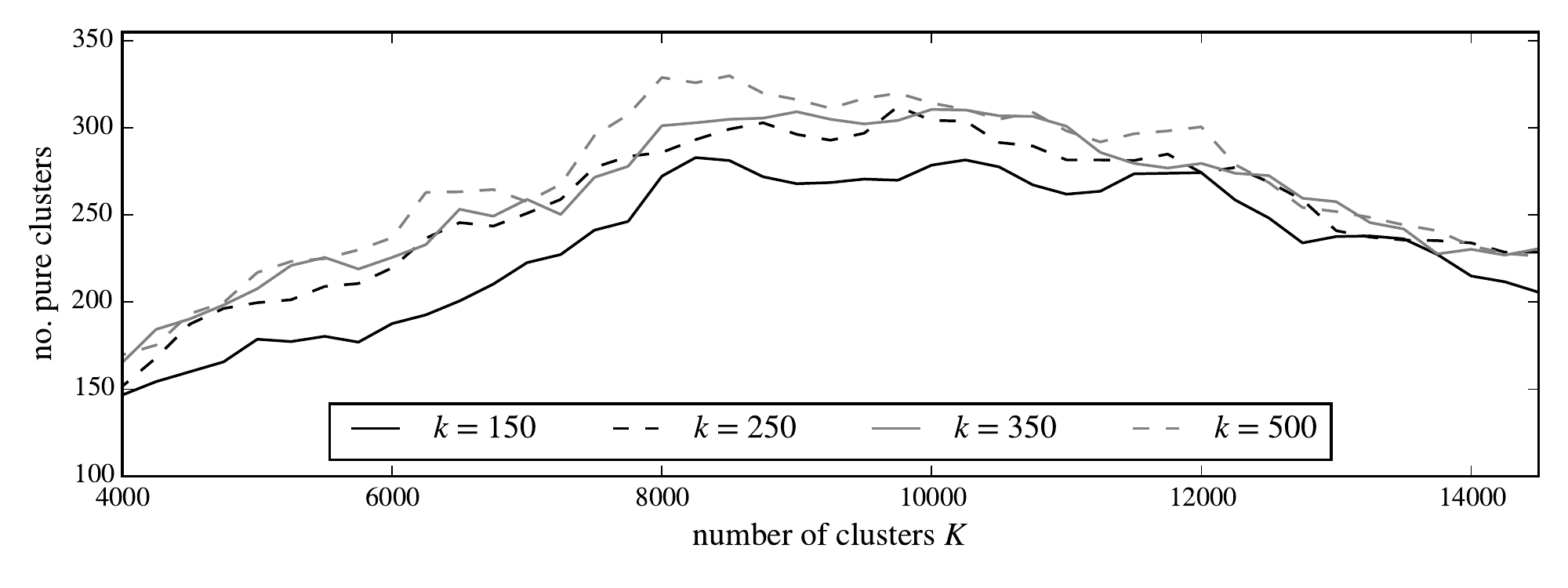}
\caption{Number of discovered pure clusters in $K$-Means and SVD for different number of clusters $K$ and rank $k$.}
\label{fig:k-vs-kmeans-len-svd-strong}
\end{figure}

Like for the soft association case, we choose not to perform usual $K$-Means,
and instead proceed directly to \textbf{MiniBatch $K$-Means} on the LSA space
reduced with \textbf{SVD}. With rank $k=500$ and number of clusters $K = 8250$
is achieves the best result of 340 clusters (see fig.~\ref{fig:k-vs-kmeans-len-svd-strong}),
which is slightly worse than in the weak association case. The purity
of obtained clustering is 0.5683.

We do not attempt to perform \textbf{Non-Negative Matrix Factorization} as we have previously
established that it usually does not give better results while taking significantly
longer time.

\subsubsection{Russian Wikipedia}  \ \\

Based on the experiments we have performed on the English Wikipedia, we see that the best
performance is obtained with weak association by using MiniBatch $K$-Means on
LSA space reduced by SVD.
We apply the same best performing technique on the Russian Wikipedia.

The identifier-document matrix has the dimensionality of $3948 \times 5319$
with 79171 non-zero elements, so the density of this matrix is 0.0038.

\begin{figure}[h!]
\centering
\includegraphics[width=0.9\textwidth]{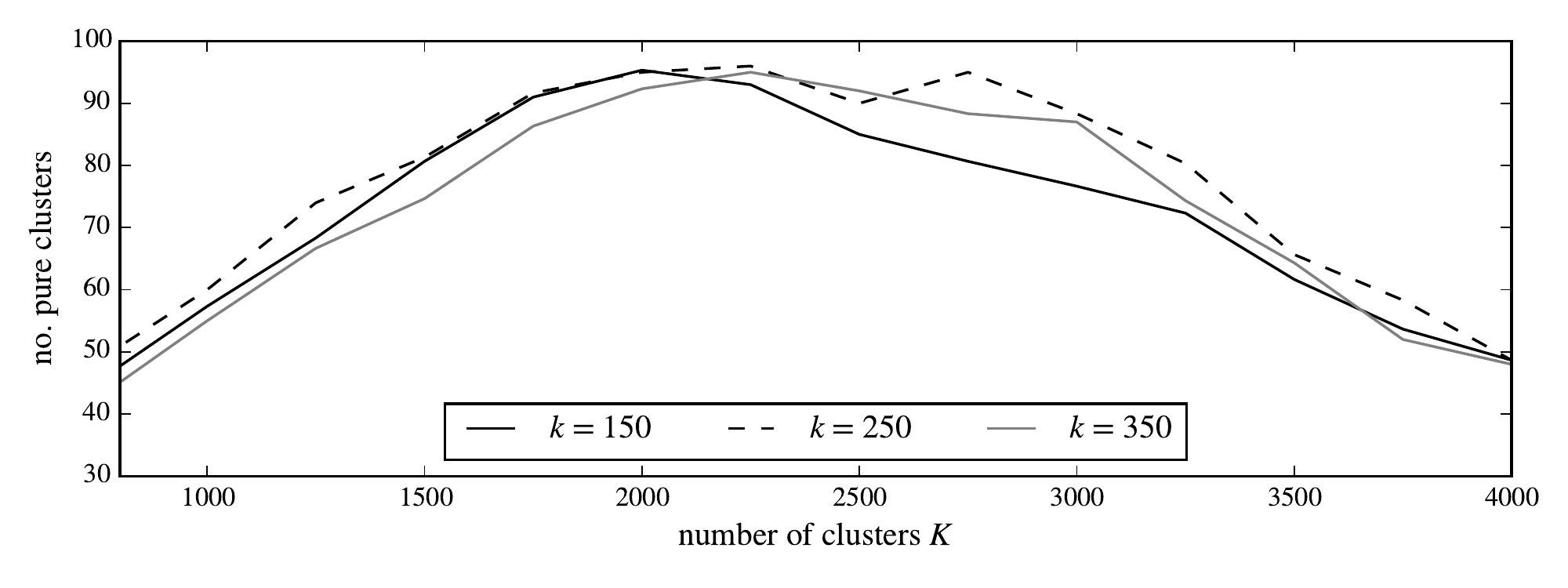}
\caption{Performace of $K$-Means with SVD for Russian Wikipedia.}
\label{fig:k-vs-kmeans-rus.pdf}
\end{figure}

As usually, we applied SVD with different values of rank $k \in \{150, 250, 350\}$,
and, similarity to no-definitions case for the English Wikipedia, we do not
observe significant differences across different values of $k$
(see fig.~\ref{fig:k-vs-kmeans-rus.pdf}). The best achieved result is
105 namespace-defining clusters.

\subsection{Result Analysis} \label{sec:result-analysis}

\subsubsection{English Wikipedia} \ \\

In the previous chapter we have established that the best way to incorporate
definitions into Intensifier Vector Space is by using soft association, and the best
clustering performing method is MiniBatch $K$-Means.

\begin{table}[h!]
\centering
\begin{tabular}{|c|c|c|}
  \hline
  Name & Size & Purity \\
  \hline
Astronomical catalogues & 53 & 0.9811 \\
Statistics & 20 & 0.8500 \\
Category theory & 16 & 0.8125 \\
Electromagnetism & 12 & 0.8333 \\
Thermodynamics & 11 & 0.8182 \\
Mathematical analysis & 11 & 0.8182 \\
Graph theory & 10 & 0.9000 \\
Graph algorithms & 10 & 0.8000 \\
Fluid dynamics & 10 & 1.0000 \\
Numerical analysis & 9 & 0.8889 \\
Group theory & 9 & 1.0000 \\
Stochastic processes & 9 & 1.0000 \\
Measure theory & 8 & 1.0000 \\
\hline
\end{tabular}
\caption{Top namespace-defining clusters.}
\label{tab:soft-kmeans-svd}
\end{table}

\begin{table}[h!]
\centering
\begin{subtable}{0.7\textwidth}
\centering
\begin{tabular}{|c|c|}
  \hline
  Article & Identifiers \\
  \hline
Diagonalizable matrix & $v_1,\lambda_1,v_k,\lambda_3,\lambda_2,\lambda_i,\lambda_k,\lambda_j,\lambda_n,...$\\
Eigenvalues and eigenvectors & $v_i,\mu_A,\lambda_i,d,\lambda_n,...$ \\
Principal axis theorem & $v_1,u_1,\lambda_1,\lambda_2,D,S,u,...$ \\
Eigendecomposition of a matrix & $\lambda,\lambda_1,\lambda,\lambda_2,\lambda_k,R,U,T,...$ \\
Min-max theorem & $\sigma,u_n,u_k,u_i,u_1,\alpha,\lambda_1,\lambda,\lambda_i,...$ \\
Linear independence & $\Lambda,v_j,u_2,v_3,u_n,\lambda_1,\lambda_3,\lambda_2,...$ \\
Symmetric matrix & $\Lambda,\lambda_1,\lambda_2,D,Q,P,\lambda_i,...$ \\
\hline
\end{tabular}
\caption{Wiki Articles in the cluster ``Linear Algebra''}
\label{tab:soft-kmeans-la}
\end{subtable}
\begin{subtable}{0.3\textwidth}
\centering
\begin{tabular}{|c|c|c|}
\hline
ID & Definition & Score\\
\hline
$D$ & diagonal matrix & 0.72 \\
$t$ & real argument & 0.46 \\
$u$ & eigenvalues & 0.42 \\
$u_i$ & eigenvector & 0.42 \\
$v_1$ & eigenvectors & 0.73 \\
$\Lambda$ & diagonal matrix & 0.87 \\
$\lambda$ & eigenvalue & 0.4  \\
$\lambda_1$ & eigenvalues & 0.95 \\
$\lambda_2$ & eigenvalues & 0.71 \\
$\lambda_3$ & eigenvalues & 0.39 \\
$\lambda_i$ & eigenvalue & 0.98 \\
\hline
\end{tabular}
\caption{Definitions in ``Linear Algebra''}
\label{tab:soft-kmeans-la-def}
\end{subtable}
\caption{A ``Linear Algebra'' cluster.}
\label{tab:soft-kmeans-lsa}
\end{table}

\begin{table}
\centering
\begin{tabular}{|c|c|c|c|}
  \hline
  \multicolumn{4}{|c|}{$\lambda$}\\
  \hline
  Size & Namespace Name & Definition & Score \\
  \hline
3 & Algebra & multiplicity & 0.43 \\
4 & Analysis of variance & marquardt & 0.69 \\
3 & Applied and interdisciplinary physics & wavelength & 0.98 \\
6 & Cartographic projections & longitude & 1.00 \\
3 & Cartography & longitude & 1.00 \\
3 & Category theory & natural isomorphisms & 0.40 \\
4 & Condensed matter physics & penetration depth & 0.44 \\
5 & Continuous distributions & affine parameter & 0.46 \\
3 & Coordinate systems & longitude & 0.88 \\
3 & Differential equations & differential operator & 0.42 \\
8 & Differential geometry & vector fields & 0.72 \\
7 & Electronic amplifiers & typical value & 0.43 \\
3 & Electrostatics & unit length & 0.43 \\
10 & Fluid dynamics & wavelength & 1.00 \\
6 & Fluid dynamics & free path & 0.43 \\
3 & Infinity & limit ordinals & 0.87 \\
7 & Linear algebra & eigenvalue & 0.4  \\
5 & Linear algebra & matrix & 0.41 \\
3 & Linear algebra & eigenvalue & 0.85 \\
3 & Liquids & relaxation time & 0.41 \\
3 & Materials science & rate & 0.44 \\
3 & Mathematical analysis & eigenvalue & 0.41 \\
3 & Mathematical theorems & poisson distribution & 0.41 \\
4 & Measure theory & lebesgue measure & 0.44 \\
3 & Measurement & order & 0.42 \\
8 & Mechanics & previous expression & 0.44 \\
4 & Mechanics & power series & 0.41 \\
3 & Metalogic & empty word & 0.45 \\
7 & Number theory & partition & 0.74 \\
4 & Number theory & modular lambda function & 0.46 \\
3 & Operator theory & algebraic multiplicity & 0.44 \\
5 & Optics & wavelength & 0.71 \\
5 & Partial differential equations & constants & 0.41 \\
4 & Physical optics & wavelength & 0.95 \\
5 & Physics & exciton state & 0.88 \\
6 & Probability distributions & references & 0.42 \\
4 & Quantum field theory & coupling constant & 0.75 \\
5 & Quantum mechanics & wavelength & 1.00 \\
5 & Quantum mechanics & state & 0.87 \\
3 & Radioactivity & decay & 0.72 \\
4 & Representation theory of Lie groups & weight & 1.00 \\
3 & Riemannian geometry & contravariant vector field & 0.45 \\
4 & Rubber properties & engineering strain & 1.00 \\
3 & Statistical data types & regularization parameter & 0.45 \\
20 & Statistics & words & 0.43 \\
3 & Statistics & expectation & 0.46 \\
3 & Stellar astronomy & mean free path & 0.43 \\
3 & Surface chemistry & ideal gas & 0.39 \\
3 & Theoretical physics & eigenvalue & 0.88 \\
5 & Theories of gravitation & dicke & 0.44 \\
3 & Wave mechanics & wavelength & 0.8 \\
\hline
\end{tabular}
\caption{Some of definitions of ``$\lambda$''.}
\label{tab:soft-kmeans-lambda}
\end{table}

The best result is 414 namespace-defining clusters, it is ten times better
than the baseline result. The result is achieved by $K$-Means with
soft association using parameters $K=9750$ and $k=350$. The purity of this
clustering is $0.63$. The largest namespace-defining clusters discovered
by this methods are presented in the table~\ref{tab:soft-kmeans-svd}.

Let us consider a ``Linear Algebra'' cluster (table~\ref{tab:soft-kmeans-lsa})
with 6 documents and some of extracted definitions in documents
of this cluster, and all these articles share identifers $\lambda_1$, $m$ and $n$.
Let us consider all definitions of identifier ``$\lambda$''. In total, there
are 93 clusters where ``$\lambda$'' is used (see table~\ref{tab:soft-kmeans-lambda}),
and in many cases it is possible to determine that the assignment is correct
(e.g. ``eigenvalue'', ``wavelength'', ``regularization parameter'').
Some cases are not correct, for example, when we have clusters with the same name
where $\lambda$ denotes different things (e.g. in two ``Quantum Mechanics'' clusters),
or in the case of ``Linear Algebra'' cluster where it denotes a matrix.

Clustering results with sort association are better than results obtained
with hard association. One of the reasons for that can the the fact that
definitions may act as keywords that describe the document and they are
better in describing the semantic content of the document.

Additionally, we see that clustering on the reduced space works better,
and in our case the best dimensionality reduction method is SVD.

We also note that we do not discover many namespace-defining clusters. The
best result identifiers 414 clusters, while the desired number of clusters $K$
is almost 10\,000. It means that information from about 9\,000 clusters is
discarded: in total there are 22\,512 documents, but identifiers from only 1\,773
are used, and the rest of the document (about 92\%) are not utilized at all.

\subsubsection{Russian Wikipedia} \ \\

For the Russian Wikipedia, the best results is 105 namespace-defining clusters
with overall purity of 0.73. It was obtained by $K=3000$ and $k = 250$.
The largest namespace-defining clusters
are shown in table~\ref{tab:rus-wiki-categories}. Interestingly, there is a cluster
``Животные'' (``Animals'') where mathematical formulae are used to describe ``tooth formula''.
Let us consider a cluster about Linear Algebra (see table~\ref{tab:rus-la})
and definitions extracted from it (see table~\ref{tab:rus-la}). Similarity to English Wikipedia,
some of the definitions are correct and valid, for example, ``сингулярный число''
(``singular value'') for ``$\sigma$''  or ``ранг'' (``rank'') for ``$r$'', while some
are not quite correct, e.g. ``скаляр'' (``scalar'') for ``$\lambda$''.
Additionally, some of the non-valid definitions seem to result from misclassifications
by the POS-tagger, for example, ``$\Sigma$'' is defined as ``вдоль'' (literally ``along''),
which do not make much sense.
Furthermore, we can look at all definitions of ``$\lambda$'' across all discovered namespaces
(see table~\ref{tab:rus-lambda}). The scores of namespace relations extracted from Russian
wikipedia are generally smaller than in English, but it is most likely because
there are fewer relations in the Russian dataset.

For the Russian part of Wikipedia, we also utilize just 382 documents, which is only 7\%,
and the rest of the documents (93\%) are filtered out.

\begin{table}[h!]
\centering
\begin{tabular}{|c|c|c|c|}
  \hline
  Article (Original name) & Article (English) & Size & Purity \\
  \hline
Общая алгебра & Algebra & 7 & 0.8571 \\
Диф. геометрия и топология & Differential geometry and topology & 7 & 1.0000 \\
Диф. геометрия и топология & Differential geometry and topology & 6 & 0.8333 \\
Функциональный анализ &  Functional analysis & 6 & 0.8333 \\
Животные & Animals & 6 & 0.8333 \\
Картография & Cartography & 6 & 1.0000 \\
Математический анализ & Mathematical analysis & 5 & 1.0000 \\
Математический анализ & Mathematical analysis & 5 & 0.8000 \\
Теория вероятностей & Probability theory & 5 & 1.0000 \\
Механика & Mechanics & 5 & 0.8000 \\
Диф. уравнения в частных производных & Partial differential equations & 5 & 0.8000 \\
Математический анализ & Mathematical analysis & 5 & 0.8000 \\
Релятивистские и гравитационные явления & Relativity and gravitation & 5 & 0.8000 \\
Линейная алгебра & Linear algebra & 5 & 1.0000 \\
Математический анализ & Mathematical analysis & 5 & 1.0000 \\
Физика твёрдого тела & Solid-state physics & 5 & 0.8000 \\
\hline
\end{tabular}
\caption{Largest namespace-defining clusters extracted from Russian Wikipedia.}
\label{tab:rus-wiki-categories}
\end{table}

\begin{table}[h!]
\centering
\begin{subtable}{\textwidth}
\centering
\begin{tabular}{|c|c|c|}
  \hline
  Article (Original name) & Article (English) & Identifiers \\
  \hline
Линейная алгебра & Linear algebra &  $V, v_n, \alpha, v_1, v_2, x, \alpha_1, \beta, \lambda, f, U,...$ \\
Спектральная теорема & Spectral theorem & $\Lambda, \lambda, H, K, P, U, T, V, X, f, y, x,...$ \\
Сингулярное разложение & Singular value decomposition & $\Sigma, V_k, \sigma, M, U, T, V, k, r, u, v,...$ \\
\begin{tabular}[x]{@{}c@{}} Ковариантность и \\ контравариантность \end{tabular}
 & Covariance and contra-variance & $S, V, d, g, f, k, m, f_i, n, u, v, x,...$ \\
Теорема Куранта -- Фишера & Courant–Fischer theorem & $k, V, S, L_k,...$ \\
\hline
\end{tabular}
\caption{A namespace-defining cluster about Linear algebra.}
\label{tab:rus-la}
\end{subtable}

\begin{subtable}{\textwidth}
\centering
\begin{tabular}{|c|c|c|c|}
  \hline
  ID & Definition & Definition (English) & Score \\
  \hline
$H$ & гильбертов & hilbert & 0.44 \\
$L$ & отображение & transformation & 0.95 \\
$L_k$ & оператор & operator & 0.41 \\
$M$ & воздействие матрица & matrix & 0.71 \\
$T$ & линейный отображение & linear transformation & 0.43 \\
$U$ & унитарный матрица & unitary matrix & 0.71 \\
$V$ & пространство & space  & 0.99 \\
$g$ & билинейный форма & bilinear form & 0.71 \\
$r$ & ранг & rank & 0.46 \\
$x$ & собственный вектор & eigenvector & 0.44 \\
$\Lambda$ & диагональный матрица & diagonal matrix & 0.88 \\
$\Sigma$ & вдоль & ``along'' & 0.44 \\
$\lambda$ & скаляр & scalar & 0.46 \\
$\sigma$ & сингулярный число & singular value & 0.45 \\
\hline
\end{tabular}
\caption{Definitions in the ``Linear Algebra'' namespace.}
\label{tab:rus-la-def}
\end{subtable}

\begin{subtable}{\textwidth}
\centering
\begin{tabular}{|c|c|c|c|c|c|}
  \hline
  \multicolumn{6}{|c|}{$\lambda$}\\
  \hline
  Size & Original name & English name & Original definition & English definition & Score \\
  \hline
3 & Алгебра & Algebra & поль & field & 0.74 \\
5 & Гидродинамика & Fluid dynamics & тепловой движение & thermal motion & 0.42 \\
4 & Гравитация & Gravitation & коэф. затухание & damping coefficient & 0.46 \\
6 & Картография & Cartography & долгота & longitude & 0.98 \\
5 & Линейная алгебра & Linear algebra & скаляр & scalar & 0.46 \\
4 & Оптика & Optics & длина & length & 0.88 \\
3 & Оптика & Optics & длина волна & wavelength & 0.44 \\
5 & \begin{tabular}[x]{@{}c@{}} Релятивистские и \\ гравитационные явления \end{tabular} & \begin{tabular}[x]{@{}c@{}} Relativity and \\ gravitation \end{tabular} & частота & frequency & 0.42 \\
3 & Статистическая физика & Statistical physics & итоговый выражение & final expression & 0.42 \\
3 & \begin{tabular}[x]{@{}c@{}} Теоремы \\ комплексного анализа \end{tabular}  & \begin{tabular}[x]{@{}c@{}} Theorems of \\ complex analysis \end{tabular}  & нуль порядок & ``zero order'' & 0.45 \\
3 & Теория алгоритмов & Algorithms & функция переход & transition function & 0.89 \\
5 & Физические науки & Physical sciences & длина & length & 0.43 \\
  \hline
\end{tabular}
\caption{Definitions of ``$\lambda$'' across all namespaces.}
\label{tab:rus-lambda}
\end{subtable}
\caption{Namespaces extracted from Russian wikipedia.}
\label{tab:rus-namespaces}
\end{table}

\subsection{Building Hierarchy} \label{sec:hierarchy}

After the namespaces are found, we need to organize them into a hierarchical
structure. It is hard to do automatically, and we choose to use
existing hierarchies for mathematical knowledge, and then map the
found namespaces to these hierarchies.

The first hierarchy that we use is ``Mathematics Subject Classification'' (MSC)
hierarchy \cite{ams2010msc} by the American Mathematical Society, and it
is used for categorizing mathematical articles. In this scheme there are
64 top-level categories such as ``Mathematical logic'', ``Number theory'',
or ``Fourier analysis''. It also includes some physics categories such
as ``Fluid mechanics'' or ``Quantum Theory''. The following top level
categories are excluded: ``General'', ``History and biography'' and
``Mathematics education''.

Each top-level category contains second-level categories and third-level
categories. In this work we exclude all subcategories those code
ends with 99: they are usually ``Miscellaneous topics'' or
``None of the above, but in this section''.

Additionally, we excluded the following second level categories because
they interfere with PACS, a different hierarchy for Physics:

\begin{itemize}
\item Quantum theory $\to$ Axiomatics, foundations, philosophy
\item Quantum theory $\to$ Applications to specific physical systems
\item Quantum theory $\to$ Groups and algebras in quantum theory
\item Partial differential equations $\to$ Equations of mathematical physics and other areas of application
\end{itemize}


The second hierarchy is ``Physics and Astronomy Classification Scheme'' (PACS)
\cite{aps2010pacs}, which is a scheme for categorizing articles about Physics.
Like in MSC, we remove the top-level category  ``GENERAL''.

Finally, we also use the ACM Classification Scheme \cite{rous2012acm}
available as a SKOS \cite{miles2005skos} ontology at their
website\footnote{\url{https://www.acm.org/about/class/2012}}. The SKOS ontology graph was
processed with RDFLib \cite{rdflib}. We use the following top level categories:
``Hardware'', ``Computer systems organization'', ``Networks'',
``Software and its engineering'', ``Theory of computation'', ``Information systems'',
``Security and privacy'', ``Human-centered computing'', ``Computing methodologies''.

After obtaining and processing the data, the three hierarchies
are merged into one.

However these categories are only good for English articles and
a different hierarchy is needed for Russian. One of such hierarchies is
``Госу\-дар\-ствен\-ный руб\-ри\-ка\-тор научно-тех\-ни\-чес\-кой инфор\-ма\-ции''
(ГРНТИ)~-- ``State categorizator of scientific and technical information'', which
is a state-recommended scheme for categorizing scientific articles published
in Russian \cite{feodosimov2000grnti}. The hierarchy  is extracted from the
official website\footnote{\url{http://grnti.ru/}}. It provides a very general categorization
and therefore we keep only the following math-related categories:
``Астрономия'' (``Astronomy''), ``Биология'' (``Biology''),
``Информатика'' (``Informatics''), ``Математика'' (``Mathematics''),
``Механика'' (``Mechanics''), ``Ста\-тис\-тика'' (``Statistics''),
``Физика'' (``Physics''), ``Химия'' (``Chemistry''),
``Экономика. Экономические Науки'' (``Economics'') and others.

Once the hierarchy is established, each found namespace is mapped to
the most suitable second-level category. This is done by keywords matching.
First, we extract all key words from the category, which includes
top level category name, subcategory name and all third level categories.
Then we also extract the category information from the namespace, but
we also use the names of the articles that form the namespace.
Finally, the keyword matching is done by using the cosine similarity
between the cluster and each category. The namespace is assigned to the
category with the best (largest) cosine score.

If the cosine score is low (below $0.2$) or there is only one
keyword matched, then the cluster is assigned to the ``OTHERS''
category.

For example, consider a namespace  derived from the cluster consisting of
``Tautology (logic)'', ``List of logic systems'', ``Regular modal logic''
``Combinational logic'' documents. Among others, these articles belong to categories
``Mathematical logic'' and ``Logic''. Then the following is the list of keywords
extracted from the cluster:
``tautology'', ``logic'', ``list'', ``systems'', ``regular'', ``modal'', ``combinational'',
``logical'', ``expressions'', ``formal'', ``propositional'', ``calculus'' and so on.
Apparently, this namespace is about mathematical logic.

Then consider a list of keywords for ``'General logic'', a subcategory of
``Mathematical logic and foundations'' from MSC:
``mathematical'', ``logic'', ``foundations'', ``general'', ``classical'', ``propositional'', ``type'', ``subsystems'' and others.

These keywords are represented as vectors in a vector space and the cosine score
between these vectors is calculated. For this example, the cosine is
approximately 0.75, and this is the largest similarity, and therefore this namespace
is mapped to the ``General logic'' subcategory.

\ \\

Let us consider the namespaces discovered from the English Wikipedia.
The majority of namespaces are mapped correctly. For example:

\begin{small}
\begin{itemize}
  \item ATOMIC AND MOLECULAR PHYSICS
    \begin{itemize}
      \item Atomic properties and interactions with photons (wiki: Atomic physics; Quantum mechanics; Atomic, molecular, and optical physics)
    \end{itemize}

  \item Algebraic geometry
    \begin{itemize}
      \item Computational aspects in algebraic geometry (wiki: Sheaf theory, Theorems in geometry, Theorems in algebraic geometry
      \item Computational aspects in algebraic geometry (wiki: Algebraic geometry, Algebraic varieties, Manifolds)
      \item Surfaces and higher-dimensional varieties (wiki: Algebraic varieties, Threefolds, Surfaces
    \end{itemize}

  \item Algebraic topology
    \begin{itemize}
      \item Fiber spaces and bundles (wiki: Differential geometry, Fiber bundles, Topology)
      \item Spectral sequences (wiki: Abstract algebra, Homological algebra, Algebraic topology)
      \item Applied homological algebra and category theory (wiki: Continuous mappings, Algebraic topology, Homotopy theory)
    \end{itemize}

  \item Biology and other natural sciences
    \begin{itemize}
      \item Mathematical biology in general (wiki: Evidence-based practices, Public health, Epidemiology)
      \item Genetics and population dynamics (wiki: Population genetics, Genetics, Subfields and areas of study related to evolutionary biology)
    \end{itemize}

  \item  Computing methodologies
    \begin{itemize}
      \item Machine learning (wiki: Machine learning, Learning, Artificial intelligence)
      \item Machine learning (wiki: Statistical data types, Multivariate statistics, Statistical classification)
    \end{itemize}

  \item Information systems
    \begin{itemize}
      \item Data management systems (wiki: Databases, Information technology management, Computer data)
    \end{itemize}
\end{itemize}
\end{small}

However, some of the mapped namespaces are only partially accurate:

\begin{small}
\begin{itemize}
  \item Computer systems organization
    \begin{itemize}
      \item Real-time systems (wiki: Computer languages, Type systems, Data types;
            matched keywords: languages computer systems architecture)
    \end{itemize}

\item  Fluid mechanics
    \begin{itemize}
      \item Biological fluid mechanics (wiki: Fluid mechanics, Soft matter, Gases;
             matched keywords: mechanics fluid)
      \item Biological fluid mechanics
             (wiki: Fluid dynamics, Fluid mechanics, Computational fluid dynamics;
             matched keywords: mechanics fluid)
    \end{itemize}
\item Functional analysis
    \begin{itemize}
      \item Distributions, generalized functions, distribution spaces
             (wiki: Probability distributions, Exponential family distributions, Continuous distributions;
             matched keywords: analytic distribution distributions generalized)
    \end{itemize}
\item $K$-theory
    \begin{itemize}
      \item Whitehead groups and $K_1$
             (wiki: Group theory, Subgroup properties, Metric geometry;
             matched keywords: group subgroup)
    \end{itemize}

\item Partial differential equations
    \begin{itemize}
      \item Close-to-elliptic equations (wiki: Differential equations, Numerical analysis, Numerical differential equations;
             matched keywords: partial differential equations)
    \end{itemize}
\end{itemize}
\end{small}

Finally, namespaces marked as ``OTHER'' are usually matched incorrectly:

\begin{small}
\begin{itemize}
  \item OTHER
    \begin{itemize}
   \item Randomness, geometry and discrete structures
             (wiki: Coordinate systems, Cartography, Cartographic projections;
             matched keywords: projections)
   \item Other generalizations
             (wiki: Electrostatics, Concepts in physics, Electromagnetism;
             matched keywords: potential)
   \item Computational methods
             (wiki: General relativity, Exact solutions in general relativity, Equations;
             matched keywords: relativity)
   \item Other classes of algebras
             (wiki: International sailing classes, Keelboats, Olympic sailboat classes;
             matched keywords: classes)
  \end{itemize}
\end{itemize}
\end{small}

\ \\

For the Russian version of Wikipedia, the majority of namespaces are also mapped correctly.
For example:

\begin{small}
\begin{itemize}

\item АСТРОНОМИЯ (ASTRONOMY)
    \begin{itemize}
      \item Звезды (Stars)
           (wiki: Физические науки (Physical sciences), Астрофизика (Astrophysics), Астрономия (Astronomy))
    \end{itemize}

\item ГЕОФИЗИКА (GEOPHYSICS)
    \begin{itemize}
      \item Океанология (Oceanology) (wiki: Океанология (Oceanology), Физическая география (Physical geography), Гидрология (Hydrology))
    \end{itemize}

\item КИБЕРНЕТИКА (CYBERNETICS)
    \begin{itemize}
      \item Теория информации (Information theory) (wiki: Цифровые системы (Digital systems), Теория информации (Information theory), Теория кодирования (Coding theory))
      \item Теория конечных автоматов и формальных языков (Finite state automata and formal languages) (wiki: Теория алгоритмов (Algorithmic theory), Теория автоматов (Automata theory), Визуализация данных (Data visualization))
    \end{itemize}

\item МАТЕМАТИКА (MATHEMATICS)
    \begin{itemize}
      \item Математический анализ (Mathematical analysis) (wiki: Математический анализ (Mathematical analysis), Разделы математики (Parts of mathematics), Функциональный анализ (Functional analysis))
      \item Теория вероятностей и математическая статистика (Probability and statistics) (wiki: Теория вероятностей (Probability), Теория вероятностей и математическая статистика (Probability and statistics), Теория меры (Measure theory))
      \item Основания математики и математическая логика (Foundation of mathematics and mathematica logics) (wiki: Логика (Logic), Разделы математики (Parts of mathematics), Математика (Mathematics))
      \item Алгебра (Algebra) (wiki: Теория колец (Rings Theory), Теория полей (Fields Theory), Теория групп (Groups Theory))
    \end{itemize}

\item МЕТРОЛОГИЯ (METROLOGY)
    \begin{itemize}
      \item Измерения отдельных величин и характеристик (Measurements of individual values and characteristics) (wiki: Единицы измерения (Units of measure), Системы мер (Measure systems), Макс Планк (Max Plank))
    \end{itemize}

\item ФИЗИКА (PHYSICS)
    \begin{itemize}
      \item Физика элементарных частиц. Теория полей. Физика высоких энергий (Particle physics. Field theory. High-energy physics) (wiki: Гравитация (Gravity), Классическая механика (Classical mechanics), Классическая физика (Classical physics))
      \item Физика твердых тел (Physics of solids) (wiki: Физика конденсированного состояния (Condensed matter physics), Кристаллография (Crystallography), Физика твёрдого тела (Physics of solids))
      \item Оптика (Optics) (wiki: Оптика (Optics), Физические науки (Physical sciences), Методы экспериментальной физики (Methods of experimental physics))
    \end{itemize}

\end{itemize}
\end{small}

\subsection{Evaluation Summary} \label{sec:evaluation-summary}

The best definition embedding technique is soft association.
The best clustering algorithm is $K$-Means with $K=9500$
on the semantic space produced by rank-reduced SVD with $k = 250$
with TF-IDF weight where TF is sublinear.

We used ``$E = mc^2$'' as a motivating example for namespace discovery.
When namespaces are discovered, we can look at how the identifiers
``$E$'', ``$m$'', ``$c$'' are used in different namespaces. Additionally,
we can look at other common identifiers such as ``$\lambda$'', ``$\sigma$'' and
``$\mu$'' (see table~\ref{tab:def-en}). We see that definitions of
``$E$'', ``$m$'', ``$c$'' are correct for the ``General relativity'' namespace,
and ``$E$'' is ``expectation'' for the ``Probability'' namespace, however
we have not observed that ``$E$'' is ``Elimination matrix'' for the ``Linear
algebra'' namespaces, but it is just ``matrix''.

We also repeat this experiment for the Russian Wikipedia (table~\ref{tab:def-ru}), but
chose slightly different namespaces.
Similarly to English Wikipedia, ``$E$'', ``$m$'', ``$c$'' are discover
correctly, but nothing is discovered for ``$E$'' in namespaces about probability
and linear algebra.

\begin{table}[h!]
\centering
\makebox[\textwidth][c]{\begin{tabular}{|x{2cm}|x{2cm}|x{2cm}|x{2cm}|x{2cm}|x{2cm}|x{2cm}|}
\hline													
	&	$E$	&	$m$	&	$c$	&	$\lambda$	&	$\sigma$	&	$\mu$	\\
\hline													
Linear algebra	&	matrix	&	matrix	&	scalar	&	eigenvalue	&	related permutation	&	algebraic multiplicity	\\
\hline
General relativity	&	energy	&	mass	&	speed of light	&	length	&	shear	&	reduced mass	\\
\hline
Coding theory	&	encoding function	&	message	&	transmitted codeword	&		&	natural isomorphisms	&		 \\
\hline
Optics	&		&	order fringe	&	speed of light in vacuum	&	wavelength	&	conductivity	&	 permeability	\\
\hline
Probability	&	expectation	&	sample size	&		&	affine parameter	&	variance	&	mean vector	\\
\hline													
\end{tabular}}
\caption{Definitions for selected identifiers and namespaces extracted
from the English Wikipedia.}
\label{tab:def-en}
\end{table}

\begin{table}[h!]
\centering
\makebox[\textwidth][c]{\begin{tabular}{|x{3cm}|x{1.8cm}|x{1.8cm}|x{1.8cm}|x{1.8cm}|x{1.8cm}|x{1.8cm}|}
\hline													
	&	$E$	&	$m$	&	$c$	&	$\lambda$	&	$\sigma$	&	$\mu$	\\
\hline													
Линейная алгебра (Linear algebra)	&		&	смена базис (change of basis)	&		&	скаляр (scalar)	&	
сингулярный число (singular value)	&		\\
\hline													
Физические науки (Physical sciences)	&	энергия (energy)	&	масса частицы (mass of particle)	&	 скорость свет (speed of light)	&	длина (length)	&		&	привести масса (reduced mass)	\\
\hline													
Вероятность (Probability)	&		&		&		&		&	дисперсия (variance)	&	среднее (mean)	\\
\hline													
\end{tabular}}
\caption{Definitions for selected identifiers and namespaces extracted
from the Russian Wikipedia.}
\label{tab:def-ru}
\end{table}

To visualize the discovered namespaces, we first map them to a hierarchy
as previously described in section~\ref{sec:hierarchy}, and then find the
most frequent categories according to this hierarchy. Then, for each
category, we rank all discovered identifier-definitions pairs, and show
only the most highly ranked ones
(see table~\ref{tab:top-namespaces}). Additionally, we show the most
frequent Wikipedia categories that the documents inside the namespaces
have, and also the most influential documents: the ones that contain
more identifiers than others. We also repeat the same for Russian Wikipedia
(see table~\ref{tab:top-namespaces-rus}).

We can note that the top discovered namespaces are quite different
across the two datasets, and there is no single namespace among the top namespaces
that both datasets share. However, ``Group theory and generalizations''
and ``Алгебра'' (``Algebra'') look a little similar and share a few identifiers.
It is probably due to the fact that the datasets may be different in the content
and have different distribution of categories. Also, the hierarchies matter
as well: in case of Russian, the hierarchy is more general, and therefore
the matched categories tend to be more general as well.

\begin{table}
\makebox[\textwidth][c]{\begin{tabular}{|c|x{3cm}|x{4cm}|x{3.5cm}|x{4cm}|}
		\hline							
Freq.	&	Namespaces	&	Definitions	&	Categories	&	Top Articles	\\
		\hline							
10	&	Physics: Fluid mechanics	&	$\rho$: density,
$p$: pressure,
$g$: acceleration,
$k$: wavenumber,
$u$: velocity,
$v$: velocity,
$\eta$: free surface,
$\omega$: angular frequency,
$z$: free surface,
$\nu$: kinematic viscosity	&	Fluid dynamics;
Fluid mechanics;
Dynamics;
Aerodynamics;
Partial differential equations	&	Navier–Stokes equations;
Stokes wave;
Airy wave theory;
Mild-slope equation;
Classic energy problem in open-channel flow	\\
		\hline							
9	&	Algebra: Differential and difference algebra	&	$R$: ring,
$k$: field,
$D$: city,
$K$: field,
$x$: polynomials,
$\Gamma$: value group,
$M$: submodule,
$n$: matrix,
$S$: ring,
$v$: valuation	&	Abstract algebra;
Algebra;
Functions and mappings;
Polynomials;
Analytic functions	&	Recurrence relation;
Spectral radius;
Levi-Civita symbol;
Perron–Frobenius theorem;
Leibniz formula for determinants	\\
		\hline							
9	&	Mathematical analysis: Partial differential equations	&	$\Omega$: domain,
$t$: time,
$L$: space,
$p$: space,
$\omega$: angular frequency,
$V$: hilbert space,
$D$: domain,
$u$: horizontal velocity,
$x$: time,
$U$: velocity profiles	&	Partial differential equations;
Differential equations;
Multivariable calculus;
Mathematical analysis;
Fluid dynamics	&	Orr–Sommerfeld equation;
Helmholtz equation;
Fictitious domain method;
Green's function for the three-variable Laplace equation;
Eikonal equation	\\
		\hline							
8	&	Economics: Mathematical economics	&	$K$: strike,
$P$: price level,
$p$: probability,
$u$: utility,
$V$: money,
$M$: money,
$\alpha$: confidence level,
$g$: distortion function,
$\sigma$: volatility,
$R_f$: return	&	Economics;
Financial economics;
Microeconomics;
Mathematical finance;
Economic theories	&	Modern portfolio theory;
Lookback option;
Binary option;
Equation of exchange;
Slutsky equation	\\
		\hline							
8	&	Probability theory	&	$X$: process,
$t$: time,
$X_t$: process,
$P$: probability measure,
$S$: state space,
$s$: stochastic processe,
$f$: measurable function,
$M$: local martingale,
$M_t$: local martingale,
$W_t$: process	&	Stochastic processes;
Probability theory;
Statistics;
Statistical data types;
Measure theory	&	Wiener process;
It\={o} calculus;
Local martingale;
Stratonovich integral;
Glivenko–Cantelli theorem	\\
		\hline							
7	&	Algebra: Group theory and generalizations	&	$G$: group,
$H$: subgroup,
$Z$: group,
$K$: subgroup,
$N$: group,
$p$: power,
$n$: root,
$F$: free group,
$Q$: extension,
$T$: homomorphism	&	Group theory;
Abstract algebra;
Metric geometry;
Algebraic structures;
Theorems in group theory	&	Free abelian group;
Center (group theory);
Holomorph (mathematics);
$P$-group;
Powerful $p$-group	\\
		\hline							
7	&	Mathematical logic: General logic	&	$w$: world,
$R$: binary relation,
$P$: predicate,
$Q$: statement,
$W$: nodes,
$\phi$: modal formula,
$n$: natural number,
$v$: world,
$T$: relation,
$k$: degree	&	Mathematical logic;
Logic;
Proof theory;
Syntax (logic);
Formal systems	&	Sequent calculus;
First-order logic;
Original proof of G\"odel's completeness theorem;
Kripke semantics;
Szpilrajn extension theorem	\\
		\hline							
\end{tabular}}
\caption{Most frequent definitions in most frequent namespaces extracted from
the English Wikipedia.}
\label{tab:top-namespaces}
\end{table}

\newgeometry{bottom=2cm}
\begin{table}
\makebox[\textwidth][c]{\begin{tabular}{|c|x{3cm}|x{5.2cm}|x{4.2cm}|x{4cm}|}
		\hline							
Freq.	&	Namespaces	&	Definitions	&	Categories	&	Top Articles	\\
		\hline							
10	&	Алгебра (Algebra)	&	$G$: группа (group),
$V$: пространство (space),
$p$: простой число (prime number),
$K$: подпространство (subspace),
$R$: кольцо (ring),
$C$: категория (category),
$n$: ``o ( n )'',
$F$: поль (field),
$L$: линейный пространство (linear space),
$D$: категория (category)	&	Действие группы (Group action),
Линейная алгебра (Linear algebra),
Векторное пространство (Vector space),
Алгебра (Algebra),
Нормированное пространств (Normed vector space)	&	Общая алгебра (General algebra),
Алгебра (algebra),
Теория групп (Group theory),
Линейная алгебра (Linear algebra),
Теория колец (Ring theory)	\\
\hline									
7	&	Топология (Topology)	&	$M$: многообразие (manifold),
$X$: топологический пространство (topologic space),
$n$: многообразие (manifold),
$U$: окрестность (neighborhood),
$C$: класс (class),
$S$: пучок (sheaf),
$x$: пространство (space),
$k$: ранг (rank),
$V$: слой (layer),
$f$: степень отображение (degree of mapping)	&	Топология (Topology);
Дифференциальная геометрия и топология (Differential geometry and topology);
Геометрия (Geometry);
Общая топология (General topology);
Общая алгебра (General algebra)	&	Параллельное поле (Parallel field);
Алгебраическая топология (Algebraic topology);
Векторное расслоение (Vector bundle);
Когомологии де Рама (De Rham cohomologies);
Структура (дифференциальная геометрия) (Structure -- differential geometry)	\\
\hline									
6	&	Теория вероятностей и математическая статистика (Probability and statistics)	&	$\Omega$: элементарный событие (event),
$P$: вероятность (probability),
$F$: алгебра событие (event algebra),
$X$: случайный величина (random variable),
$\omega$: множество элемент (element of set),
$g$: интегрировать функция (integrable function),
$n$: стремление (convergence),
$N$: счётный мера (countable measure),
$\sigma$: событие (event),
$p$: момент (moment)	&	Теория вероятностей (Probability);
Теория вероятностей и математическая статистика (Probability and statistics);
Теория меры (Measure theory);
Математические теоремы (Mathematical theorems);
Теоремы теории вероятностей и математической статистики (Theorems of probability and statistics)	&	 Случайная величина (Random variable);
Аксиоматика Колмогорова (Kolmogorov axioms);
Пространство элементарных событий (Sample space);
Теорема Лебега о мажорируемой сходимости (Lebesgue's dominated convergence theorem);
$t$-Критерий Стьюдента (Student's $t$-test)	\\
\hline									
5	&	Физика элементарных частиц. Теория полей. Физика высоких энергий (Particle physics. Field theory. High energy physics)	&	$G$: гравитационный постоянный (gravitation constant),
$M$: масса (mass),
$c$: скорость свет (speed of light),
$m_2$: масса (mass),
$m$: масса (mass),
$\Psi$: волновой функция (wave function),
$m_1$: материальный точка масса (mass of material point),
$t$: время (time),
$r$: расстояние (distance),
$R$: масса (mass)	&	Гравитация (Gravitation);
Астрономия (Astronomy);
Общая теория относительности (General relativity);
Теория относительности (Relativity theory);
Физическая космология (Physical cosmology)	&	Чёрная дыра (Black hole);
Гравитационное красное смещение (Gravitational redshift);
Квантовый компьютер (Quantum computer);
Метрика Шварцшильда (Schwarzschild metric);
Гравитация (Gravitation)	\\
\hline									
5	&	Оптика (Optics)	&	$\Phi_0$: поток излучение (Radiant flux),
$\lambda$: длина (length),
$\Phi$: поток излучение (Radiant flux),
$l$: расстояние (distance),
$\varepsilon$: приёмник (``receiver''),
$n_1$: показатель преломление (refractive index),
$K_m$: световой эффективность излучение (luminous efficacy),
$n$: показатель преломление (refractive index),
$n_2$: преломление среда (``refractive environment''),
$r$: рассеяние (dispersion)	&	Оптика (Optics);
Физические науки (Physical sciences);
Волновая физика (Wave mechanics);
Фотометрия (Photometry);
Методы экспериментальной физики (Methods of experimental physics)	&	Энергия излучения (оптика) (Radiant energy);
Облучённость (фотометрия) (Irradiance);
Фотонный кристалл (Photonic crystal);
Свет (Light);
Сила излучения (фотометрия) (Radiant intensity) \\
\hline									
\end{tabular}}
\caption{Most frequent definitions in most frequent namespaces extracted from
the Russian Wikipedia.}
\label{tab:top-namespaces-rus}
\end{table}
\restoregeometry

\section{Conclusions} \label{sec:conclusions}

The goal of this work was to discover namespaces in mathematical notation
given a collection of documents with mathematical formulae. 
This problem could not be performed manually: this task it too time consuming
and requires a lot of effort. 

To achieve the goal we proposed an automatic method based on cluster analysis. 
We noted that document representation in terms of identifiers is similar 
to the classic Vector Space Model. This allowed us to apply traditional 
document clustering techniques to the namespace discovery problem. 

We expected to discover namespaces, that are homogenies and corresponded
to the same area of knowledge. The clusters that we discovered are homogenous, 
but not all corresponded to the same category. This is why we 
additionally used the category information to recognize the namespaces-defining
clusters amount all clusters, and then we built namespaces from them.

We also initially expected that there would be more namespace-defining clusters, 
but in our results the majority of clusters are not ``pure'': documents inside these 
clusters do not belong to the same category. These clusters are only homogenous in 
the cluster analysis sense: the within-cluster distances is minimal. 

To prove that namespace discovery is possible in principle, we first applied 
the proposed  method to a dataset where the ``gold standard'' is known: to Java source code,
and we were able to partially recover the namespaces using only the information
about identifiers. 



Then we used the method to extract namespaces from the English Wikipedia,
and we were able to discover 414 namepaces from this dataset. This result 
is better than random guessing by the factor of ten. 
We observed that dimensionality reduction techniques
are very helpful, and clustering algorithms work better on the reduce space. 
MiniBatch $K$-Means algorithms shows the best results for discovering 
namespace-defining clusters.

Additionally, we applied the same method to the Russian version of Wikipedia, 
and, although the distribution of most frequent namespaces are different 
in these two datasets, the overall results are consistent.

The problem of namespace discovery has not been studied before, and there was 
no dataset where identifiers were assigned to namespaces. 
In this work we showed that the automatic namespace discovery is possible,
and it is a good start. 

However, there are many ways in which the present approach can be improved further.
In the next section we discuss possible directions.

\section{Outlook and Future Work} \label{sec:future-work}

\subsection{Implementation and Other Algorithms}  

We use the Probabilistic approach to extracting definitions for identifiers, and
it is good because it requires almost no parameter tuning.
While this approach works well most of the time, sometimes we observe
some false positives, most likely due to the fact that the dataset is
quite noisy. It potentially can be improved by using some Machine Learning
method, which, however, may require creating a hand-labeled dataset
with identifier-definition relations. To facilitate the creation of such a
dataset it is possible to pre-generate some data using using the current approach
for further labeling.

In the experiments section we have observed that cluster algorithms that produce
many clusters tend to have good performance. However, they also tend to create related
clusters from the same category and with same or similar identifiers and
definitions. Therefore such results can be refined further and merged.
This can be done, for example, by using the join operation from
the Scatter/Gather algorithm \cite{cutting1992scatter}, which finds the most
similar clusters and merges them.

We were not able to apply hierarchical agglomerative clustering algorithms because
their time complexity is prohibitive, but they may produce good clusters.
For these algorithms we are usually interested in the nearest neighbors of a given
data point, and therefore we can use approximation algorithms for computing nearest
neighbors such as Locality-Sensitive Hashing (LSH) \cite{leskovec2014mining}.
The LSH algorithms can be used for text clustering \cite{ravichandran2005randomized},
and therefore they should work well for identifiers. Additionally, LSH is also a
dimensionality reduction technique, and we have observed that generally
reducing dimensionality helps to obtain better clusters.

In this work we use hard assignment clustering algorithms, which means, that a document
can import only from one namespace. This assumption does not necessarily always hold true
and we may model the fact that documents may import from several namespaces by
using Fuzzy Clustering (or Soft Clustering) algorithms \cite{baraldi1999survey}.

In Latent Semantic Analysis other dimensionality reduction techniques
can be used, for example, Local Non-Negative Matrix Factorization \cite{li2001learning}.
There is also a randomized Non-Negative Matrix Factorization algorithm that uses
random projections \cite{wang2010efficient} \cite{damle2014random},
which potentially can give a speed up while not significantly losing
in performance. Another dimensionality reduction technique useful for
discovering semantics is Dynamic Auto-Encoders \cite{mirowski2010dynamic}.

Additionally, we can try different approaches to clustering such as
Spectral Clustering \cite{ng2002spectral} or Micro-Clustering \cite{uno2015micro}.

Finally, topic modeling techniques such as Latent Dirichlet Allocation
\cite{blei2003latent} can be quite useful for modeling namespaces. It can be
seen as a ``soft clustering'' technique and it can naturally model the fact that
a document may import from several namespaces.

\subsection{Other Concepts} 

In this work we assume that document can import only from one namespace,
but in reality is should be able to import from several namespaces. As discussed,
it can be modeled by Fuzzy Clustering. But it also can be achieved by
dividing the document in parts (for example, by paragraphs)
and then treating each part as an independent document.

For document clustering we only use identifiers, extracted definitions
and categories. It is possible to take advantage of additional information from
Wikipedia articles. For example, extract some keywords from the articles
and use them to get a better cluster assignment.

The Wikipedia data set can be seen as a graph, where two articles have
an edge if there is an interwiki link between them. Pages that describe
certain namespaces may be quite interconnected, and using this idea it is possible
to apply link-based clustering methods (such as ones described in
\cite{botafogo1991identifying} and \cite{johnson1996adaptive}) to find namespace
candidates. There are also hybrid approaches that can use both textual representation
and links \cite{oikonomakou2005review}.

Vector Space Model is not the only possible model to represent textual
information as vectors. There are other ways to embed textual information
into vector spaces like word2vec \cite{mikolov2013efficient} or
GloVe \cite{pennington2014glove}, and these methods may be useful
for representing identifers and definitions as well.

Tensors may be a better way of representing
identifier-definition pairs. For example, we can represent the data set
as a 3-dimensional tensor indexed by documents, identifiers and definition.
Tensor Factorization methods for revealing semantic information
are an active area of research in NLP and linguistics \cite{anisimov2014semantic},
so it is also possible to apply these methods to the namespace discovery problem.

Finally, while running experiments, we observed that sometimes results of
clustering algorithms with the same parameters produce quite different
results, or some algorithms produce a small amount of good quality namespaces,
while others produce many namespaces which may be less coherent.
Therefore it can be interesting to investigate how to combine the results
of different cluster assignments such that the combined result is better
in terms of the number of namespace-defining clusters. One way of
achieving this can be building ensembles of clustering algorithms \cite{strehl2003cluster}.
Alternatively, a special approach for optimizing for the number of pure clusters
can be proposed, for example, partially based on the ideas from
Boosting \cite{freund1996experiments}: apply a clustering algorithm,
remove the discovered pure clusters, and run the algorithm again on the remaining
documents until no new clusters are discovered.

\subsection{Other Datasets} 

In this work we use Wikipedia as the data source and extract namespaces from the
English part of Wikipedia. Additionally, we also apply the methods to the Russian
part, and therefore it shows that it is possible to extract namespaces from
Wikipedia in any other available language.

But we can also apply to some other larger dataset, such as arXiv\footnote{\url{http://arxiv.org/}}, a repository of over one million
of scientific papers in many different areas. The source code of these
articles are available in \LaTeX, and it can be processed automatically.

There are many scientific Q\&A websites on the Internet. The stack
exchange\footnote{\url{http://stackexchange.com/}} is one of the largest Q\&A networks,
and there are many sites on this network that contain mathematical formulae, such as
``mathematics'', ``mathoverflow'', ``cross validated'', ``data science'',
``theoretical computer science'', ``physics'', ``astronomy'', ``economics'' and many others.
This network makes their data available for download and it also can be a good
data source for namespace discovery. In addition to content, the questions contain
a lot of potentially useful metadata such as related questions and tags.

\subsection{Unsolved Questions} 

The most important question is how to extend this method to situations when
no additional information about document category is known. To solve
it, we need to replace the notion of purity with some other objective
for discovering namespace-defining clusters.

Also, a metric for evaluating the quality of a namespace is needed.
Now we assume that pure clusters are namespace-defining clusters. But the namespace
candidates should adhere to the namespace definition as much as possible,
and therefore a good criteria is needed to quantify to what extent the definition
is satisfied. This will help to define whether a cluster defines a good namespace
or not.

After namespaces are discovered we organize them into hierarchies.
To do that we use existing hierarchies, but they are not always complete
and there are mismatches. What is more, when this technique is applied to some
other language, a different hierarchy is needed for this language, and we experienced
it when processing the Russian part of Wikipedia: for that we needed to obtain
a special hierarchy. There should be a way of building these hierarchies
automatically, without the need of external dataset.
Potentially it should be possible to use hierarchical clustering
algorithms, but it may result in very deep and unnatural hierarchies, and
therefore some additional investigation in this direction may be needed.

\section{Bibliography}

\bibliographystyle{unsrt}
\bibliography{bibliography}

\end{document}